\documentclass[cernpreprint,texlive=2011,txfonts,UKenglish]{latex/atlasdoc} 
\usepackage[subfigure=true,lineno=false,biblatex=false]{latex/atlaspackage} 
\usepackage[BSM]{latex/atlasphysics}
\usepackage{cite}
\graphicspath{{logos/}{figures/}}
\usepackage{subfigure}
\usepackage{mathrsfs}
\usepackage{placeins}
\usepackage{multirow}
\usepackage{accents}
\PreprintIdNumber{CERN-PH-EP-2015-090}
\newcommand{\kt}{\ensuremath{k_{t}}}
\newcommand{\dsplit}{\ensuremath{\sqrt{d_{12}}}}

\newcommand{\mttbar}{\ensuremath{m_{\ttbar}}}
\newcommand{\mttbarreco}{\ensuremath{m_{\ttbar}^{\mathrm{reco}}}}

\newcommand{\Jimmy}{{\sc Jimmy}}
\newcommand{\Pythia}{{\sc Pythia}}
\newcommand{\Herwig}{{\sc Herwig}}
\newcommand{\AcerMC}{{\sc AcerMC}}
\newcommand{\Powheg}{{\sc Powheg-Box}}
\newcommand{\Geant}{{\sc Geant}}
\newcommand{\Madgraph}{{\sc MadGraph}}
\newcommand{\Alpgen}{{\sc Alpgen}}
\newcommand{\Sherpa}{{\sc Sherpa}}
\newcommand{\MCNLO}{{\sc MC@NLO}}
\newcommand{\aMCNLO}{{\sc MadGraph\_aMC@NLO}}

\newcommand{\Zpxseclow}{4.2~pb}
\newcommand{\Zpxsechigh}{0.03~pb}
\newcommand{\kkgxseclow}{4.8~pb}
\newcommand{\kkgxsechigh}{0.09~pb}
\newcommand{\kkGxseclow}{2.5~pb}
\newcommand{\kkGxsechigh}{0.03~pb}
\newcommand{\scalarxseclow}{3.0~pb}
\newcommand{\scalarxsechigh}{0.03~pb}

\newcommand{\ZpexcludedMass}{1.8~\tev}
\newcommand{\ZpexcludedMassExpected}{2.0~\tev}
\newcommand{\kkgexcludedMass}{2.2~\tev}
\newcommand{\kkgexcludedMassExpected}{2.3~\tev}

\newcommand{\totlumi}{20.3\,\mbox{fb$^{-1}$}}
\newcommand{\delayedlumi}{17.4\,\mbox{fb$^{-1}$}}

\newcommand{\NdataResolvedEle}{\ensuremath{114{,}377}} 
\newcommand{\NdataResolvedMuo}{\ensuremath{108{,}953}}
\newcommand{\NdataResolvedTot}{\ensuremath{223{,}330}}

\newcommand{\NdataBoostedEle}{\ensuremath{4{,}148}}
\newcommand{\NdataBoostedMuo}{\ensuremath{4{,}058}}
\newcommand{\NdataBoostedTot}{\ensuremath{8{,}206}}

\AtlasTitle{A search for $\mathbf{t\bar{t}}$ resonances using lepton-plus-jets events  in proton--proton collisions at $\sqrt{s} = 8$~\TeV\ with the ATLAS detector }

\author{The ATLAS Collaboration}

\AtlasJournal{JHEP}
\date{\today}

\AtlasAbstract{
A search for new particles that decay into top quark pairs is reported. The search is performed with the ATLAS experiment at the LHC using an integrated luminosity of 
20.3\,fb$^{-1}$ of proton--proton collision data collected at a centre-of-mass energy of \mbox{$\sqrt{s}=8$~\TeV}. 
The lepton-plus-jets final state is used, where the top pair decays to $W^+bW^-\bar{b}$, with one $W$ boson decaying leptonically and the other hadronically. 
The invariant mass spectrum of top quark pairs is examined for local excesses or deficits that are inconsistent with the Standard Model predictions. 
No evidence for a top quark pair resonance is found, and 95\% confidence-level limits on the production rate are determined for massive states in benchmark models. 
The upper limits on the cross-section times branching ratio of a narrow \Zprime\ boson decaying to top pairs range from \Zpxseclow\ to \Zpxsechigh\ for resonance masses
 from 0.4 \TeV\  to 3.0~\TeV. 
A narrow leptophobic topcolour \Zprime\ boson with mass below \ZpexcludedMass\ is excluded. 
Upper limits are set on the cross-section times branching ratio for a broad colour-octet resonance with $\Gamma/m =$~15\% decaying to $t\bar{t}$. These range from \kkgxseclow\ to \kkgxsechigh\     for masses from 0.4\,TeV to 3.0\,TeV. A Kaluza--Klein excitation of the gluon in a Randall--Sundrum model is excluded for masses below \kkgexcludedMass.
}

\begin{document}

\section{Introduction}
\label{sec:introduction}

Many models of physics beyond the Standard Model (SM) predict production at the LHC of additional particles with masses near the TeV scale. This paper presents a search for such heavy particles decaying to top quark pairs ($t\bar{t}$) using data from proton--proton collisions collected at a centre-of-mass energy $\sqrt{s}=8$~TeV with the ATLAS detector, corresponding to an integrated luminosity of \totlumi.

Searches for production of heavy particles that decay to $t\bar{t}$  are of high interest at the LHC due to the  role that the top quark plays
in many models of physics beyond the SM (BSM). The top quark is the 
most massive of the fundamental particles in the SM,
and it can have a large coupling to heavy Higgs bosons. 
Thus, heavy Higgs bosons in two-Higgs-doublet models~\cite{Lee:1973iz,Branco:2011iw} 
can have a large branching ratio to $t\bar{t}$ final states.  Furthermore, many models that propose alternative
mechanisms for electroweak symmetry breaking (EWSB) incorporate new heavy particles with a larger coupling to 
$t\bar{t}$ than to lighter quarks. Examples include strong EWSB models such
as topcolour-assisted technicolour~\cite{Hill:1994hp} (TC2) and Composite Higgs~\cite{Kaplan:1983fs,Kaplan:1983sm,Georgi:1984ef,Banks:1984gj,Georgi:1984af,Dugan:1984hq,Georgi:1985hf,Bellazzini:2014yua} scenarios. 
Models with warped extra dimensions~\cite{Randall:1999ee,Agashe:2003zs,Davoudiasl:1999tf,Pomarol:1999ad} 
form an additional class of models that predict heavy particles that decay to $t\bar{t}$ pairs.
In such models, the heavy particles are the counterparts of the gluon and graviton.

The search starts by selecting events with one isolated charged lepton (electron or muon), missing transverse momentum (whose magnitude is denoted by \met{}) and hadronic jets, which are compatible with
 $t\bar{t} \rightarrow W^{+}bW^{-}\bar{b}$, with one $W$ boson decaying leptonically and the other hadronically.
At least one of the hadronic jets is required to be consistent with having originated from a $b$-quark. 
An estimator of the $t\bar{t}$ invariant mass (\mttbarreco) is constructed
 with the events divided into two orthogonal classes by topology: the {\em{boosted}} topology, where the decay products of the hadronically decaying top quark are expected to be fully enclosed within one large-radius jet, and the {\em{resolved}} topology where four 
small-radius jets are reconstructed and attributed to the $b\bar{b}q\bar{q'}$ quarks. The \mttbarreco\ spectrum is scanned for localised deviations relative to the expectations from the background processes.
The compatibility of the data and expectations is assessed, and limits on
the production cross-section for new particles are set.

This search is designed to be sensitive to the production of any new particle that decays to $t\bar{t}$. 
Nonetheless the selection efficiency and acceptance can differ between particular model choices. 
Hence the sensitivity to a variety of different new particles was evaluated to quantify the performance of the search.
The benchmark models adopted in this search include colour-singlet and colour-octet bosons with spin 0, 1 and 2 and  masses from 0.4 to 3 TeV. 
The resonance width for the specific models varies from very narrow (1\%) to a size similar to that of the experimental resolution ($15\%$). 
Furthermore, the dependence of the limits on the resonance width is explored for heavy gluons up to a width of $40\%$. 
With these results, it is possible to interpret the cross-section limits in the context of a search for other new particles with the same production modes.

In addition to making use of 8\ TeV data, this search incorporates several improvements with respect to the previous most sensitive ATLAS collaboration search for the same signature\cite{Aad:2013nca}, which used a similar strategy. Trimming~\cite{Krohn:2009th} is now used to mitigate pile-up effects on large-radius jets, a new procedure is applied to recover efficiency loss from removal of overlap between electron and jet objects in the detector, and a $\chi^2$ variable is exploited to remove events from non-$t\bar{t}$ background processes. Furthermore, the limits on production of spin 2 bosons and on heavy gluons of varying width represent an extension compared to previous searches by the ATLAS and CMS~\cite{Chatrchyan:2013lca} Collaborations.

\section{Models tested}
\label{sec:theory}

The details of the benchmark models considered in this search are reviewed below.
Interference between these processes and SM $t\bar{t}$ production is not considered in this search.

\subsection{Spin-1 colour singlet}

The first class of models explored produces spin-1 colour-singlet vector bosons, $Z'$. 
This search uses topcolour-assisted technicolour $Z'_{\mathrm{TC2}}$~\cite{Hill:1994hp,topcolor2,Harris:2011ez} as a benchmark. This is a leptophobic boson, with
couplings only to first- and third-generation quarks, referred to as Model IV in ref.~\cite{topcolor2}. The properties of the boson are controlled by three parameters:  $\cot \theta_\mathrm{H}$, which controls the width and the production cross-section, and $f_1$ and $f_2$, which are related to the coupling to up-type and down-type quarks 
respectively. Here $f_1=1$ and $f_2=0$, which maximises the fraction of $Z'_\mathrm{TC2}$ that decay to $t\bar{t}$. 
The parameter $\cot \theta_\mathrm{H}$ is tuned for each mass point such that the resonance has a width of 1.2\% of its mass. 
To account for higher-order contributions to the cross-section, the leading-order calculation is multiplied by a factor of 1.3 based on calculations performed at next-to-leading order (NLO) in QCD~\cite{Gao:2010bb,Caola:2012rs}.

Constraints on $Z'_{\mathrm{TC2}}$ have been set by the CDF~\cite{Aaltonen:2011vi,Aaltonen:2012af} and D0~\cite{Abazov:2011gv} collaborations using data from proton--antiproton collisions at the Tevatron. Previous constraints at the LHC were set using  proton--proton collisions at $\sqrt{s}=7$ TeV with an integrated luminosity of 5\,fb$^{-1}$ by the ATLAS~\cite{Aad:2012raa,Aad:2013nca} and CMS~\cite{Chatrchyan:2012yca,Chatrchyan:2012cx,Chatrchyan:2012ku} collaborations, and using  20\,fb$^{-1}$  of \mbox{$\sqrt{s}=8$\,TeV} data by the CMS Collaboration~\cite{Chatrchyan:2013lca}. For narrow (wide) $Z'_{\mathrm{TC2}}$ of width 1.2\% (10\%) the strongest lower bound on the allowed mass is 2.1\,TeV (2.7\,TeV) from the search performed by CMS at \mbox{$\sqrt{s}=8$\,TeV}.

\subsection{Spin-1 colour octet}

The second class of models considered produces spin-1 colour-octet vector bosons. Specifically, heavy Kaluza--Klein gluons, $g_\mathrm{KK}$, as produced in Randall--Sundrum (RS) models with a single warped extra dimension~\cite{Lillie:2007yh}, are used as a benchmark in this search. In this model, the $g_\mathrm{KK}$ has a nominal width of 15.3\% of its mass. Previous searches using $\sqrt{s}=7$ TeV ATLAS data~\cite{Aad:2013nca} exclude a $g_\mathrm{KK}$ with a mass less than 2.1 TeV. The CMS Collaboration searched for similar resonances~\cite{Chatrchyan:2013lca}, using a slightly different benchmark model~\cite{Agashe:2006hk}. The CMS choice leads to a larger natural width of 20\% and a larger production cross-section,
and, for such a scenario, CMS excludes the existence of a $g_\mathrm{KK}$ with mass
less than 2.5 TeV. 
In the analysis presented here, the sensitivity to the width of the colour octet is also tested for widths from 10\% to 40\% of the resonance mass. 
 
\subsection{Spin-2 colour singlet}

The third class of models explored in this search produces spin-2
colour singlets, such as Kaluza--Klein excitations of the graviton, $G_\mathrm{KK}$.
The search  uses a Randall--Sundrum model with extra dimensions where the SM fields are in the warped bulk and the fermions are localised appropriately to explain the flavour structure of the SM~\cite{Randall:1999ee, Agashe:2007zd, Fitzpatrick:2007qr}. This kind of graviton is commonly referred to as a ``Bulk'' RS graviton and is characterised by a dimensionless coupling constant $k/\bar{M}_{\mathrm{Pl}} \sim 1$, where $k$ is the curvature of the warped extra dimension and $ \bar{M}_{\mathrm{Pl}} = M_{\mathrm{Pl}}/\sqrt{8 \pi}$ is the reduced Planck mass. For such gravitons, decays to light fermions are suppressed, and the branching ratio to photons is negligible. 
The branching ratios to $t\bar{t}$, $WW$, $ZZ$ and $HH$ are significant. In the model used, $k/\bar{M}_{\mathrm{Pl}}$ is chosen to be 1, and the $G_\mathrm{KK}$ width varies from 3\% to 6\% in the  mass range 400--2000 GeV.
The branching ratio of $G_\mathrm{KK}$ decay into a $t\bar{t}$ pair rapidly increases from 18\% to 50\%  between 400 and 600~\GeV, plateauing at 68\% for masses larger than 1~\TeV.
There have been no previous direct searches for such gravitons in the $t\bar{t}$ decay channel.
The ATLAS Collaboration used the same model to explore the $G_\mathrm{KK} \rightarrow ZZ$ channel~\cite{Aad:2014xka} and excluded Bulk RS $G_\mathrm{KK}$ with mass less than 740 GeV. The CMS Collaboration  performed
searches in the $G_\mathrm{KK} \rightarrow ZZ$ and $G_\mathrm{KK} \rightarrow WW$ decay channels~\cite{Khachatryan:2014gha} but 
did not consider the case of Bulk RS gravitons with $k/\bar{M}_{\mathrm{Pl}} > 0.5$.

\subsection{Spin-0 colour singlet}

The last class of models examined here produces  colour-singlet scalar particles via gluon fusion which decay to $t\bar{t}$. 
The approach previously adopted by the CMS Collaboration~\cite{Chatrchyan:2013lca} is followed, 
in which narrow scalar resonance benchmarks are generated while the interference with SM $t\bar{t}$ production is neglected.
Even though such signals with negligible interference are not predicted by any particular BSM model, they can be used to evaluate the experimental sensitivities and set upper limits on the production cross-sections. 
The CMS Collaboration excluded such resonances with production cross-sections greater than  0.8 pb and 0.3 pb for masses of 500 and 750 GeV, respectively.

\section{The ATLAS detector}
\label{sec:atlas}

The ATLAS experiment~\cite{Aad:2008zzm} is a multipurpose particle physics detector with forward-backward symmetric cylindrical geometry.\footnote{ATLAS uses a right-handed coordinate system with its origin at the nominal interaction point (IP) in the centre of the detector and the $z$-axis along the beam pipe. The $x$-axis points from the IP to the centre of the LHC ring, and the $y$-axis points upward.  Cylindrical coordinates $(r,\phi)$ are used in the transverse plane, $\phi$ being the azimuthal angle around the beam pipe. The pseudorapidity is defined in terms of the polar angle $\theta$ as $\eta=-\ln\tan(\theta/2)$. The distance in $\eta$--$\phi$ space is commonly referred to as $\Delta R= \sqrt{(\Delta \phi)^2+(\Delta \eta)^2 }$. }
The inner detector (ID) consists of multiple layers of silicon pixel and microstrip detectors  
and a straw-tube transition radiation tracker and covers a pseudorapidity
range of $|\eta | <2.5$.
The ID is surrounded by a superconducting solenoid that provides a 2 T axial magnetic field.
The calorimeter system, surrounding the ID and the solenoid, covers the pseudorapidity range \mbox{$|\eta|< 4.9$}. 
It consists of high-granularity lead and liquid-argon (LAr) electromagnetic 
calorimeters, 
a steel and scintillator-tile hadronic calorimeter covering $|\eta| < 1.7$ and 
two copper/LAr hadronic endcap calorimeters covering $1.5 < |\eta| < 3.2$. 
Forward copper/LAr and tungsten/LAr calorimeter modules complete the solid-angle coverage out to
 \mbox{$|\eta|=4.9$}. 
The muon spectrometer resides outside the calorimeters.  It consists of multiple layers of 
trigger and tracking chambers within a system of air-core toroids, which enables an independent, precise measurement 
of muon track momenta for $|\eta| < 2.7$. The muon trigger chambers cover $|\eta| < 2.4$.

\section{Data and Monte Carlo samples}
\label{sec:data}

This search is performed in proton--proton collision data at $\sqrt{s} = 8$~\TeV\ 
collected with the ATLAS detector in 2012. 
The data are only used if they were recorded during stable beam conditions and with
all relevant subdetector systems operational. 
Lepton-plus-jets events were
 collected using single-electron and single-muon triggers  with thresholds chosen in each case
 such that the efficiency is uniform for leptons satisfying offline selections including transverse momentum $p_{\mathrm{T}} > 25$~\GeV.
 The ATLAS muon trigger system suffers from a 20\% inefficiency, relative to the offline event selection used in this analysis, largely due to a lack of geometrical coverage 
 by muon chambers owing to support structures in those regions~\cite{Aad:2014zya}. 
To mitigate this loss of efficiency, a large-radius-jet trigger, which triggers on anti-$k_t$ jets (see section \ref{sec:selection}) with radius parameter $R=1.0$,  was used to collect muon-plus-jets events which failed the muon trigger.
This large-radius-jet trigger recorded \delayedlumi{} data.
The chosen trigger threshold yields a uniform efficiency exceeding 99\% for events containing a large-radius jet with reconstructed $p_{\mathrm{T}} > 380$~\GeV.
For \ttbar{} events with invariant masses above 1.5~\tev, this addition increased the overall trigger efficiency in the muon channel to 96\%\footnote{The 4\% loss here is relative to the full \totlumi data set.}.

Simulated Monte Carlo (MC) samples were used for signal processes, as well as background processes producing jets and prompt leptons. The MC samples are employed to develop the event selection, provide SM background estimates, and evaluate signal efficiencies.
Background contributions from  processes in which there are no genuine prompt isolated leptons were estimated directly from the data as described in section~\ref{sec:backgrounds}. The MC samples were processed through the full ATLAS detector simulation~\cite{Aad:2010ah} based on \Geant 4~\cite{Agostinelli:2002hh} or through a faster simulation making use of parameterised showers in the calorimeters~\cite{ATLAS:2010bfa}. Additional simulated proton--proton collisions generated using \Pythia{} v8.1~\cite{Sjostrand:2007gs}
were overlaid to simulate the effects of additional collisions from the same and nearby bunch crossings (pile-up).
All simulated events were then processed using the same reconstruction algorithms and analysis chain as
the data. The simulated trigger and selection efficiencies were corrected to agree with the performance observed in data.

Production of Bulk RS gluon and graviton signals was modelled using \Madgraph 5~\cite{Alwall:2011uj} interfaced with \Pythia{} v8.1. For the gluon, the MSTW2008LO parton
distribution function (PDF) set~\cite{Martin:2009iq} was used, while for the graviton, the CTEQ6L1~\cite{Pumplin:2002vw} PDF set was used. The $Z'$ signal was modelled using \Pythia{} v8.1 with the MSTW2008LO PDF set. Heavy scalar signal samples were generated using \aMCNLO{}~\cite{Alwall:2014hca} with LO matrix elements and the CTEQ6L1 PDF set.

Pair production of top quarks is the dominant background in this search. It was simulated using the \Powheg{}~\cite{Frixione:2007nw,Nason:2004rx,Frixione:2007vw,Alioli:2010xd} generator r2330.3 interfaced with \Pythia{} v6.427~\cite{Sjostrand:2006za} with the Perugia 2011C~\cite{Skands:2010ak} tune and the CT10~\cite{Lai:2010vv} next-to-leading-order PDF set. In the MC generation of \Powheg{} 
events, the parameter {\it{hdamp}} was set to the top quark mass, corresponding to a damping of high-$p_\text{T}$ radiation, in order to achieve good agreement with the differential cross-section measurements~\cite{ATL-PHYS-PUB-2015-002}.
Alternative samples of $t\bar{t}$ events, used to evaluate uncertainties on $t\bar{t}$ modelling, were generated using the  \Powheg{} and \MCNLO{} v4.1~\cite{Frixione:2002jk,Frixione:2003ei,Frixione:2010wd} generators interfaced with \Herwig{} v6.5~\cite{HerwigGC,Corcella:2002jc} with \Jimmy{}~\cite{Butterworth:1996zw} for the modelling of the underlying event. For these samples the CT10 PDF set and the ATLAS-AUET2~\cite{ATLAS:2011gmi} tune were used. In all cases, the top quark mass used was 172.5 GeV. The cross-sections for these  samples were normalised to the calculation from {\sc{Top++ v2.0}}~\cite{Czakon:2011xx} at next-to-next-to-leading-order (NNLO) accuracy in the strong coupling constant $\alpha_{\mathrm{s}}$,  including resummation of next-to-next-to-leading-logarithmic (NNLL) soft gluon terms~\cite{Beneke:2011mq,Cacciari:2011hy,Baernreuther:2012ws,Czakon:2012zr,Czakon:2012pz,Czakon:2013goa}. The top quark kinematics in all SM $t\bar{t}$ samples were corrected to account for electroweak higher-order  effects~\cite{Kuhn:2013zoa}. This correction is applied after generating the samples, by applying a weight that depends on the flavour of the initial-partons, the centre-of-mass energy of the initial partons, and the decay angle of the tops in the centre-of-mass frame of the initial partons.

Production of $W$ bosons in association with jets ($W$+jets) is also a significant background process. Samples of $W$+jets events were generated using the \Alpgen{} v2.13 generator~\cite{Mangano:2002ea} interfaced with \Pythia{} v6.426, including  up to five extra partons in the matrix element. 
Configurations with additional heavy quarks (a single $c$-quark, a $c\bar{c}$ pair  or a $b\bar{b}$ pair in the hard process) were included, with the masses of heavy quarks taken into account.
The CTEQ6L1
 PDF set and the Perugia 2011C tune were used. The samples were normalised using data as described in section~\ref{sec:backgrounds}. Additional samples were generated with different choices of \Alpgen{} matrix element/parton shower matching parameters in order to estimate modelling uncertainties in the production of $W$+jets events.

Production of single top quarks can yield events that satisfy the analysis event selection. The \Powheg{} generator interfaced with \Pythia{} v6.246 was used to estimate the $s$- and $Wt$-channels~\cite{Alioli:2009je,Frederix:2012dh,Re:2010bp} with the same configuration as for the $t\bar{t}$ samples. The $t$-channel was also generated with   \Powheg{} but in a four-flavour scheme, hence  the CT10 NLO four-flavour PDF set was used. Overlap between the $Wt$ sample and $t\bar{t}$ samples was handled using the diagram removal scheme~\cite{Frixione:2008yi}. These samples were normalised to approximate NNLO cross-sections~\cite{Kidonakis:2010tc,Kidonakis:2011wy,Kidonakis:2010ux}.

Other minor background processes producing prompt isolated leptons include heavy diboson production, production of $Z$ bosons in association with jets ($Z$+jets) and production of heavy gauge bosons in association with $t\bar{t}$ ($t\bar{t}V$). Production of $Z$+jets was modelled using  \Alpgen{}
 interfaced with \Pythia{} v6.426, in the same configuration used for the $W$+jets samples described above. The samples were normalised to the inclusive $Z$ boson production cross-section calculated at NNLO in QCD using {\sc Fewz}~\cite{Gavin:2012sy}. Diboson production was modelled using the {\Sherpa}~\cite{Gleisberg:2008ta,Hoeche:2009rj,Gleisberg:2008fv,Schumann:2007mg} generator, with up to three extra partons in the matrix element and taking into account the mass of the $b$- and $c$-quarks. The diboson samples were normalised to calculations at NLO in QCD performed using {\sc Mcfm}~\cite{Campbell:2010ff} v5.8. The $t\bar{t}V$ production was modelled using \Madgraph 5 interfaced with \Pythia{} v6.426 and normalised to NLO cross-section predictions~\cite{Garzelli:2012bn}.

\section{Event selection}
\label{sec:selection}

Events consistent with $t\bar{t}$ decaying to a single charged lepton together with hadronic jets and missing transverse momentum are selected.
Electron candidates are required to have a transverse energy $E_{\mathrm{T}} > 25$~\GeV\ and $|\eta_{\mathrm{cluster}}| < 2.47$, where $\eta_{\mathrm{cluster}}$ is the pseudorapidity 
of the cluster of energy deposited in the electromagnetic calorimeter, computed  with respect to the centre of ATLAS detector and matched to the candidate~\cite{Aad:2014fxa}.  Electron candidates in the
transition region  $1.37 < |\eta_{\mathrm{cluster}}| < 1.52$ between calorimeter barrel and endcap are excluded.
  
In order to reduce backgrounds from non-prompt sources and hadronic showers with a high electromagnetic 
energy fraction in the calorimeter, electron isolation is imposed using a {\it{mini-isolation}} variable 
$MI_{10}$ defined as the $\sum\limits_{\mathrm{tracks}} {\pt^{\mathrm{track}}}$ for all tracks (except the matched lepton track) 
with $p_{\mathrm{T}} > 1$~\GeV\, satisfying quality selection criteria, and within a cone of size $\Delta R < 10$~\GeV $/ E_{\mathrm{T}}$~\cite{Rehermann:2010vq,Aad:2013nca} centred on the barycentre of the cluster.  
Electrons are defined to be isolated if $MI_{10} / E_{\mathrm{T}} < 0.05$. 
The $E_{\mathrm{T}}$ here is the transverse energy of the reconstructed electron.
The isolation variable $MI_{10}$ is particularly useful in the case of boosted top quark decays
since the \pt-dependent cone size 
reflects the \pt\ dependence of the separation between objects with a common boosted parent.
The matching of the electron to the collision vertex~\cite{ATLAS-CONF-2010-069} is imposed by requiring that the longitudinal impact parameter relative to it be less than $ 2$~mm.    

Muon candidates are required to have $p_{\mathrm{T}} > 25$~\GeV\ and $|\eta|
< 2.5$.  The matching of the muon to the collision vertex is 
imposed by the requirements that the longitudinal impact parameter relative to the collision
vertex be less than $ 2$~mm and that the transverse impact parameter relative to the collision vertex divided by its
uncertainty, $|d_0/\sigma_{d_0}|$, be less than $3.0$.  Muons are also required to satisfy the same $MI_{10}$ requirement as electrons, with the cone centred on the inner-detector track associated with the muon.

Jets are reconstructed using the anti-$k_t$ algorithm~\cite{Cacciari:2008gp} applied to clusters of calorimeter cells 
that are topologically connected and calibrated to the hadronic energy scale~\cite{Aad:2011he} using a local calibration scheme~\cite{ISS0501}. 
Small-radius jets (radius parameter $R=0.4$) as well as large-radius jets ($R=1.0$)  are used.
The energies of all jets and the masses of the large-radius jets have been calibrated to their values at particle level~\cite{Aad:2014bia,Aad:2012ef,ATLAS-CONF-2012-065}.
Small-radius jets are required to satisfy 
$p_{\mathrm{T}} > 25$~\GeV{} and $|\eta| < 2.5$, while large-radius jets are required to satisfy $p_{\mathrm{T}} > 300$~\GeV{} and $|\eta| < 2.0$.  
Low-$p_{\mathrm{T}}$ central small-radius jets ($p_{\mathrm{T}} < 50$\,GeV, $|\eta|<2.4$) are required to 
have a jet vertex fraction~\cite{ATLAS-CONF-2013-083} greater than $ 0.5$.  
The jet vertex fraction is defined as the total transverse momentum (using a scalar sum) of tracks in the jet that are associated with the primary vertex
divided by the scalar sum of the transverse momentum of all tracks in the jet.  
This variable suppresses jets arising from pile-up effects. 
Large-radius jets have jet trimming~\cite{Krohn:2009th} applied. In trimming, subjets are formed by applying a jet algorithm with smaller radius parameter, $R_{\mathrm{sub}}$, and then soft subjets with less than a certain fraction, $f_{\mathrm{cut}}$,  of the original jet $p_{\mathrm{T}}$ are removed. The properties of the trimmed jet are then calculated using the surviving subjets.
 This procedure mitigates the effect of pile-up~\cite{Aad:2013gja}. The trimming parameters used in this search are $ f_{\mathrm{cut}}= 0.05$ and $R_{\mathrm{sub}} = 0.3$, and the inclusive $k_t$~\cite{Ellis:1993tq} algorithm is used to form the subjets.

Only small-radius jets are considered for $b$-jet identification ($b$-tagging).  The $b$-tagging algorithm uses
a multivariate approach with inputs taken from the results of separate impact parameter, 
secondary vertex and decay topology algorithms~\cite{ATLAS-CONF-2014-046}.
The operating point of the algorithm is 
chosen such that the $b$-tagging efficiency for simulated \ttbar\ events is 70\%.
In MC simulation, factors are applied to correct for the differences between the $b$-tagging efficiency in simulated events and that measured in data. 
The factors are adapted to be appropriate for $b$-jets from high-$p_{\mathrm{T}}$ top quarks, for which the $b$-tagging efficiencies are lower.

The \MET{} is calculated from the vector sum of the transverse energy of topological clusters in the calorimeter~\cite{Aad:2012re}.  
The clusters associated with the reconstructed electrons and small-radius jets are replaced by the calibrated energies of these objects.
Muon transverse momenta determined from the ID and the muon spectrometer are also included in the calculation.

Overlap in identification of the relevant physics objects is possible and a procedure 
is implemented to remove duplication.   Electrons and small-radius jets are considered 
for overlap removal if the cluster associated with the electron is within $\Delta R = 0.4$ of the nearest jet.  
In such cases, the jets have their four-momentum and jet vertex fraction recalculated after subtracting the electron 
four-momentum and then are removed if the recalculated values do not satisfy the original jet selection criteria.  
If the distance $\Delta R$ between the electron and the recalculated jet is $< 0.2$, 
the electron candidate is likely to be from the hadronic jet. 
Therefore the electron is removed from the electron candidate list and its four-momentum is added to that of the recalculated jet.
Muons are removed from the muon candidate list if the distance $\Delta R$ between the muon and small-radius jet is less than $ 0.04 + 10~\GeV/p_{\mathrm{T}}$.  
This criterion exploits the anti-correlation between the muon \pt{} and its angular distance from the $b$-quark, in an approach similar to
 the isolation variable.
The parameters are tuned based on signal MC simulations in order to provide a constant high efficiency as a function of resonance mass.

A set of common preselection criteria is used for events to be considered for the boosted and resolved topologies.  
Events are required to have exactly one lepton (electron or muon) plus multiple jets.
Events recorded by the lepton triggers are required to have lepton-trigger objects that match the selected lepton. 
Additionally, events must have
\met\ $> 20$~\GeV\ and \met\ $ + $ $m_{\mathrm{T}} > 60$~\GeV, where $m_{\mathrm{T}}$ is the transverse mass calculated
as $m_{\mathrm{T}} = \sqrt{2 p_{\mathrm{T}} \met(1-\cos{\phi_{\ell\nu}})}$ where $p_{\mathrm{T}}$ is the transverse momentum of the lepton
and $\phi_{\ell\nu}$ is the angle between the $p_{\mathrm{T}}$ and \met\ vectors.

Events are next checked against the {\em{boosted-topology selection}}.  
The selected lepton is required to have at least one small-radius jet within a distance of $\Delta R(\ell,j) = 1.5$  and, of these,
the jet with highest $p_{\mathrm{T}}$  is termed $j_{\mathrm{sel}}$.
Boosted-topology events must have 
at least one large-radius jet with $p_{\mathrm{T}} > 300$~\GeV\ (380~\GeV\ for the muon-plus-jets events selected by the large-radius-jet trigger), $|\eta| < 2.0$, mass $m_{\mathrm{jet}} > 100$~\GeV, 
first $k_t$ splitting scale~\cite{Ellis:1993tq} $\sqrt{d_{12}} > 40$~\GeV, 
$\Delta R > 1.5$ between the large-radius jet and $j_{\mathrm{sel}}$, 
and $\Delta \phi > 2.3$ between the large-radius jet and lepton. 
The jet mass is calculated using the four-momenta of its constituent clusters, which are taken as massless.
If multiple large-radius jets satisfy these criteria, the
highest-$p_{\mathrm{T}}$ jet is chosen as the hadronically decaying top quark candidate.  Finally, at least one of the
small-radius jets in the event must be $b$-tagged and matched to either of the top quark candidates, as described in section~\ref{sec:reconstruction}.  

Events that do not satisfy the boosted-topology selection are then tested against the criteria for the {\em{resolved-topology} selection}.  
These events are required to have at least four small-radius jets, with at least one of them $b$-tagged.
A $\chi^2$ algorithm is used to reconstruct the \ttbar\ system, as described in section~\ref{sec:reconstruction}.
The lowest $\chi^2$ value is required to satisfy $\log_{10}{(\chi^2)} < 0.9$.

\begin{figure}[htbp]
\centering
\subfigure[\Zprime{}. ]{
  \includegraphics[width=0.45\textwidth]{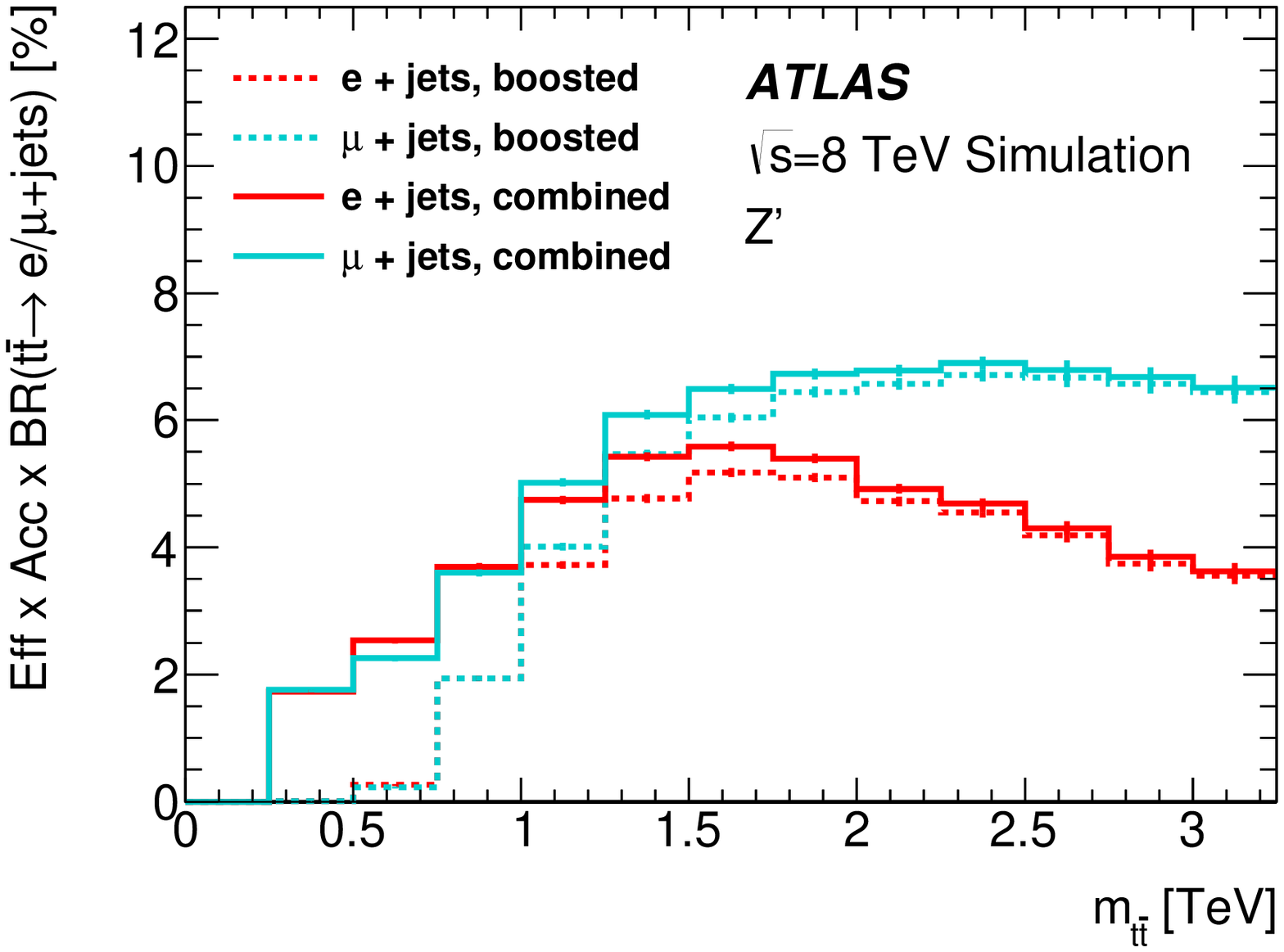}
}
\subfigure[$g_\mathrm{KK}$.]{
  \includegraphics[width=0.45\textwidth]{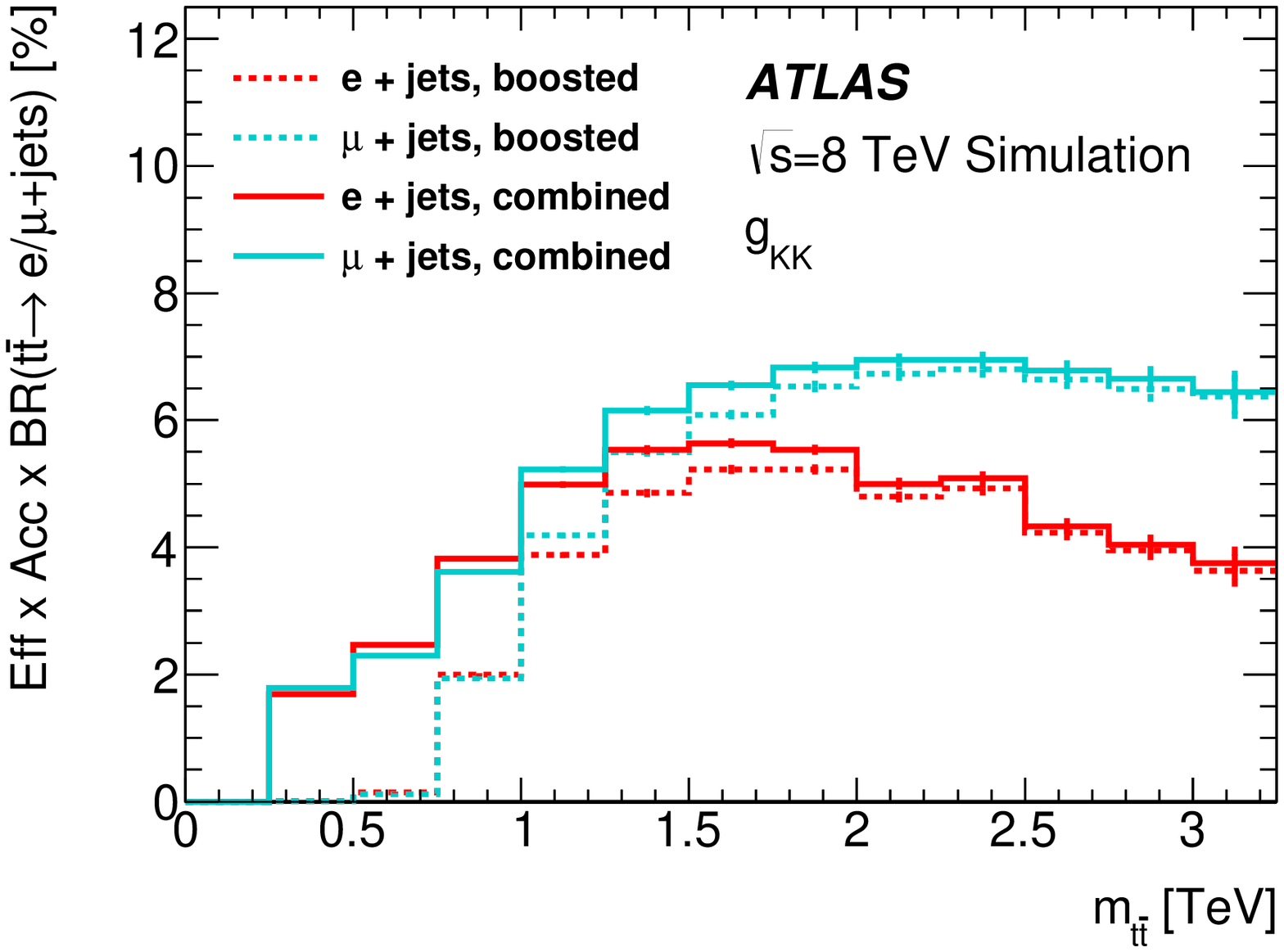}
}
\subfigure[$G_\mathrm{KK}$.]{
  \includegraphics[width=0.45\textwidth]{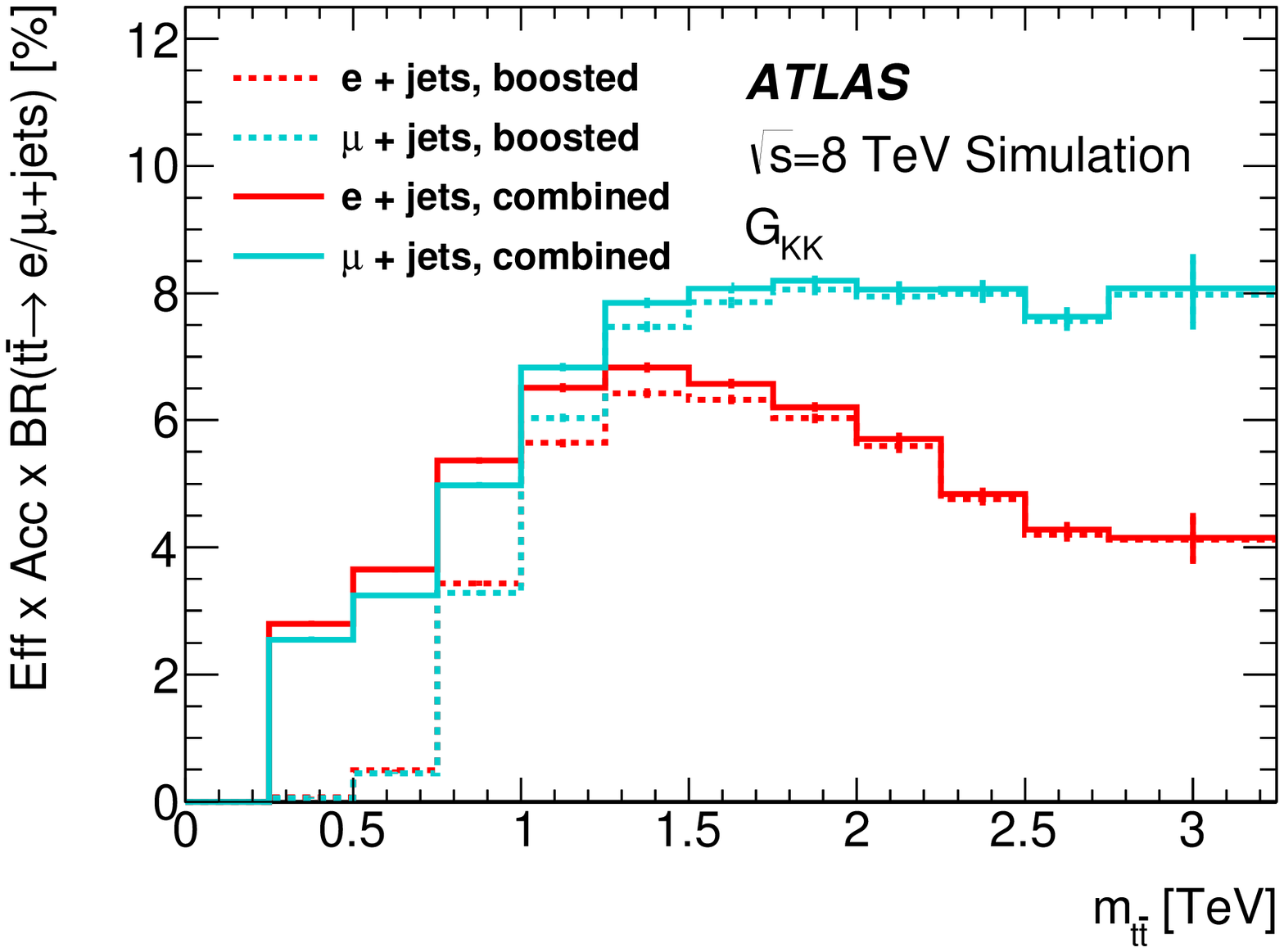}
}
\subfigure[Scalar resonance.]{
  \includegraphics[width=0.45\textwidth]{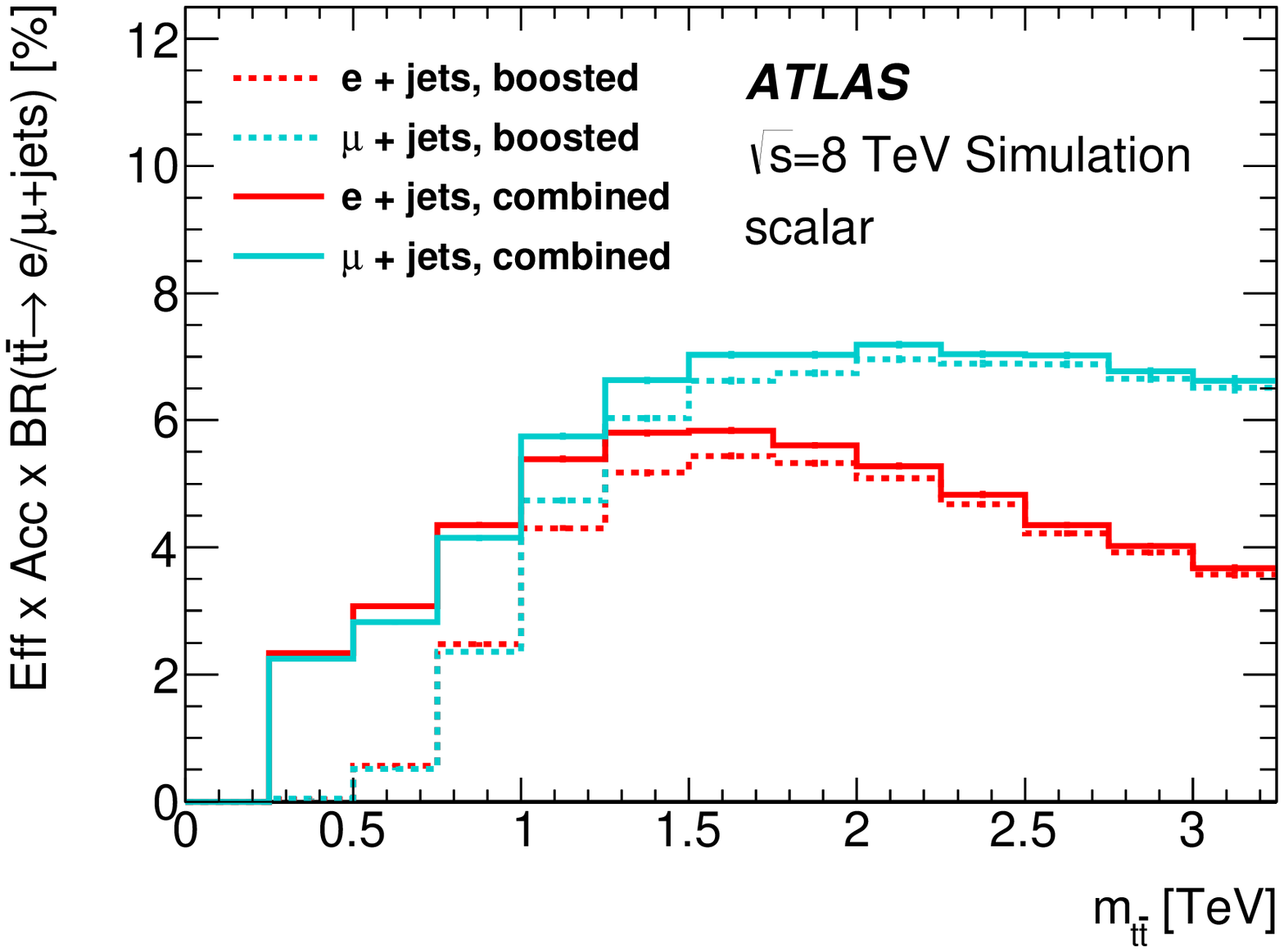}
}
\caption{Selection efficiency times acceptance times branching ratio 
as a function of the top-antitop quark invariant mass \mttbar{} at Monte Carlo generator level for the different signals in the models considered: (a) \Zprime{}, (b) $g_\mathrm{KK}$, (c) $G_\mathrm{KK}$, (d) scalar resonance. 
The dashed lines show the boosted-topology selection and the unbroken lines show the combined selection.
\label{fig:BSMSelectionEff}}
\end{figure}

The selection efficiencies for MC simulated signal events are given in figure~\ref{fig:BSMSelectionEff},
for different models of interest.
For reference, the branching ratio for $t\bar{t}$ to electron- or muon-plus-jets is about 17\% for each lepton flavour  taking into account leptonic tau decays~\cite{Agashe:2014kda}.
There are efficiency losses from both the large-radius jet requirements and $b$-tagging 
requirement for the boosted-topology selection. 
For resonance masses above $1.5$\,TeV, the efficiencies of the resolved selections are relatively insignificant due to the $\chi^2$ requirements and the veto of boosted selections.
It can also be seen that efficiency times acceptance is smaller for isolated electrons than isolated muons above the same resonance mass point, 
due to the inefficiency of electron identification and overlap removal in the boosted environment.

\section{Event reconstruction}
\label{sec:reconstruction}

Signal \ttbar\ resonances should appear in the \mttbarreco\ 
spectrum as an excess of events over the SM expectation clustered around the resonance mass. 
Events are reconstructed assuming the final state originated from a \ttbar\
decay.
To calculate \mttbarreco{}, the neutrino four-momentum must be determined.
The neutrino transverse momentum is taken to be the \MET\ vector.
The longitudinal component of the neutrino momentum, $p_z$, is calculated by constraining
the lepton plus missing momentum system to have the $W$ boson mass and solving the resulting quadratic equation in the neutrino's longitudinal momentum $p_z$~\cite{Aaltonen:2010jr,Aad:2012ux}. 
If no real solution exists, the \MET\ vector is varied by the minimal amount required to produce exactly one real solution.
If two real solutions are found, the one with the smallest $|p_z|$ is used for the boosted-topology reconstruction,
while the choice is made by the $\chi^2$ algorithm described below for the resolved topology.
 
For the boosted topology, \mttbarreco\ is computed from the four-momenta of 
the neutrino, lepton, the previously selected small-radius jet, $j_{\mathrm{sel}}$, 
and the highest-$p_{\text{T}}$ large-radius jet.  In this case the assignment of jets to
the semileptonically decaying top quark and hadronically decaying 
top quark is unambiguous.

In calculating \mttbarreco\ for the resolved topology, a $\chi^2$ algorithm is employed
to find the best assignment of jets to the semileptonically and hadronically
decaying top quarks.    Using the four-momenta of the neutrino, lepton and all small-radius jets in the event,
a $\chi^2$ is defined using the expected top quark and $W$ boson masses:
\begin{eqnarray}
\chi^2 &=& \left[ \frac{m_{jj}-m_W}{\sigma_W} \right]^2 + 
         \left[ \frac{m_{jjb}-m_{jj}-m_{t_\mathrm{h}-W}}{\sigma_{t_\mathrm{h}-W}} \right]^2 +
         \left[ \frac{m_{j\ell\nu}-m_{t_\ell}}{\sigma_{t_\ell}} \right]^2 \nonumber \\ 
        & &+ \left[ \frac{(p_{\mathrm T,jjb}-p_{\mathrm T,j\ell\nu}) - 
                (p_{\mathrm T,t_\mathrm{h}}-p_{\mathrm T,t_\ell})}{\sigma_{\mathrm{diff}p_{\mathrm T}}}  \right]^2.
\end{eqnarray}
The first term is a constraint using the mass of the hadronically decaying $W$ boson.
The second term is a constraint using the mass difference between the hadronically decaying top quark and the hadronically decaying $W$ boson.   
Since the mass of the hadronically decaying $W$ boson, $m_{jj}$,  
and the mass of the hadronically decaying top quark, $m_{jjb}$,
are highly correlated, the mass of the hadronically decaying $W$ boson
is subtracted from the second term so as to decouple it from the first term. 
The third term is a constraint using the mass of the semileptonically decaying top quark. 
The last term arises as a constraint on the expected transverse momentum balance between the two decaying top quarks. 
In the $\chi^2$ definition above, $t_\mathrm{h}$ and $t_\ell$
refer to the hadronically and semileptonically decaying top quarks.
The values of the $\chi^2$ central-value parameters $m_W$, $m_{t_\mathrm{h}-W}$, $m_{t_\ell}$ and $p_{\mathrm T,t_\mathrm{h}}-p_{\mathrm T,t_\ell}$, and the values of the width parameters $\sigma_W$,  $\sigma_{t_\mathrm{h}-W}$, $\sigma_{t_\ell}$ and $\sigma_{\mathrm{diff}p_{\mathrm T}}$ are found through Gaussian fits to the distributions of relevant reconstructed variables, using reconstructible\footnote{
Reconstructible events are those where there is a reconstructed object within $\Delta R=0.4$  of each visible parton from top decays, and the $\Delta \phi$ between the neutrino and the \MET is smaller than 1.
Among the events satisfying the final selection critera, 70\% (55\%) are reconstructible for \Zprime\ mass of 750 GeV (400 GeV). 
} 
MC events with \Zprime{} masses from 0.5 to 2.0 \TeV{}.
All possible neutrino $p_z$ solutions and jet permutations with at least one $b$-quark candidate satisfying the $b$-tagging requirement are tested, and the one with the lowest $\chi^2$ value is adopted.
About 80\% of these reconstructible events have the correct assignment of reconstructed objects to the hadronically and semileptonically decaying top quarks.

\begin{figure}[htbp]
\centering
\subfigure[\Zprime{}. ]
{
\includegraphics[width=0.45\textwidth]{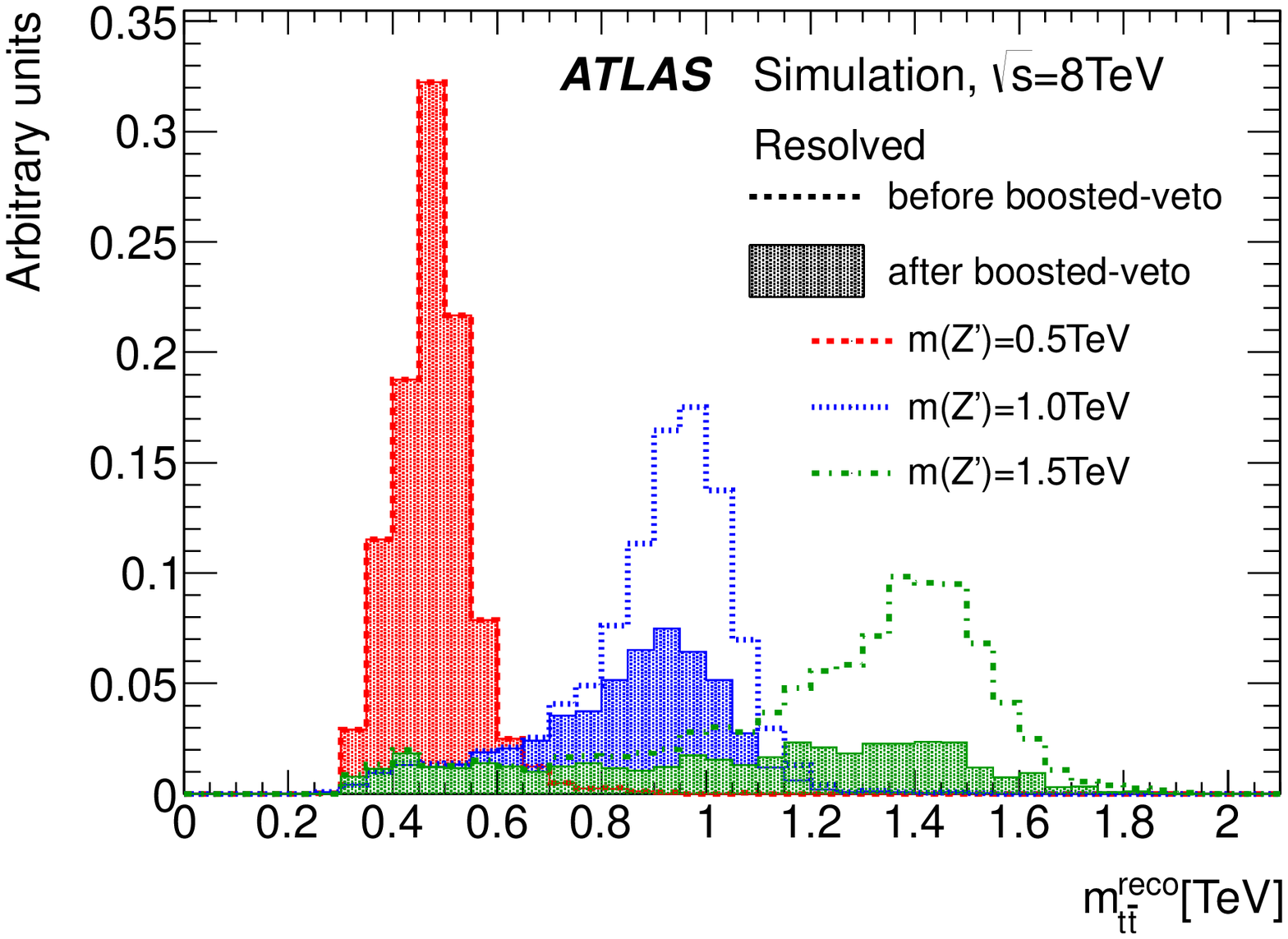}
}
\subfigure[$g_\mathrm{KK}$, width 15.3\%. ]
{
\includegraphics[width=0.45\textwidth]{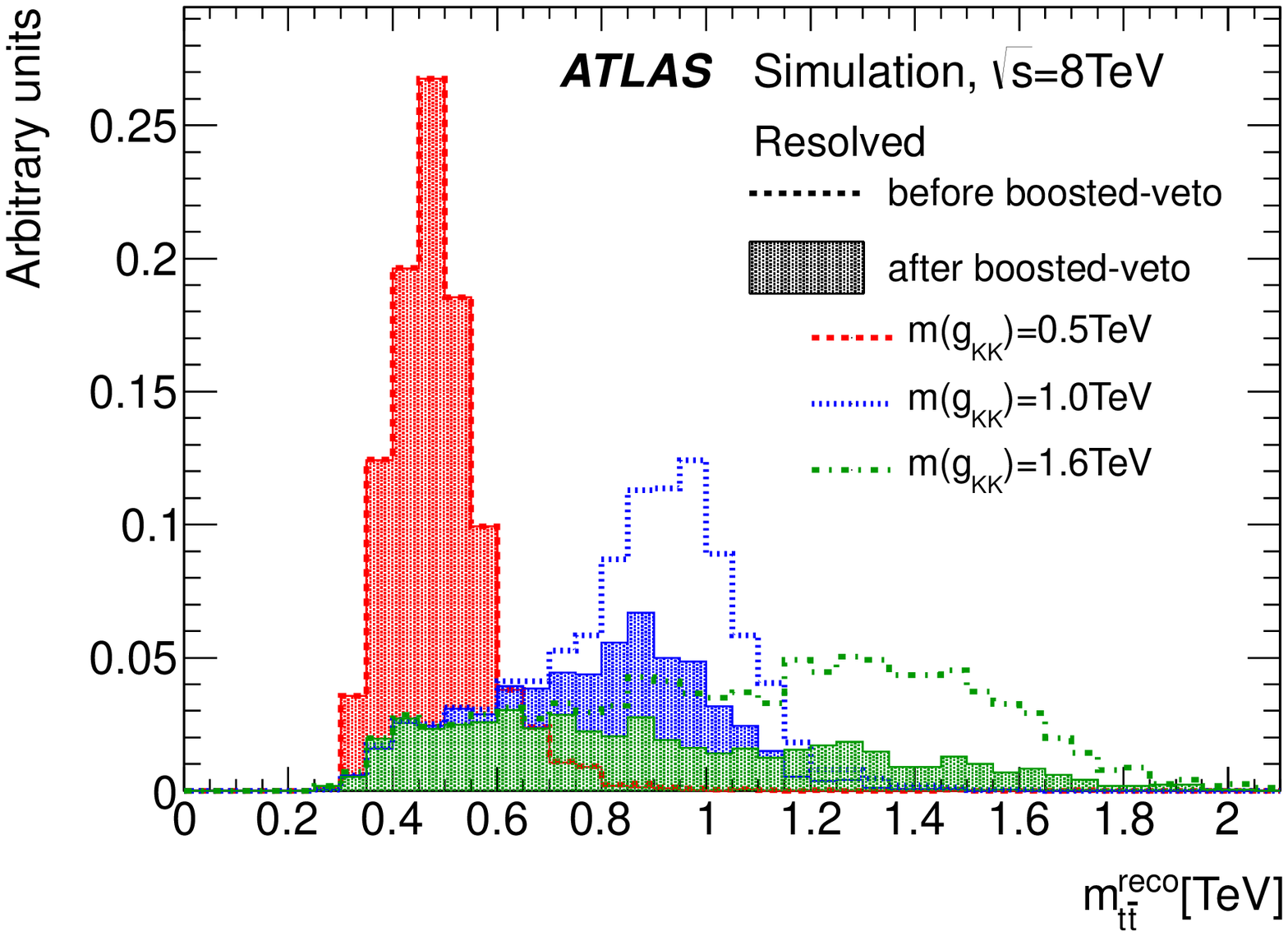}
}\\
\subfigure[$G_\mathrm{KK}$. ]
{
\includegraphics[width=0.45\textwidth]{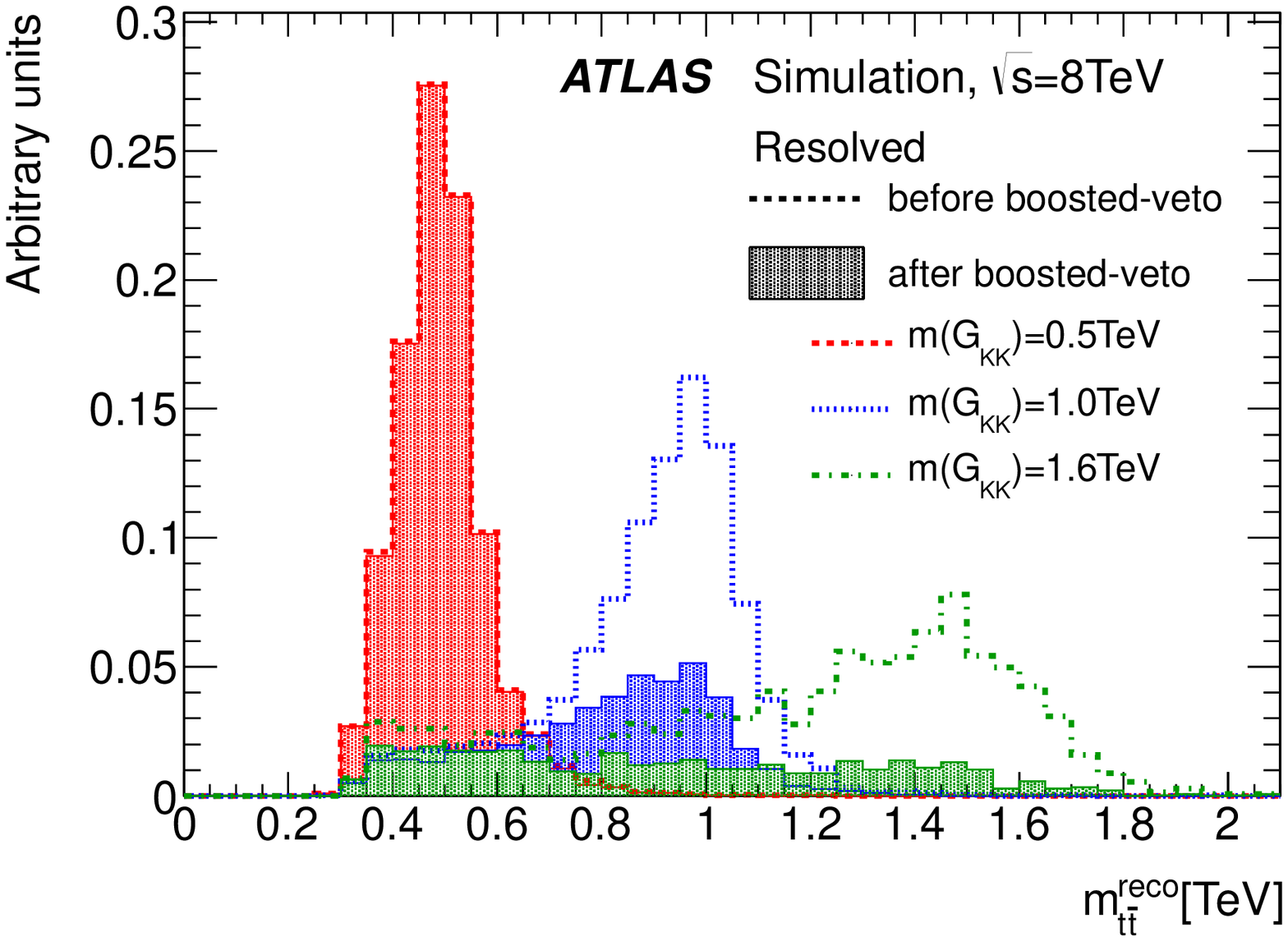}
}
\subfigure[Scalar resonance. ]
{
\includegraphics[width=0.45\textwidth]{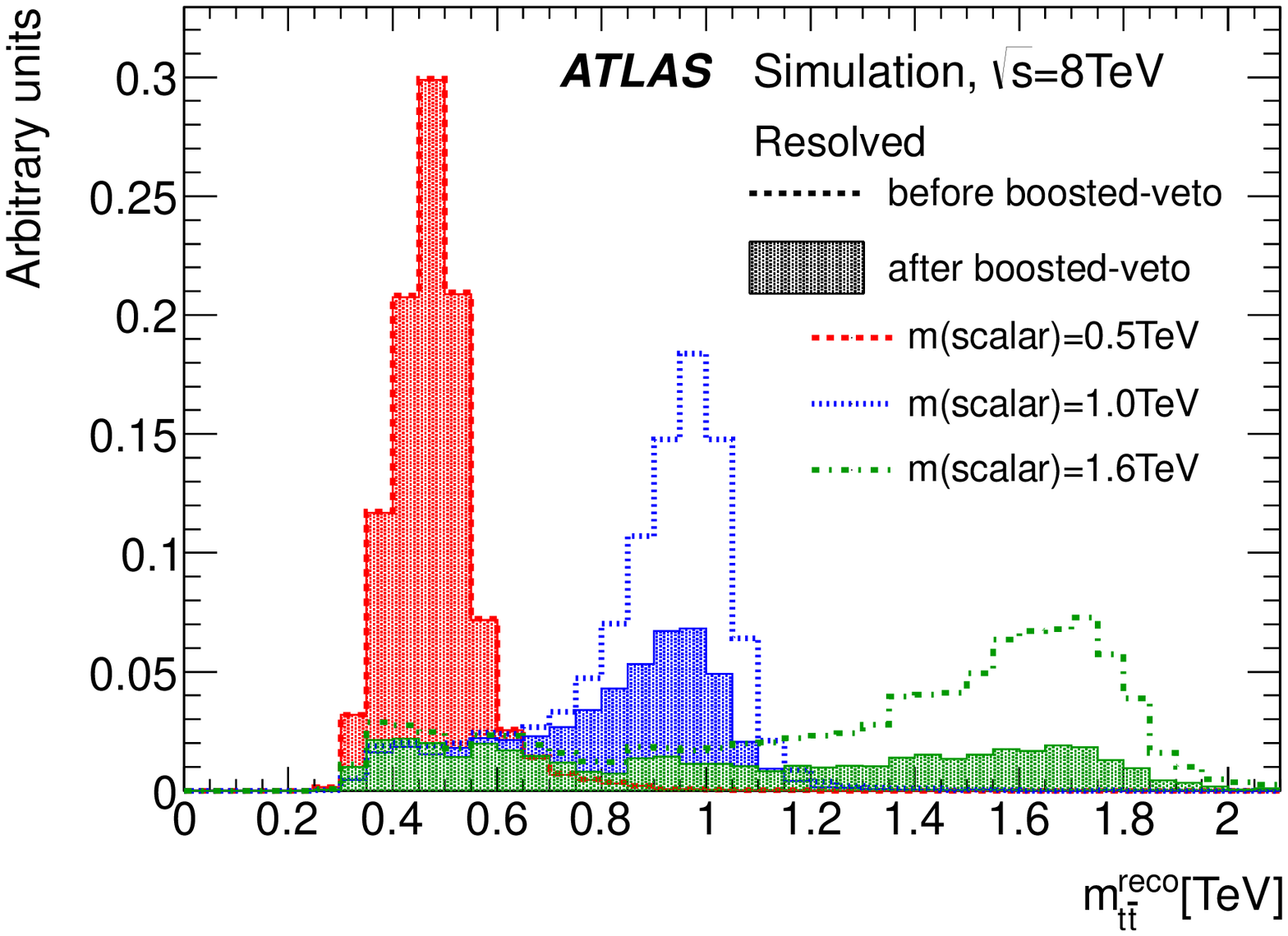}
}
\caption{Reconstructed top quark pair invariant mass, \mttbarreco{} for the different signal models for events satisfying the resolved-topology
selection and using the resolved-topology reconstruction: (a) \Zprime{}, (b) $g_\mathrm{KK}$, (c) $G_\mathrm{KK}$, (d) scalar resonance, before and after the veto on the boosted selection.  
For each signal, the two histograms are normalised to the same arbitrary luminosity.
}
\label{fig:mtt}
\end{figure}

\begin{figure}[htbp]
\centering
\subfigure[\Zprime{}.]
{
\includegraphics[width=0.45\textwidth]{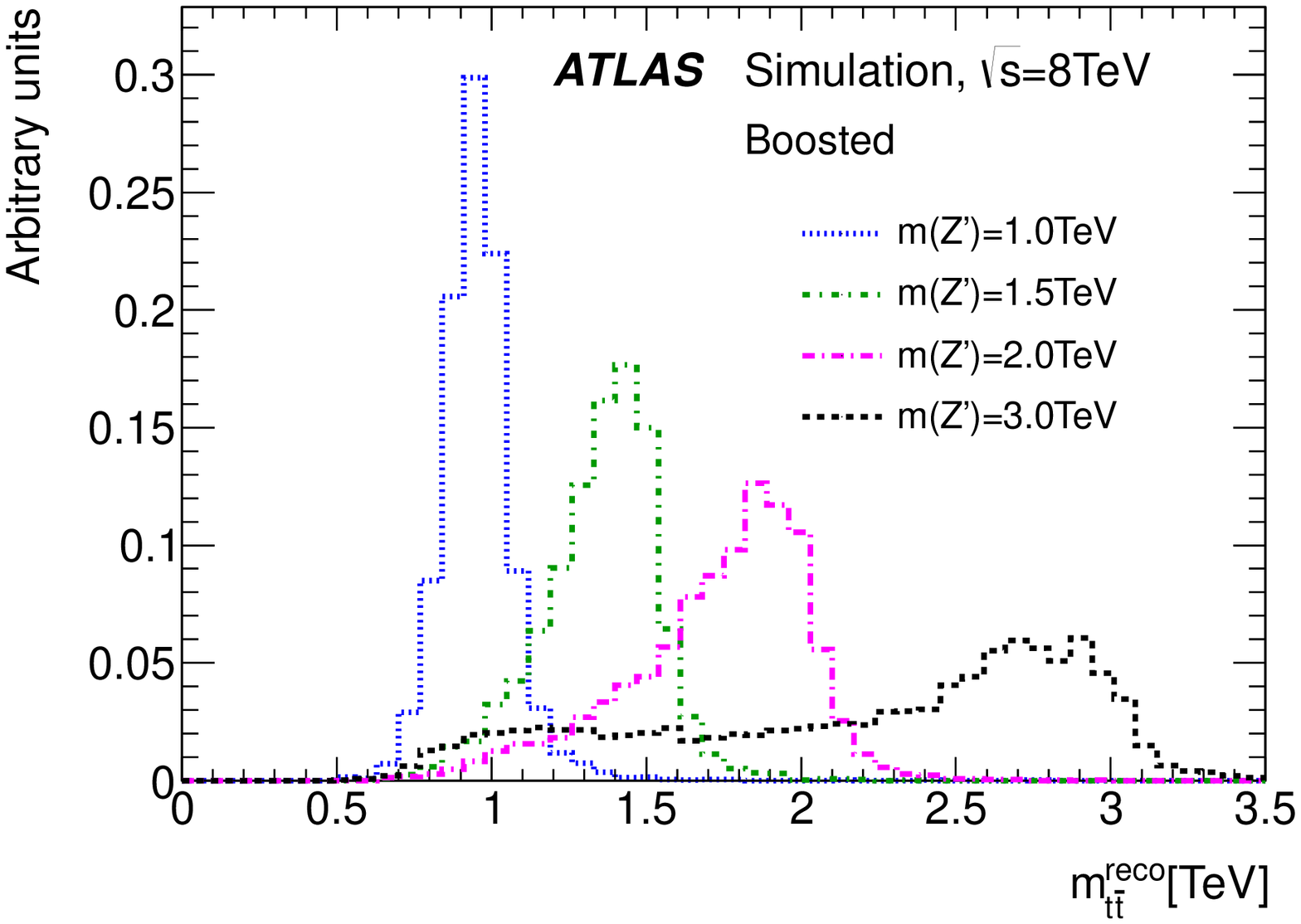}
}
\subfigure[$g_\mathrm{KK}$, width 15.3\%. ]
{
\includegraphics[width=0.45\textwidth]{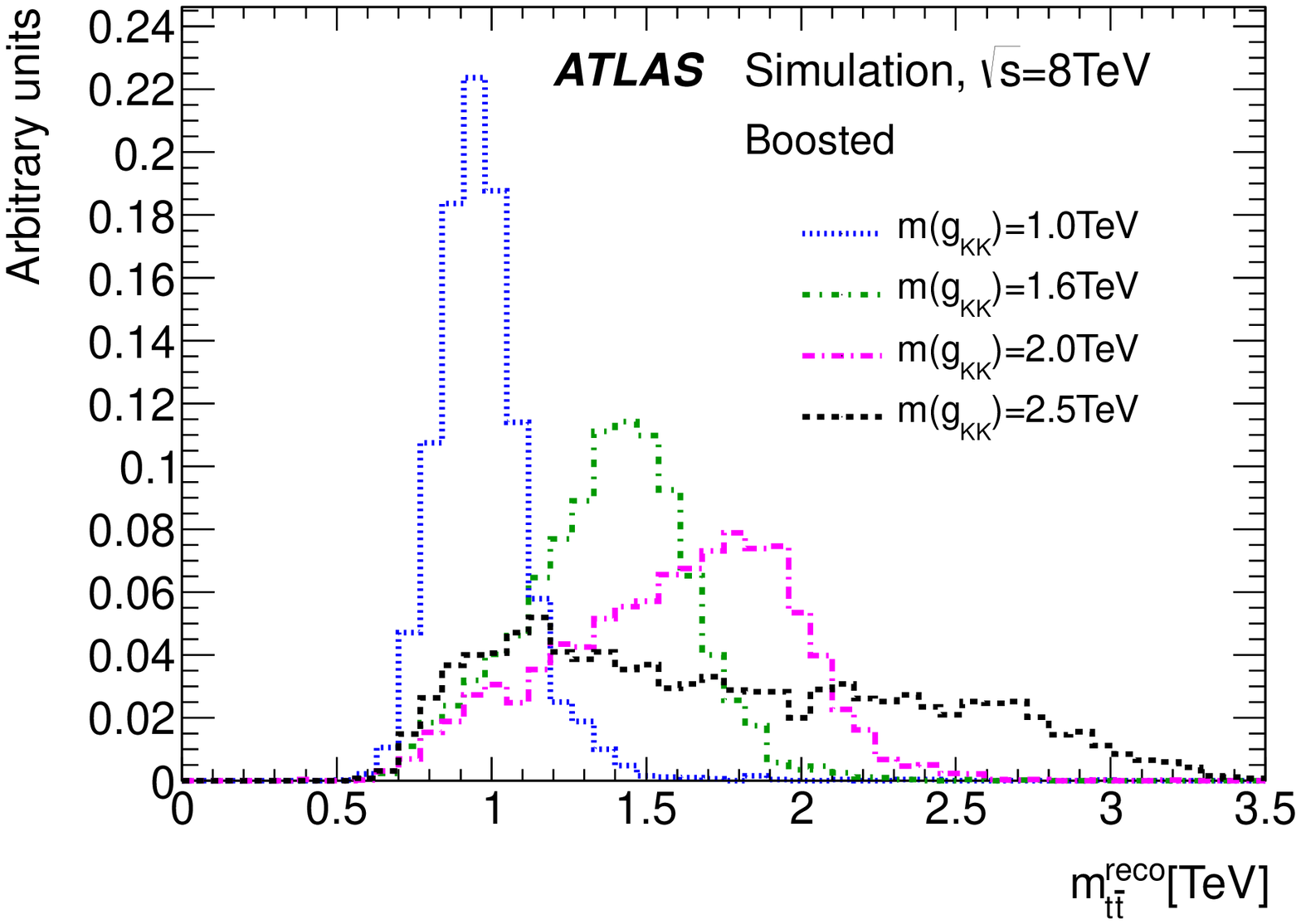}
}\\
\subfigure[$G_\mathrm{KK}$.]
{
\includegraphics[width=0.45\textwidth]{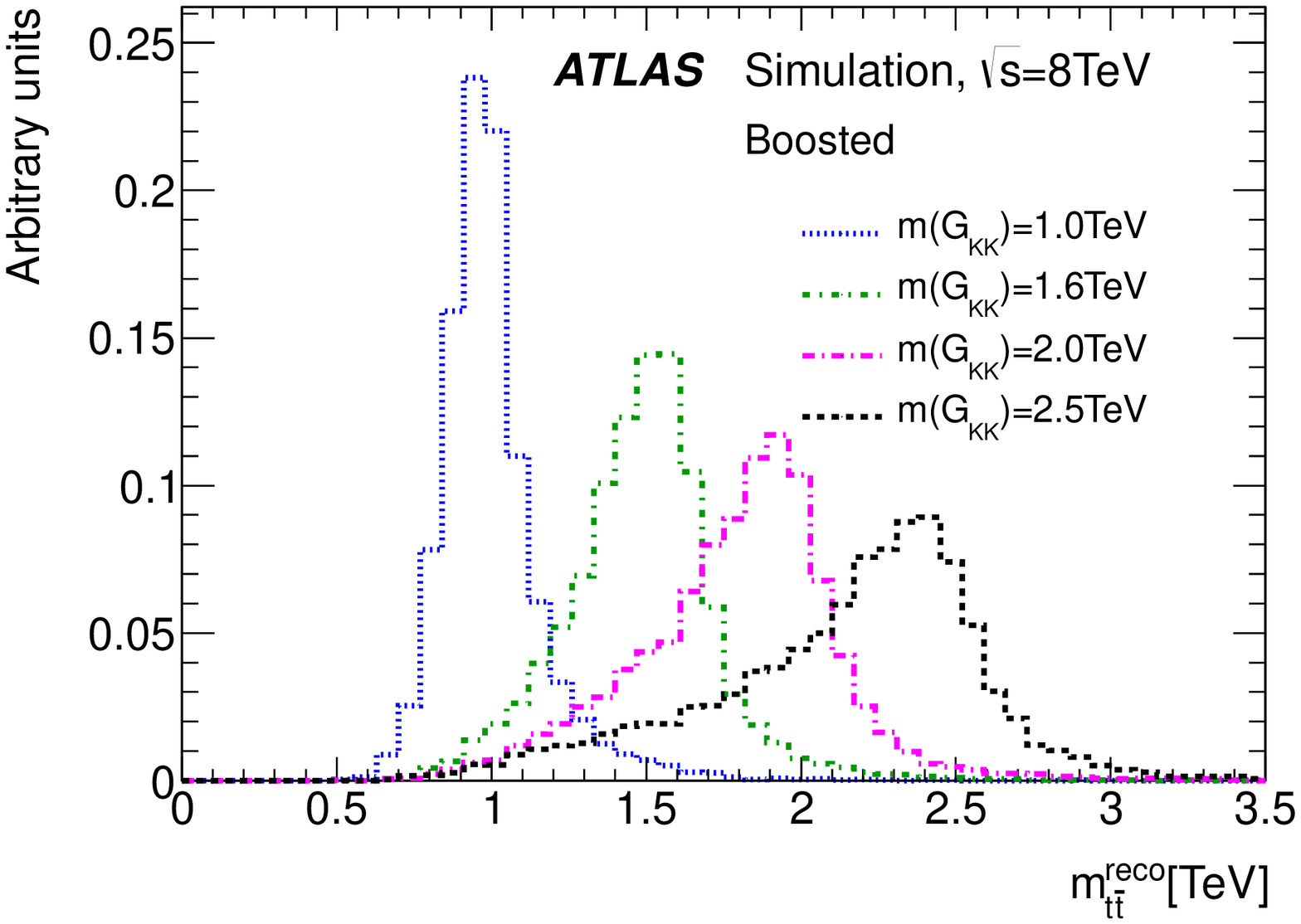}
}
\subfigure[Scalar resonance.]
{
\includegraphics[width=0.45\textwidth]{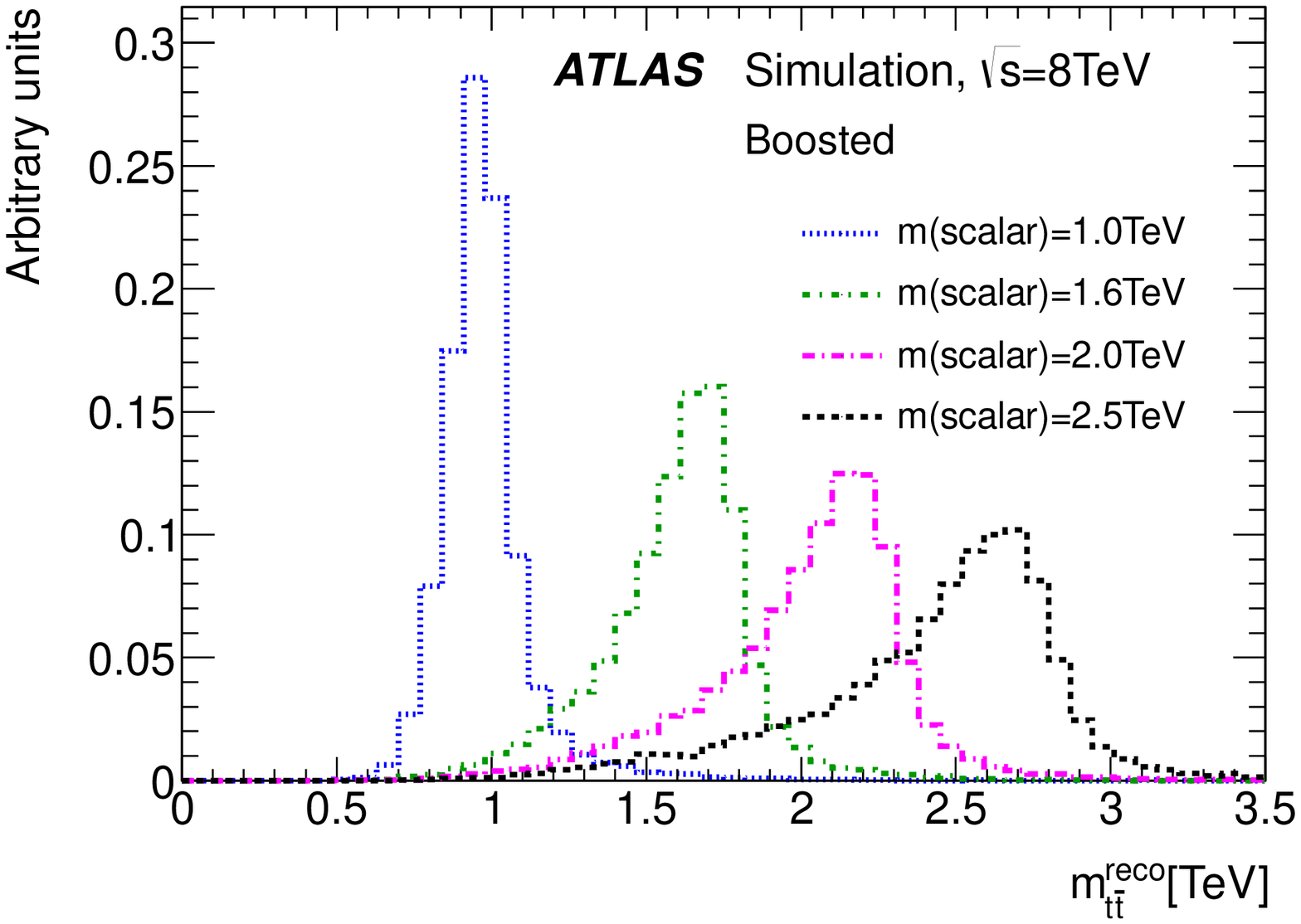}
}
\caption{Reconstructed $t\bar{t}$ invariant mass for the different signal models for events satisfying the
boosted-topology selection and using the boosted-topology reconstruction: (a) \Zprime{}, (b) $g_\mathrm{KK}$, (c) $G_\mathrm{KK}$, (d) scalar resonance.
}
\label{fig:mtt_boosted}
\end{figure}

The resulting \mttbarreco\ distributions for several signal masses 
from the resolved- and boosted-topology
reconstruction are shown in figures~\ref{fig:mtt} and ~\ref{fig:mtt_boosted}.
For these figures, all events satisfying the resolved- or boosted-topology selection
criteria are used. The low-mass tails arise from two effects: firstly, extra radiation from the $t\bar{t}$ system that is not included in the reconstruction can shift the reconstructed mass to lower values; secondly, before reconstruction the Breit--Wigner signal shape in \mttbar{} has a tail at lower values due to the steep fall in parton luminosity with increasing partonic centre-of-mass energy. 
The former is particularly true for high-mass resonances, while the latter has a larger effect on broad resonances.
The experimental resolution for the invariant mass of the $t\bar{t}$ system\footnote{The experimental resolution is extracted from a Gaussian fit of the relative difference $\frac{m_{t\bar{t}}^{\mathrm{reco, true-matched}}- m_{t\bar{t}}} {m_{t\bar{t}}}$ from reconstructible events; $m_{t\bar{t}}^{\mathrm{reco, true-matched}}$ is the reconstructed \mttbarreco{} computed with the correct combination of jets identified with the parton-level information and \mttbar{} is the true mass of the $t\bar{t}$ system in the MC simulation.}  is 8\%  for the resolved-topology selection at a resonance mass of 400 \GeV{} improving to 6\% for 1\ TeV{}. It is 6\% in the boosted-topology selection, independent of resonance mass. 
   
With hadronically and semileptonically decaying top quarks 
identified for both the boosted- and resolved-topology selections, 
three categories of $b$-tagged events are defined:
those in which both decaying top quark candidates have
a matching $b$-jet, those in which only the hadronically decaying
top quark candidate has a matching $b$-jet and those in which
only the semileptonically decaying top quark candidate has a 
matching $b$-jet.  In the boosted-topology selection, the match
is defined as either the selected jet, $j_{\mathrm{sel}}$, 
or a small-radius jet within
$\Delta R = 1.0$ of the large-radius jet as being a $b$-tagged
jet.  In the resolved-topology 
selection, the matching is determined by the $\chi^2$ 
algorithm.

\section{Background contributions estimated from data}
\label{sec:backgrounds}

Data are used to estimate the magnitudes and
uncertainties of two important background contributions: $W$+jets
and multi-jet production.  

\subsection{$W$+jets scale factors}
For the $W$+jets background, data are used to derive scale factors that are applied to correct the normalisation and flavour fractions given by \Alpgen{} MC simulations of this background. 

For both the resolved- and boosted-topology event selection criteria, the normalisation scale factors
are determined by comparing the measured 
$W$ boson charge asymmetry in data~\cite{ATLAS:2012an,Aad:2012hg}
with that predicted by \Alpgen{} MC simulation.  For the resolved topology,
all
selection criteria
are applied to the data except for the $b$-tagging requirement. For the boosted topology, in order to decrease the statistical 
uncertainty on the scale factors, a relaxed set of selection criteria 
that does not include the $b$-tagging,
 $\Delta \phi(\mathrm{jet}, \ell)>2.3$, jet mass and \dsplit{} requirements is
used.  Any bias induced by relaxing the  criteria for the boosted selection is found to be
negligible compared to the statistical uncertainty in the scale factor determination.
The total number of $W$+jets events in data, $N_{W^
+} + N_{W^-}$, is given by:
\begin{equation}
N_{W^+} + N_{W^-} = \left(\frac{r_\mathrm{MC} + 1}{r_\mathrm{MC} - 1}\right)(D_{\mathrm{corr}+} - D_{\mathrm{corr}-}), 
\label{eq:Wchargeasymm}
\end{equation}
where $r_\mathrm{MC}$ is the ratio given by MC simulation 
of the number of $W$+jets events with a positively charged lepton to that
with a negatively charged lepton
and $D_{\mathrm{corr}+(-)}$ is
the number of observed events with a positively (negatively) charged lepton.  
Contributions to $D_{\mathrm{corr}+(-)}$ from charge-asymmetric processes such as single 
top, $WZ$ and \ttbar{}+$W$ boson production are estimated from MC simulation and are subtracted. 
Contributions from charge-symmetric processes such as $t\bar{t}$ production cancel in the 
difference on the right-hand side of eq.~(\ref{eq:Wchargeasymm}). A scale factor, $C_{\mathrm{A}}$, applied to the 
MC simulated samples of $W$ + jets events 
is then calculated as 
the ratio of $N_{W^+} + N_{W^-}$ evaluated from data to that predicted from MC simulation. 
The value and statistical uncertainty obtained for $C_{\mathrm{A}}$ in the electron (muon) channel are $1.026 \pm 0.011$ ($0.978 \pm 0.010$) with the resolved selection, 
and $0.89 \pm 0.06$ ($0.81 \pm 0.05$) with the boosted selection.

Scale factors for the relative fraction of heavy-flavour contributions from 
$W+b\bar{b}$, $W+c\bar{c}$, $W+c$ are also determined from data~\cite{Aad:2012qf,Aad:2011kp,Aad:2012hg}.  
In determining these scale factors, 
events are required to satisfy all selection criteria common to the boosted- and resolved- topology selections. 
Exactly two small-radius jets are required without any $b$-tagging requirement.

The flavour fractions for $W+b\bar{b}$, $W+c\bar{c}$, $W+c$ and $W+$light-quark flavours
are first determined from MC simulation for this sample. 
The ratio
of the $W+b\bar{b}$ to $W+c\bar{c}$ contribution is taken from MC simulation
and fixed at that value.
A system of three equations is used to fit to the two-jet data sample
with at least one $b$-tagged jet
in order to determine correction
factors for each of the flavour fractions determined from MC
simulation:

\begin{equation} 
  \left( 
  \begin{matrix} 
    C_{\mathrm{A}} \cdot (N_{\mathrm{MC},W^-}^{b\bar{b}} + N_{\mathrm{MC},W^-}^{c\bar{c}}) & C_{\mathrm{A}} \cdot N_{\mathrm{MC},W^-}^{c} & C_{\mathrm{A}} \cdot N_{\mathrm{MC},W^-}^{\text{light}} \\ 
             (      f_{b\bar{b}} +       f_{c\bar{c}}) &                f_{c} &                f_{\text{light}} \\ 
    C_{\mathrm{A}} \cdot (N_{\mathrm{MC},W^+}^{b\bar{b}} + N_{\mathrm{MC},W^+}^{c\bar{c}}) & C_{\mathrm{A}} \cdot N_{\mathrm{MC},W^+}^{c} & C_{\mathrm{A}} \cdot N_{\mathrm{MC},W^+}^{\text{light}} \\ 
  \end{matrix} 
  \right)  
  \cdot 
  \left(\begin{matrix} K_{b\bar{b},c\bar{c}} \\ K_c \\ K_{\text{light}} \end{matrix}\right)   
  = 
  \left(\begin{matrix} D_{W^-} \\ 1.0 \\ D_{W^+} \end{matrix}\right) 
  \label{eqMM}
\end{equation}  
where  $D_{W^\pm}$ is the expected number of $W$+jets events with a positively 
charged or negatively charged 
lepton in the data. The flavour fraction is  $f_\text{flavour}$,  and the correction factor for a given flavour component is $K_\text{flavour}$. The different flavour labels are $b\bar{b}$, $c\bar{c}$, $c$, and $\text{light}$ corresponding to $W+b\bar{b}$, $W+c\bar{c}$, $W+c$, and $W+$light-jets respectively.
The numbers of positively charged  
and negatively charged leptons in the data are found by subtracting 
all non-$W$+jets contributions, which are determined from MC simulations as 35\% (15\%) of the selected events for the electron (muon) channel. 
An iterative process is used to find the  
$K_\text{flavour}$ factors, which are then used to correct the corresponding 
flavour fractions $f_\text{flavour}$ that are applied during the $C_{\mathrm{A}}$ factor calculation. 
In this iterative process, only the $K_\text{flavour}$ and $C_{\mathrm{A}}$ factors are allowed to vary. 
The $K_\text{flavour}$ factors are initially set to unity, thus altering the $C_{\mathrm{A}}$ factor 
calculation.  New correction factors $K_\text{flavour}$ are calculated by inverting   
eq.~(\ref{eqMM}), and then the process is repeated ten times, with 
each repetition using the correction factors determined from the previous one. It was checked that using more than ten iterations produces 
only negligible changes in the extracted correction factors.

The correction factors found for the two-jet sample are extrapolated to higher jet multiplicities 
by keeping the same ratios between them while conserving the normalisation in each jet multiplicity bin.
The $K_\text{flavour}$ factors thus obtained are different from unity. For events containing electrons (muons), the extracted values and statistical uncertainties of $K_{b\bar{b}}$ and $K_{c\bar{c}}$ are $1.36 \pm 0.07 $ ($1.51 \pm 0.08$), $K_{c}$ is  $0.71 \pm 0.03$ ($0.66 \pm 0.03$), and $K_{\text{light}}$ is $0.934 \pm 0.005$ ($0.873 \pm 0.004$).

This  method reduces the systematic
uncertainties (from the jet energy scale, $b$-tagging and
other uncertainties) compared to using $W$+jets MC simulation alone.
Systematic uncertainties in the $W+$jets normalisation and flavour-fraction corrections are determined by rederiving these scale factors
when a given systematic effect is applied. The new scale factors
are then used in producing the \mttbarreco\ mass spectrum for that
particular systematic uncertainty.

\subsection{Multijet estimate}
The multi-jet background in events satisfying the resolved
or boosted selection criteria consists of events with a 
jet that is misreconstructed 
as a lepton or with a non-prompt lepton that satisfies the identification
criteria.
The normalisation, \mttbarreco\ shape,  and statistical and systematic uncertainties 
associated with the multi-jet background are estimated 
from data using a {\em{matrix method}}~\cite{Aad:2012qf, ATLAS-CONF-2014-058}.
 
The matrix method utilises efficiencies for leptons produced by prompt and non-prompt sources.  
The efficiency $f$ is defined as the probability that a non-prompt lepton from multi-jet production that satisfies the loose identification criteria~\cite{ATLAS-CONF-2014-058} also satisfies the tight identification criteria.
It is derived from data in control regions dominated by multi-jet events, with prompt-lepton contributions subtracted based on MC simulations.
The efficiency $\epsilon$ is defined as the probability that a lepton from prompt sources ($W$ or $Z$ bosons) that satisfies the loose identification criteria also satisfies the tight identification criteria.
It is determined using SM  MC samples with a mixture of processes similar to that in the signal region, 
corrected using data versus MC correction factors derived from $Z \rightarrow \ell\ell$ events.

The number of multi-jet background events satisfying the resolved or boosted selection criteria is estimated using
data events that satisfy all selection criteria, except that the loose lepton identification criteria are used.
This sample contains prompt as well as non-prompt leptons.  

The number of events with leptons satisfying the loose identification criteria, $N_\mathrm{L}$ is defined as
\begin{equation} 
N_\mathrm{L}=N_\text{prompt}+N_\text{multi-jet} 
\end{equation} 
where $N_\text{prompt}$ is the number of events with prompt leptons satisfying the loose identification criteria and $N_\text{multi-jet}$ is the number of events satisfying the loose identification criteria with leptons from other sources. The number of events satisfying the tight identification criteria, $N_\mathrm{T}$ is then
\begin{equation} 
N_\mathrm{T}=\epsilon{}\times{}N_\text{prompt}+f\times{}N_\text{multi-jet}. 
\end{equation} 
Solving these two equations for $N_\text{prompt}$ and $N_\text{multi-jet}$ gives the multi-jet
contribution from events satisfying all the selection criteria.

Good shape modelling of the \mttbarreco\ distributions is achieved by parameterising the efficiencies as functions of relevant kinematic variables, and validated in the multi-jet control regions.
Systematic uncertainties are evaluated using several different definitions of multi-jet control regions that result in slightly different $f$ estimations.
Systematic uncertainties associated with object reconstruction and MC simulation are also considered, resulting in a total normalisation uncertainty of 20\%.

\section{Systematic uncertainties}
\label{sec:systematics}

The systematic uncertainties can be broadly divided into two categories: uncertainties that affect reconstructed objects (such as jets) and uncertainties that affect the modelling of certain background or signal processes. Some of the uncertainties affect both the shape and the normalisation of the \mttbarreco\ spectra, while others affect the normalisation only. 
In table~\ref{tab:systimpact}, an overview of the effects of the dominant systematic uncertainties on the background and signal yields is given. 
Only the impact on the overall normalisation is shown in the table, but some of the systematic uncertainties have a significant dependence on the reconstructed \ttbar\ mass, which is fully taken into account in the analysis for all of these uncertainties. The systematic uncertainties with the strongest dependence are those from the jet energy scale, parton distribution functions and $b$-jet identification.

\begin{table}[tbh]
\begin{center}
\caption{
Average impact of the dominant systematic uncertainties on the total background yield and on the estimated yield for a \Zprime\ sample with $m=1.75$ \TeV. 
The electron and muon channel spectra are added. 
Any shift in yield is given in percent of the nominal value. 
Certain systematic uncertainties are not applicable to the \Zprime\ samples, which is indicated with a bar ($-$) in the table.
}
\vspace{2mm}

\footnotesize{
\begin{tabular}{l|r|r|r|r}
\hline
\hline
                  & \multicolumn{2}{c|}{Resolved selection} & \multicolumn{2}{c}{Boosted selection} \\
                  & \multicolumn{2}{c|}{yield impact [\%]} & \multicolumn{2}{c}{yield impact [\%]}  \\
Systematic Uncertainties  & total bkg. & \Zprime  \hspace{1ex}& total bkg. & \Zprime \hspace{1ex}\\
\hline
\hline
Luminosity                         &  $2.5$  	&  $2.8$  	&  $2.6$  	&  $2.8$  	\\ \hline                 
PDF                                &  $2.4$  	&  $3.6$  	&  $4.7$  	&  $2.3$  	\\ \hline 
ISR/FSR                            &  $3.7$  	&  $-$  	&  $1.2$  	&  $-$  	\\ \hline 
Parton shower and fragmentation    &  $4.8$  	&  $-$  	&  $1.5$  	&  $-$  	\\ \hline     
\ttbar\ normalisation              &  $5.3$  	&  $-$  	&  $5.5$  	&  $-$  	\\ \hline 
\ttbar\ EW virtual correction      &  $0.2$  	&  $-$  	&  $0.5$  	&  $-$  	\\ \hline 
\ttbar\ generator                  &  $0.3$  	&  $-$  	&  $2.6$  	&  $-$  	\\ \hline 
\ttbar\ top quark mass             &  $0.6$  	&  $-$  	&  $1.4$  	&  $-$  	\\ \hline    
$W$+jets generator                 &  $0.3$  	&  $-$  	&  $0.1$  	&  $-$  	\\ \hline 
Multi-jet normalisation, $e$+jets           &  $0.5$  	&  $-$  	&  $0.2$  	&  $-$  	\\ \hline 
Multi-jet normalisation, $\mu$+jets         &  $0.1$  	&  $-$  	&  $<0.1$  	&  $-$  	\\ \hline 
JES+JMS, large-radius jets         &  $0.1$  	&  $2.1$  	&  $9.7$  	&  $2.8$  	\\ \hline 
JER+JMR, large-radius jets         &  $<0.1$  	&  $0.3$  	&  $1.0$  	&  $0.2$  	\\ \hline 
JES, small-radius jets             &  $5.6$  	&  $2.6$  	&  $0.4$  	&  $1.4$  	\\ \hline 
JER, small-radius jets             &  $1.8$  	&  $1.4$  	&  $<0.1$  	&  $0.2$  	\\ \hline 
Jet vertex fraction                &  $0.8$  	&  $0.8$  	&  $0.2$  	&  $<0.1$  	\\ \hline 
$b$-tagging $b$-jet efficiency     &  $1.1$  	&  $2.0$  	&  $2.9$  	&  $17.1$  	\\ \hline 
$b$-tagging $c$-jet efficiency     &  $0.1$  	&  $0.7$  	&  $0.1$  	&  $2.1$  	\\ \hline 
$b$-tagging light-jet efficiency   &  $<0.1$  	&  $<0.1$  	&  $0.5$  	&  $0.2$  	\\ \hline 
Electron efficiency                &  $0.3$  	&  $0.6$  	&  $0.6$  	&  $1.3$  	\\ \hline 
Muon efficiency                    &  $0.9$  	&  $1.0$  	&  $1.0$  	&  $1.1$  	\\ \hline 
MC statistical uncertainty         &  $0.4$  	&  $6.0$  	&  $1.3$  	&  $1.8$  	\\ \hline         
\hline                                                                                                       
All systematic uncertainties       &  $10.8$  	&  $8.8$  	&  $13.4$  	&  $18.0$  	\\ \hline 
\hline
\end{tabular}
}
\label{tab:systimpact}
\end{center}
\end{table}

The dominant uncertainty on the normalisation of the total background estimate is the NNLO+NNLL \ttbar\ cross-section uncertainty of 6.5\%.   
This uncertainty includes renormalisation and factorisation scale uncertainties, combined PDF and strong-coupling uncertainties evaluated following the PDF4LHC~\cite{Botje:2011sn} recommendations, and uncertainties associated with the value of the top quark mass.   
The combined PDF and strong-coupling uncertainties are extracted for each of the three PDF sets: MSTW2008 68\% confidence level (CL) NNLO~\cite{Martin:2009iq,Martin:2009bu}, CT10 NNLO~\cite{Lai:2010vv,Gao:2013xoa} and NNPDF2.3 NNLO~\cite{Ball:2012cx}; the total uncertainty associated
with the PDF and strong-coupling uncertainties is one half of the size of the envelope of the three resultant error bands, with the central prediction being the midpoint of the envelope. 
Variations from changing the top quark mass by $\pm 1.0\ $GeV are added in quadrature to the scale uncertainties, and combined
PDF and strong-coupling uncertainties.  

The $W$+jets normalisation, determined using data, is separately evaluated for each experimental source of systematic uncertainty. An additional systematic uncertainty on the prediction is evaluated by using the simulated samples generated with varied \Alpgen{} matching parameters.
The normalisation uncertainty on the multi-jet background is $20$\%, as described in section~\ref{sec:backgrounds}.

The single-top-quark background normalisation uncertainty is $7.7$\%~\cite{Kidonakis:2011wy,Kidonakis:2010ux, Kidonakis:2010tc}. 
The normalisation uncertainty on the $Z$+jets sample is $48$\%, estimated using Berends--Giele scaling~\cite{Aad:2010ey}. 
The diboson normalisation uncertainty is 34\%, which is a combination of the NLO PDF and scale uncertainties, and additional uncertainties from the requirements on the jet multiplicity.

The uncertainty on the integrated luminosity is $2.8$\%. It is
derived, following the same methodology as that detailed in
Ref.~\cite{AtlasLumi}, from a preliminary calibration of the
luminosity scale derived from beam-separation scans performed in
November 2012.
This uncertainty is applied to all signal and 
background samples except multi-jet and $W$+jets, which are estimated from data.

The effect of the PDF uncertainty on all MC samples is estimated by taking the envelope of the NNPDF2.3, MSTW2008NLO 
 and CT10 PDF set uncertainties at 68\% CL\footnote{The CT10 PDF uncertainties are scaled down by a factor 1.645 to reach an approximate 68\% CL.} following the PDF4LHC recommendation
  and normalising to the nominal cross-section. 
The PDF uncertainty on the \ttbar\ mass spectrum has a much larger effect on the boosted sample than on the resolved sample.
The effect on the total background yield is 2.1\% (4.2\%) after the resolved (boosted) selection. The size  of the uncertainty grows with reconstructed mass, attaining values of 50\% above 2 TeV in the boosted selection.

One of the dominant uncertainties affecting reconstructed objects is the jet energy scale (JES) uncertainty, especially for large-radius jets~\cite{Aad:2012ef,ATLAS-CONF-2012-065}. 
Uncertainties on the jet mass scale (JMS) and the \kt\ splitting scales~\cite{ATLAS-CONF-2012-065} are also important for this analysis. 
These uncertainties have an impact of 10\% on the overall background yield in the boosted selection. 
The impact on the background estimates falls with increasing \mttbarreco{}, varying from 22\% at lowest masses to about 7\% above $1.5$ TeV.
The impact is smaller for the resolved selection, since the large-radius jets are only used indirectly there via the vetoing of events that satisfy the boosted selection. 
For large-radius jets, uncertainties on the jet energy resolution (JER) and jet mass resolution (JMR) are also considered. These are less significant than the JES.
For small-radius jets, the uncertainties on the JES, the jet reconstruction efficiency and the jet energy resolution are considered~\cite{Aad:2011he,Aad:2014bia}. The small-radius JES uncertainty is one of the most significant systematic uncertainties in the resolved selection, affecting the overall expected yield by 6\%.  The effect of uncertainties associated with the jet vertex fraction is also considered. The $b$-tagging uncertainty is modelled through simultaneous variations of the uncertainties on the efficiency and rejection~\cite{ATLAS-CONF-2012-043,ATLAS-CONF-2012-040}. 
An additional $b$-tagging uncertainty\footnote{The additional $b$-tagging uncertainty is an extrapolation of the uncertainty from regions of lower \pt. 
Depending on the \pt\ of the jet, it is approximately 12\% to 33\% for $b$-jets and 17\% to 30\% for $c$-jets, 
added in quadrature to the uncertainty on the jet $b$-tagging efficiency correction factor for the \mbox{200--300 GeV\ }region. } 
is applied for high-momentum jets ($\pt > 300$~\GeV) to account for uncertainties on the modelling of the track reconstruction in high-\pt\ environments, 
which is one of the dominant uncertainties for high-mass signals.  
 
For the leptons, the uncertainties on the isolation efficiency, the single-lepton trigger and the reconstruction efficiency are estimated using $Z \rightarrow ee$ and $Z \rightarrow \mu\mu$ events. 
In addition, high-jet-multiplicity $Z \rightarrow \ell \ell $ events are studied, from which extra uncertainties on the isolation efficiency are assigned to account for the difference between $Z$ and \ttbar\ events.
Uncertainties on the \MET\ reconstruction, as well as on the energy scale and resolution of the leptons, are also considered, and generally have a smaller impact on the yield and 
the search sensitivity than the uncertainties mentioned above.

The uncertainty on the \ttbar\ background due to uncertainties on the modelling of QCD initial- and final-state radiation (ISR/FSR) is estimated using \AcerMC{}  v3.8~\cite{SAMPLES-ACER} plus \Pythia{} v6.426 MC samples by varying the \Pythia{} ISR and FSR parameters 
within ranges allowed by a previous ATLAS 
measurement of \ttbar\ production with a veto on additional central jet activity~\cite{ATLAS:2012al}. 
The QED ISR/FSR uncertainty is negligible at this level. 
The dependency of the \mttbarreco{} shape on the choice of NLO generator is accounted for by using the difference between samples generated with \MCNLO{} and \Powheg{}+\Herwig{} as a systematic uncertainty. The parton showering and fragmentation uncertainty on the \ttbar\ background is estimated by comparing the results from a sample generated with \Powheg{}~\cite{Frixione:2007vw}, interfaced with \Pythia{} or \Herwig{}. 
The uncertainty on the \mttbarreco{} distribution arising from the uncertainty in the top quark mass is 
evaluated by comparing the \mttbarreco{} spectrum using the nominal sample to those generated with top quark masses of 170 and 175\ GeV, 
and multiplying the difference by 0.4 (to approximate a one standard deviation uncertainty, corresponding to $\pm$1.0\ GeV). 
The uncertainty on the electroweak corrections to top quark pair production is modelled by changing the difference of each correction factor from unity by $\pm10\%$.

For the $W$+jets background, the uncertainty on the \mttbarreco{} distribution is estimated by reweighting the events to the kinematics of MC samples generated 
with a different matching scale or functional form of the factorisation scale~\cite{Mangano:2002ea}.

\section{Comparison of data with expected background contributions}
\label{sec:expectations}

After all event selection criteria are applied, \NdataResolvedTot\ events remain for the resolved 
topology and \NdataBoostedTot\ events are selected for the boosted topology.  
The event yields from data and from expected background processes are listed in table~\ref{tab:datamcyield} together with the associated systematic uncertainties.

\begin{table}[tbh!]
\begin{center}
\caption{Data and expected background event yields after the resolved and boosted selections. The 
sum in quadrature of all systematic uncertainties on the expected background yields is also given. }
\vspace{2mm}
\begin{tabular}{lr@{$~\pm~$}lr@{$~\pm~$}lr@{$~\pm~$}l}
\hline\hline
\multicolumn{7}{c}{Resolved-topology selection}\\
Type          & \multicolumn{2}{c}{$e$+jets}  & \multicolumn{2}{c}{$\mu$+jets} & \multicolumn{2}{c}{Sum} \\  \hline 
\hline
\ttbar & 93,000 &  11,000 & 91,000 &  11,000 & 184,000 &  22,000 \\ \hline
Single top & 3,800 &  500 & 3,800 &  500 & 7,600 &  1,000 \\ \hline
$t\bar{t}V$ & 274 &  40 & 267 &  40 & 541 &  80 \\ \hline
Multi-jet $e$ & 5,300 &  1,100 & \multicolumn{2}{c}{--} & 5,300 &  1,100 \\ \hline
Multi-jet $\mu$ & \multicolumn{2}{c}{--} & 1,050 &  240 & 1,050 &  240 \\ \hline
$W$+jets & 6,600 &  800 & 7,100 &  800 & 13,700 &  1,500 \\ \hline
$Z$+jets & 1,400 &  750 & 650 &  340 & 2,000 &  1,080 \\ \hline
Dibosons & 320 &  120 & 310 &  120 & 620 &  240 \\ \hline
Total & 110,000 &  12,000 & 105,000 &  12,000 & 215,000 &  24,000 \\ \hline
\hline
Data & \multicolumn{2}{c}{\NdataResolvedEle{}} & \multicolumn{2}{c}{\NdataResolvedMuo{}} & \multicolumn{2}{c}{\NdataResolvedTot{}} \\
\hline\hline
\multicolumn{7}{c}{Boosted-topology selection}\\
Type          & \multicolumn{2}{c}{$e$+jets}  & \multicolumn{2}{c}{$\mu$+jets} & \multicolumn{2}{c}{Sum} \\  \hline\hline  
\ttbar & 4,100 &  600 & 4,000 &  600 & 8,100 &  1,200 \\ \hline
Single top & 138 &  20 & 154 &  20 & 290 &  40 \\ \hline
$t\bar{t}V$ & 37 &  6 & 38 &  7 & 75 &  13 \\ \hline
Multi-jet $e$ & 91 &  18 & \multicolumn{2}{c}{--} & 91 &  18 \\ \hline
Multi-jet $\mu$ & \multicolumn{2}{c}{--} & 8.6 &  1.6 & 8.6 &  1.6 \\ \hline
$W$+jets & 260 &  50 & 290 &  50 & 550 &  100 \\ \hline
$Z$+jets & 31 &  16 & 17 &  9 & 48 &  25 \\ \hline
Dibosons & 21 &  8 & 20 &  8 & 41 &  16 \\ \hline
Total & 4,700 &  600 & 4,500 &  600 & 9,200 &  1,200 \\ \hline
\hline
Data & \multicolumn{2}{c}{\NdataBoostedEle{}} & \multicolumn{2}{c}{\NdataBoostedMuo{}} & \multicolumn{2}{c}{\NdataBoostedTot{}} \\\hline
\hline
\end{tabular}
\label{tab:datamcyield}
\end{center}
\end{table} 

\begin{figure}
\centering
\subfigure[~$e$+jets channel.]{
\includegraphics[width=0.45\textwidth]{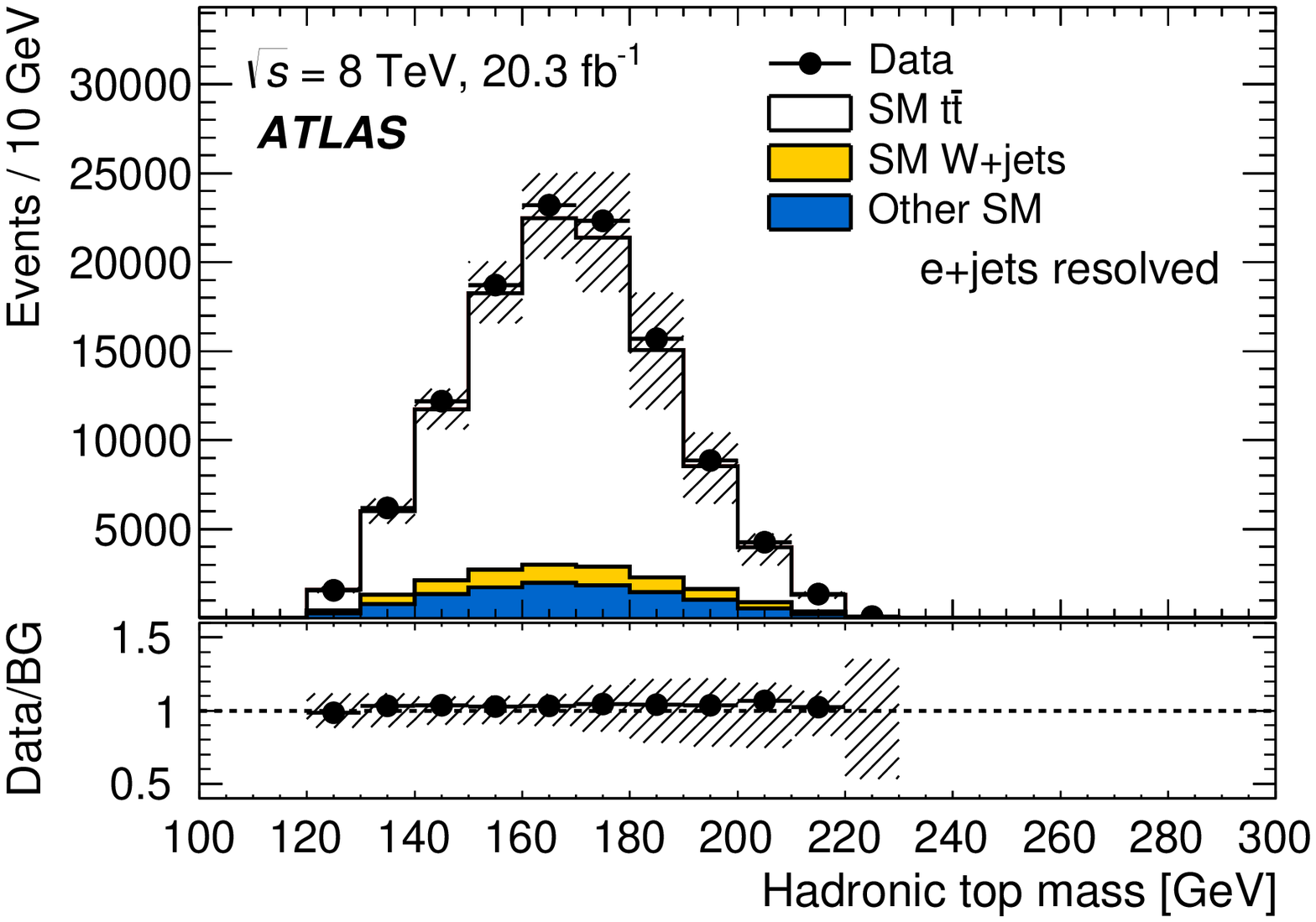}
}
\subfigure[~$\mu$+jets channel.]{
\includegraphics[width=0.45\textwidth]{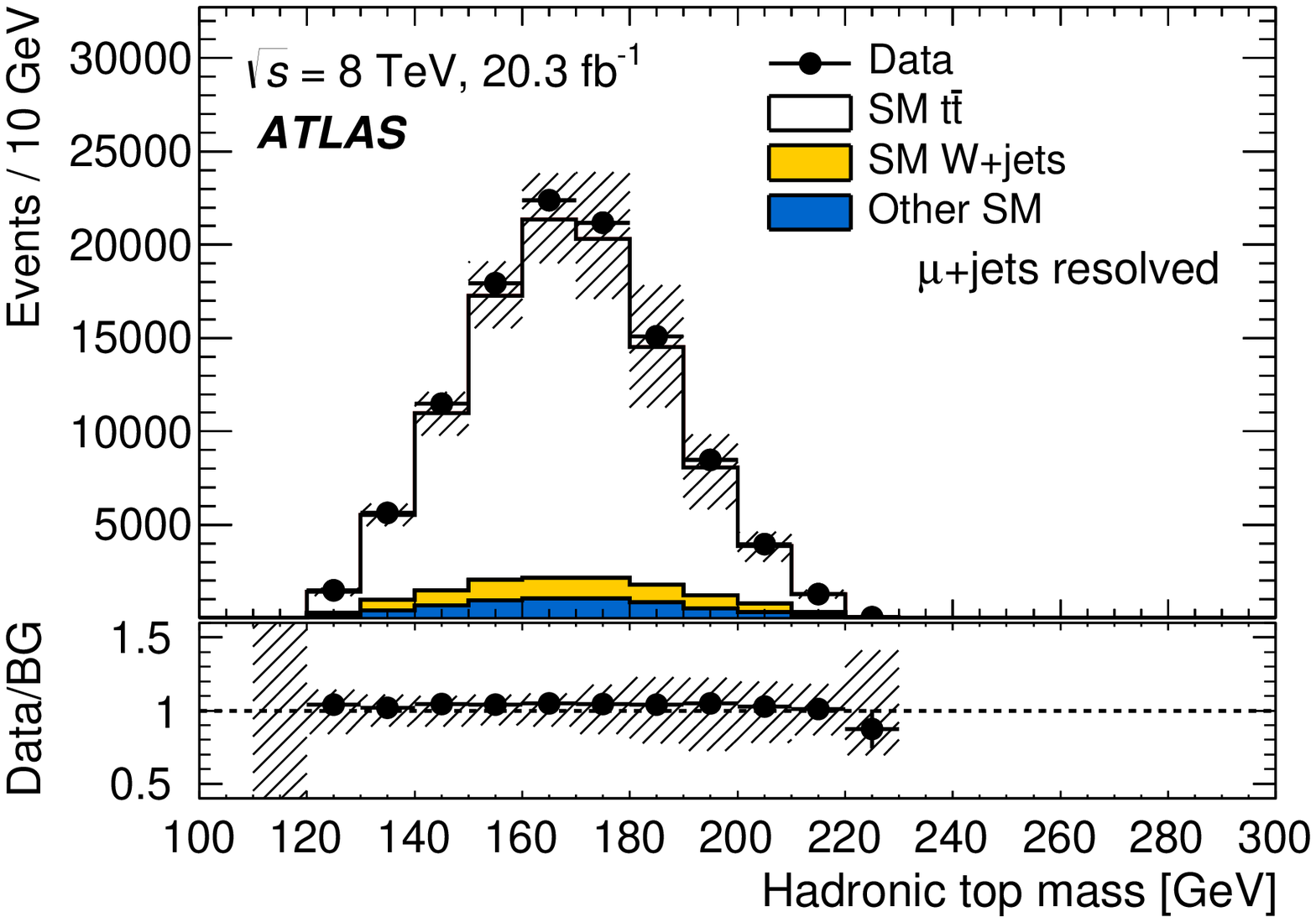}
}
\subfigure[~$e$+jets channel.]{
\includegraphics[width=0.45\textwidth]{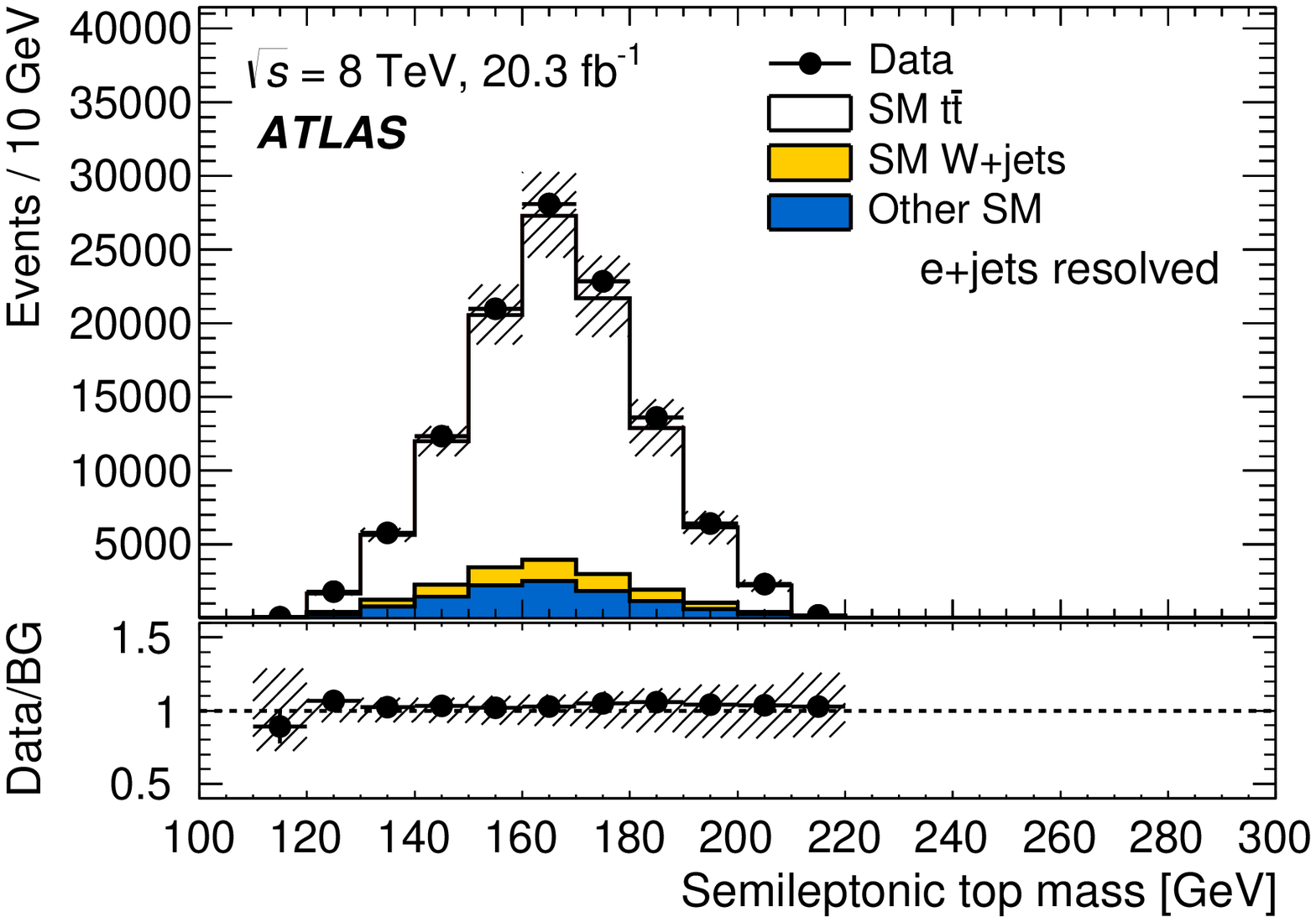}
}
\subfigure[~$\mu$+jets channel.]{
\includegraphics[width=0.45\textwidth]{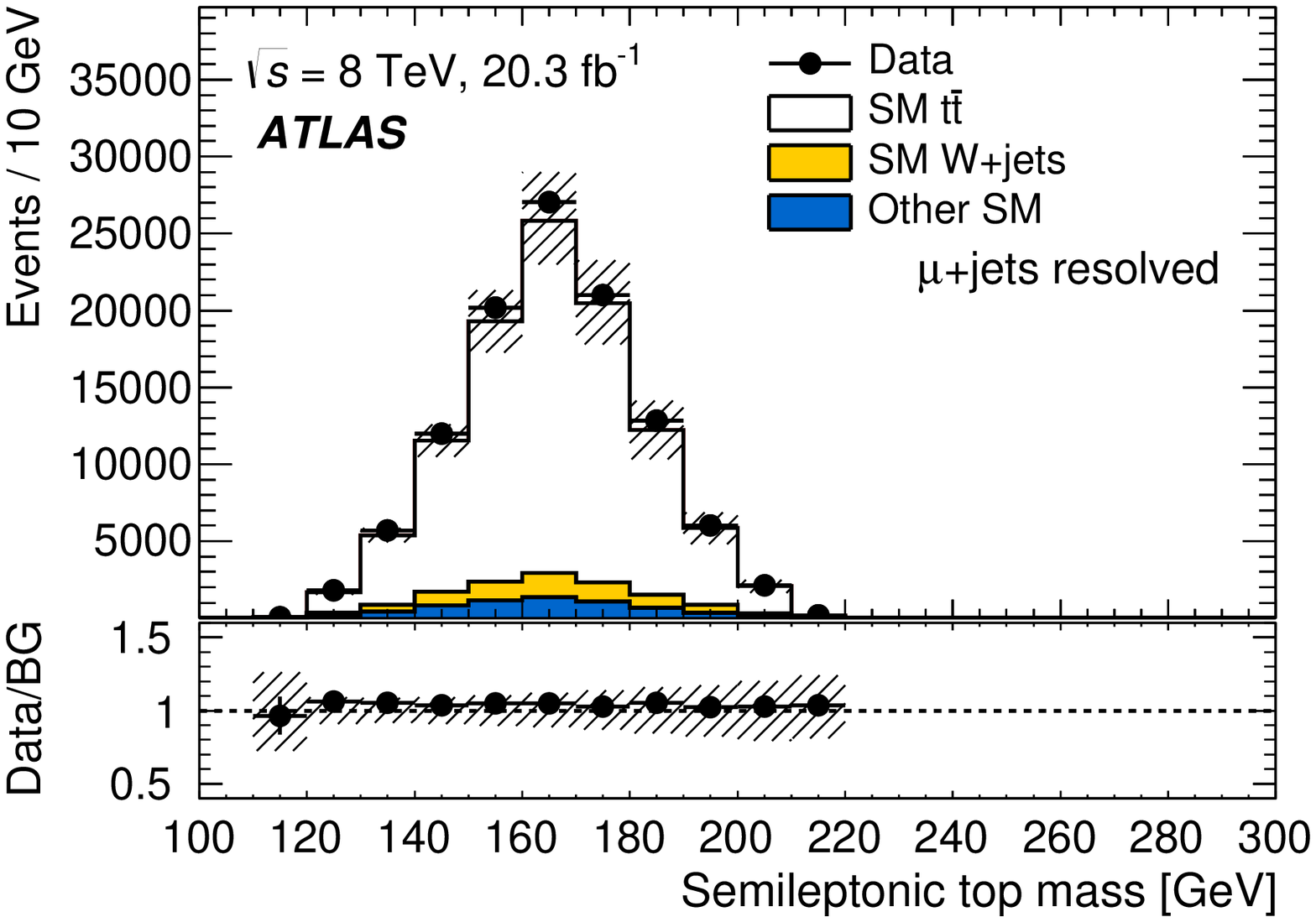}
}
\subfigure[~$e$+jets channel.]{
\includegraphics[width=0.45\textwidth]{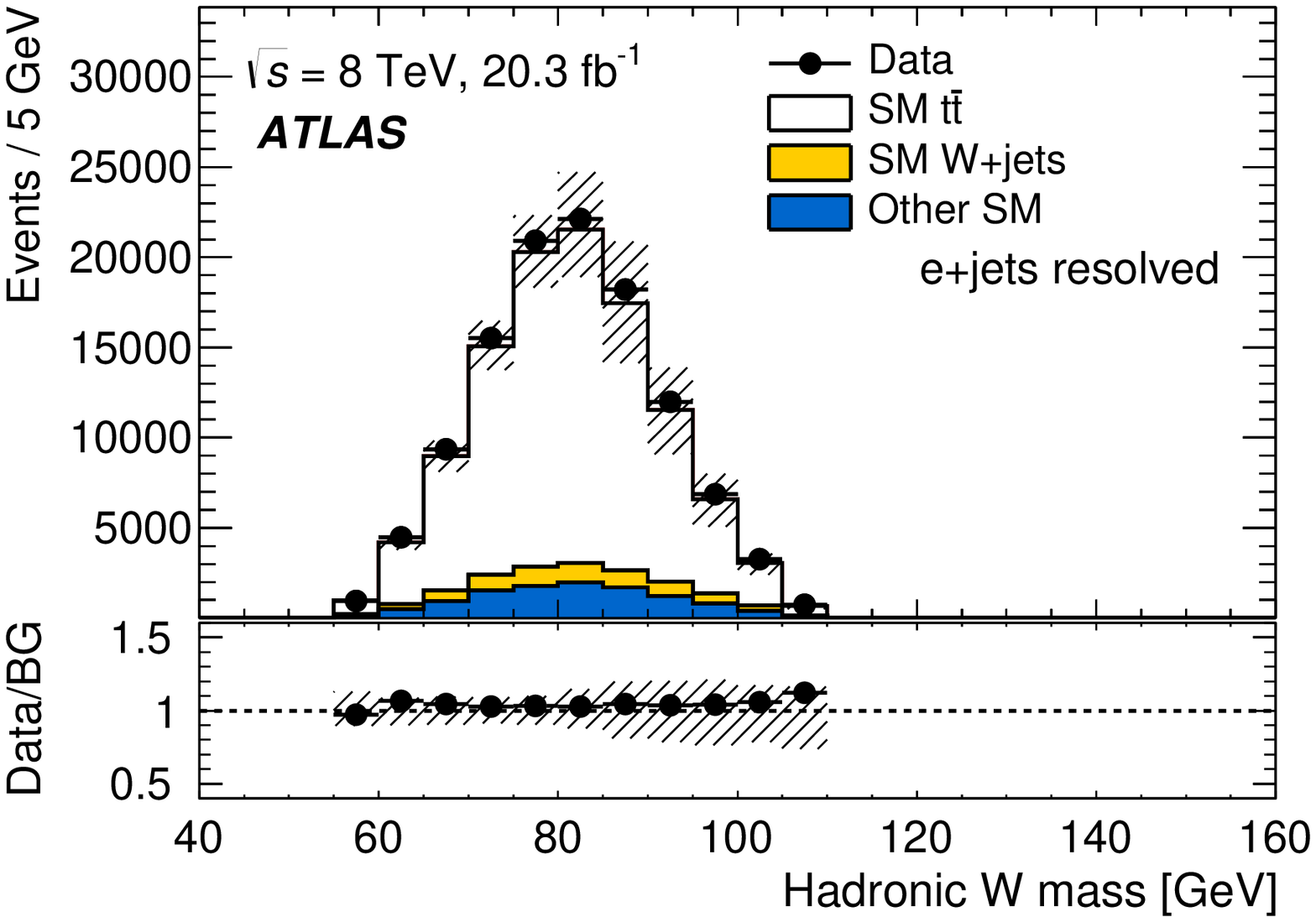}
}
\subfigure[~$\mu$+jets channel.]{
\includegraphics[width=0.45\textwidth]{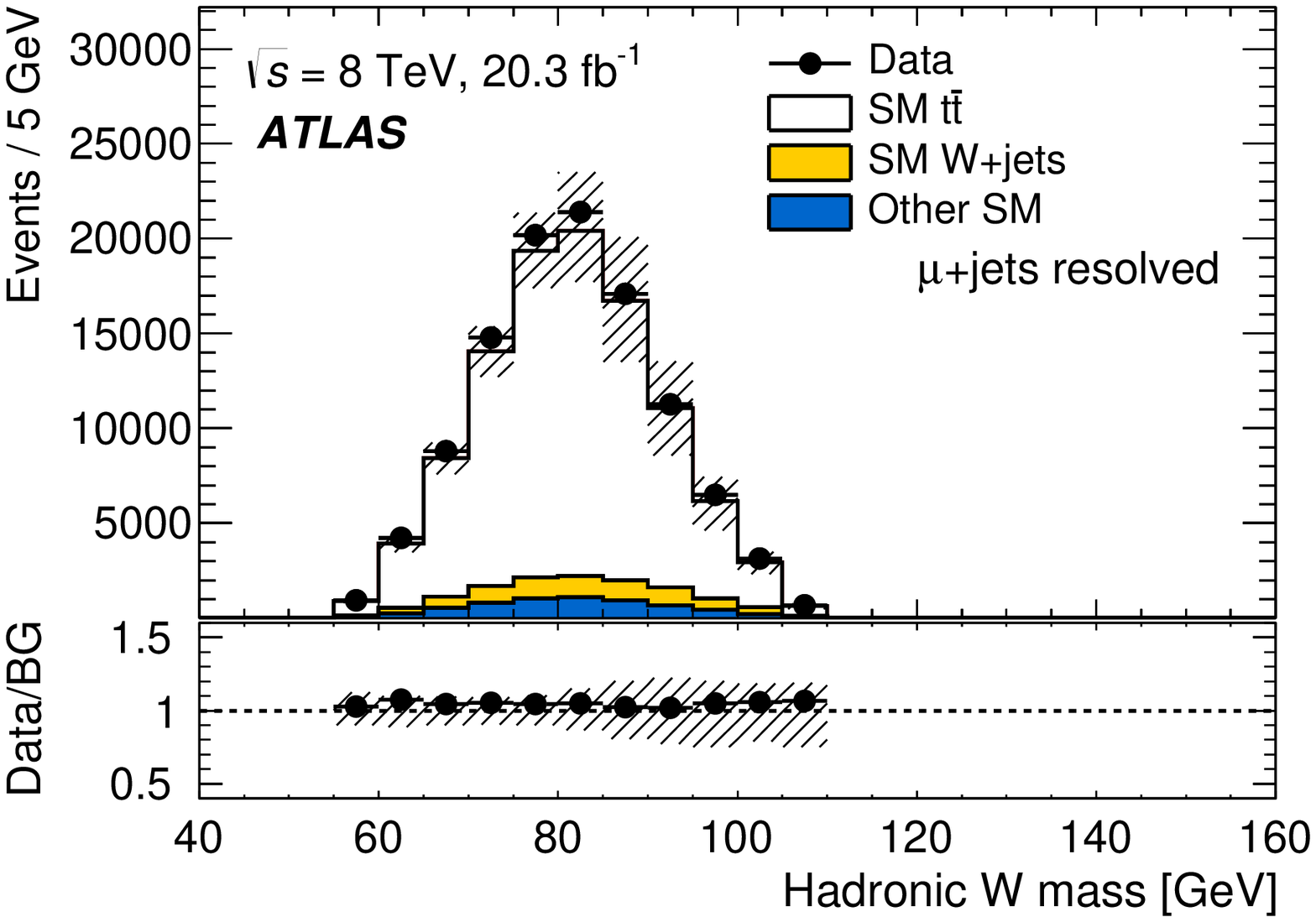}
}
\caption{Reconstructed  mass of the hadronically decaying top quark candidate, 
semileptonically decaying top quark candidate, 
and hadronically decaying $W$-boson candidate
after the resolved-topology selection in the electron and muon channels. The SM background components are shown as stacked histograms. The shaded areas indicate the total systematic uncertainties. 
 \label{fig:topmass_resolved}}
\end{figure}

\begin{figure}[tbp]
\begin{center}
\subfigure[~$e$+jets channel.]{
\label{F:hadtop_mass_ele}
\includegraphics[width=0.45\textwidth]{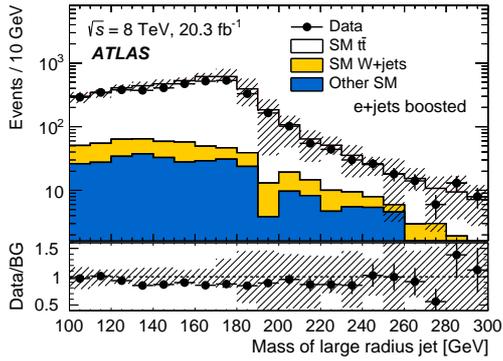}
}
\subfigure[~$\mu$+jets channel.]{
\label{F:hadtop_mass_muo}
\includegraphics[width=0.45\textwidth]{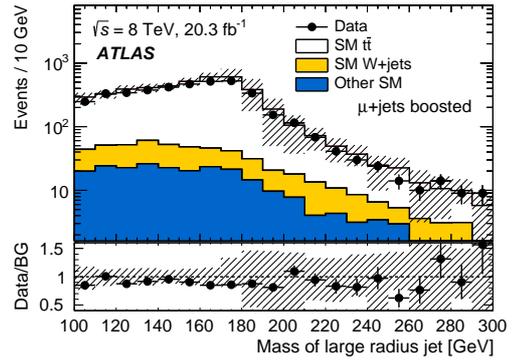}
}

\subfigure[~$e$+jets channel.]{
\label{F:leptop_mass_el}
\includegraphics[width=0.45\textwidth]{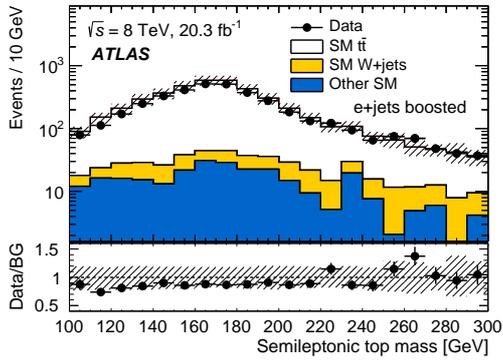}
}
\subfigure[~$\mu$+jets channel.]{
\label{F:leptop_mass_mu}
\includegraphics[width=0.45\textwidth]{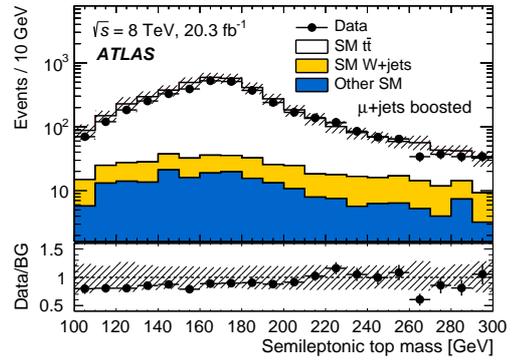}
}
\caption{
The invariant mass of the large-radius jets and the invariant mass of the semileptonically decaying top quark candidate,  
after the boosted selection.  
The SM background components are shown as stacked histograms. The shaded areas indicate the total systematic uncertainties.  
\label{fig:topmass_boosted} }
\end{center}
\end{figure}

\begin{figure}[tbp]
\begin{center}
\subfigure[~$e$+jets channel.]{
\label{F:top_pt_el}
\includegraphics[width=0.45\textwidth]{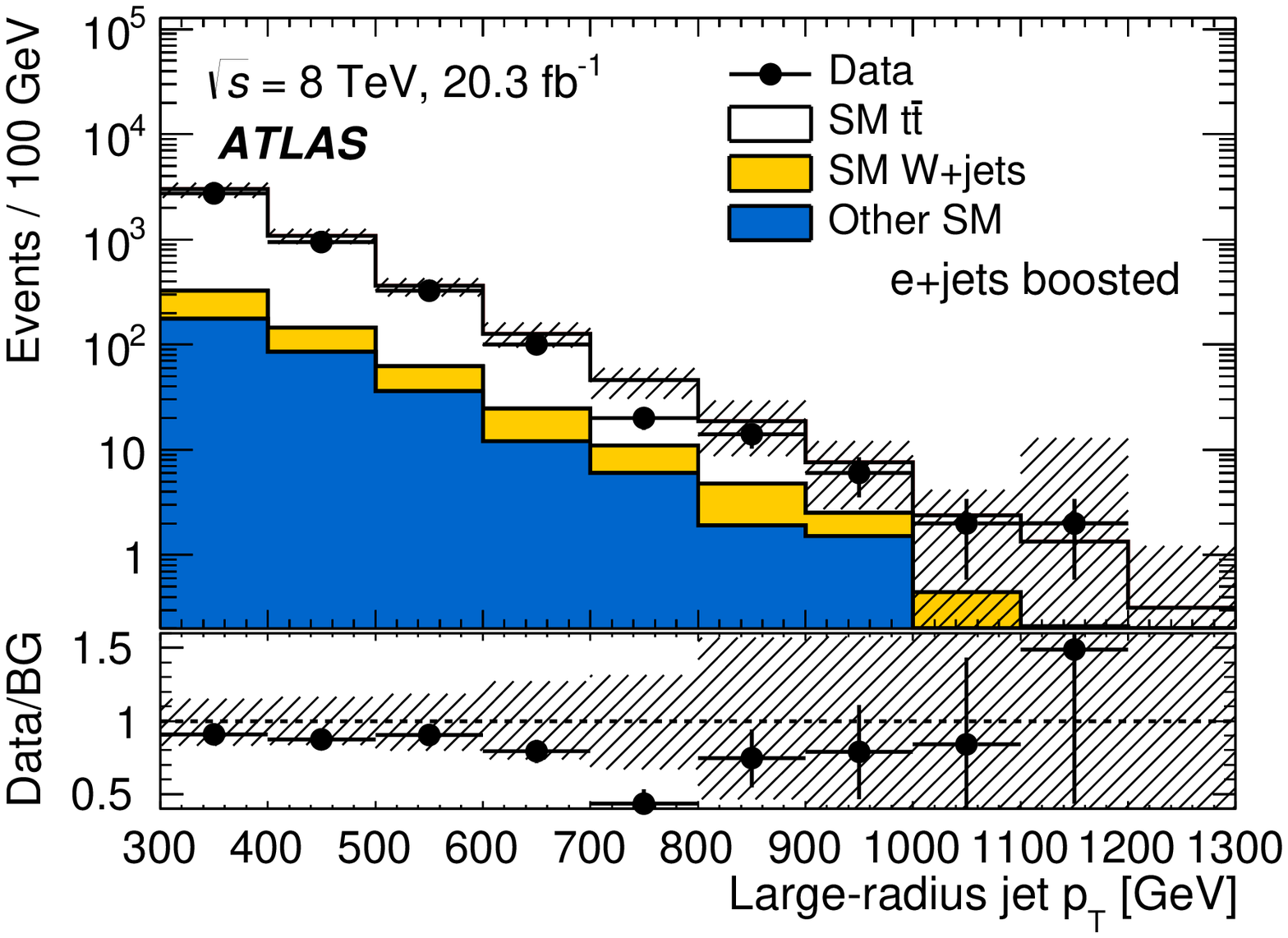}
}
\subfigure[~$\mu$+jets channel.]{
\label{F:top_pt_mu}
\includegraphics[width=0.45\textwidth]{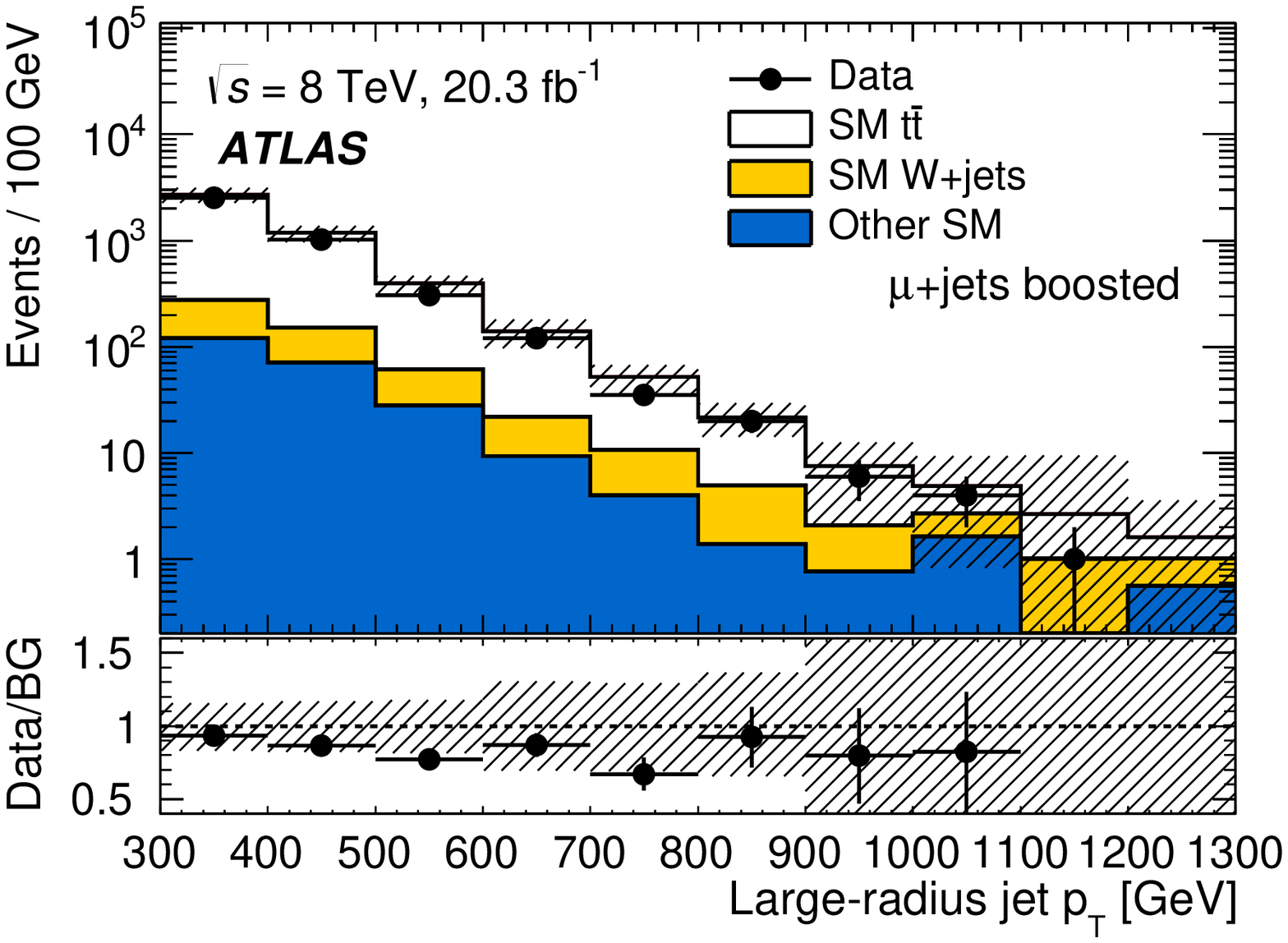}
}
\label{fig:top_mass_pt_boosted} 
\subfigure[~$e$+jets channel.]{
\label{F:hadtop_split_ele}
\includegraphics[width=0.45\textwidth]{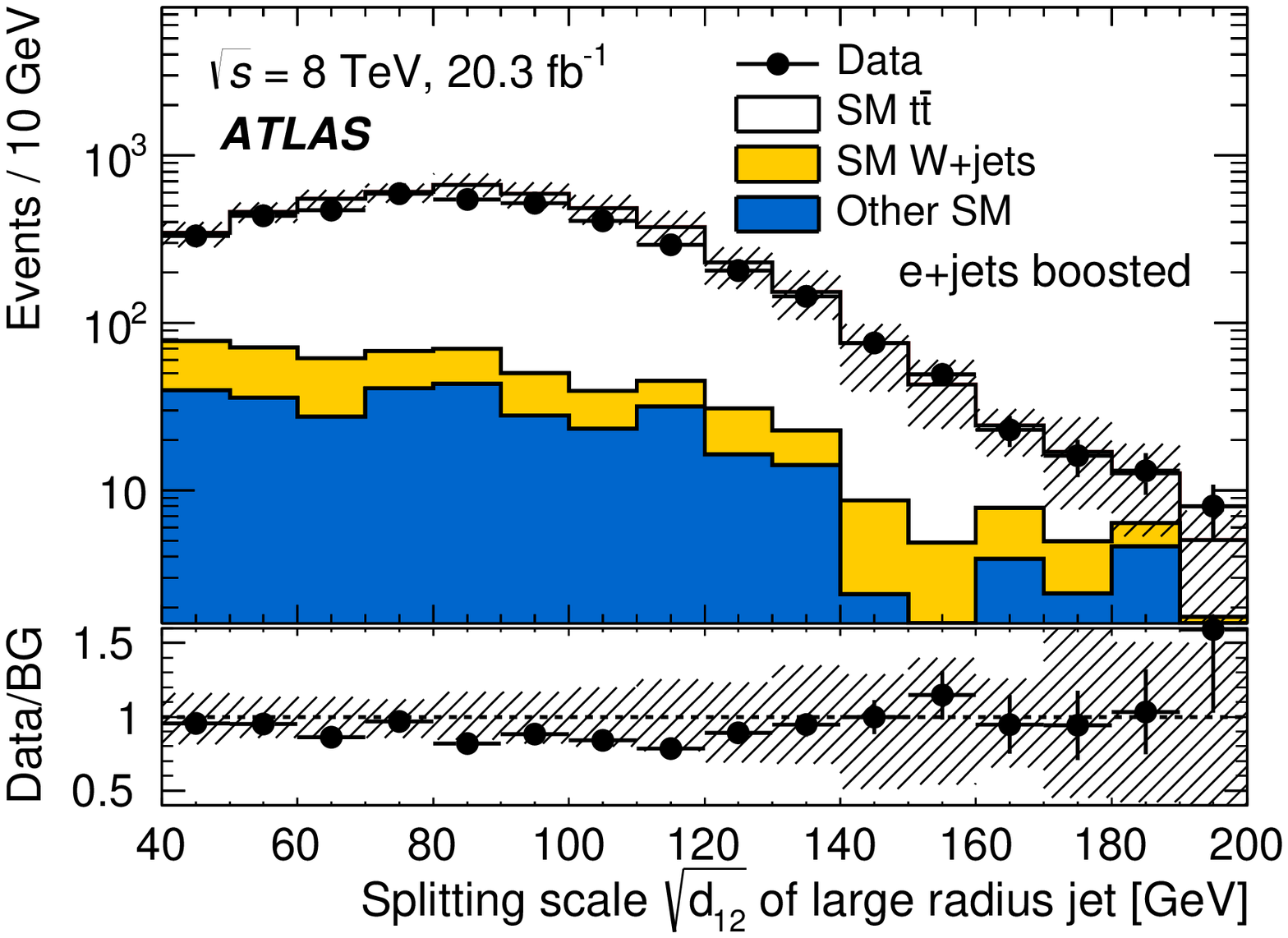}
}
\subfigure[~$\mu$+jets channel.]{
\label{F:hadtop_split_muo}
\includegraphics[width=0.45\textwidth]{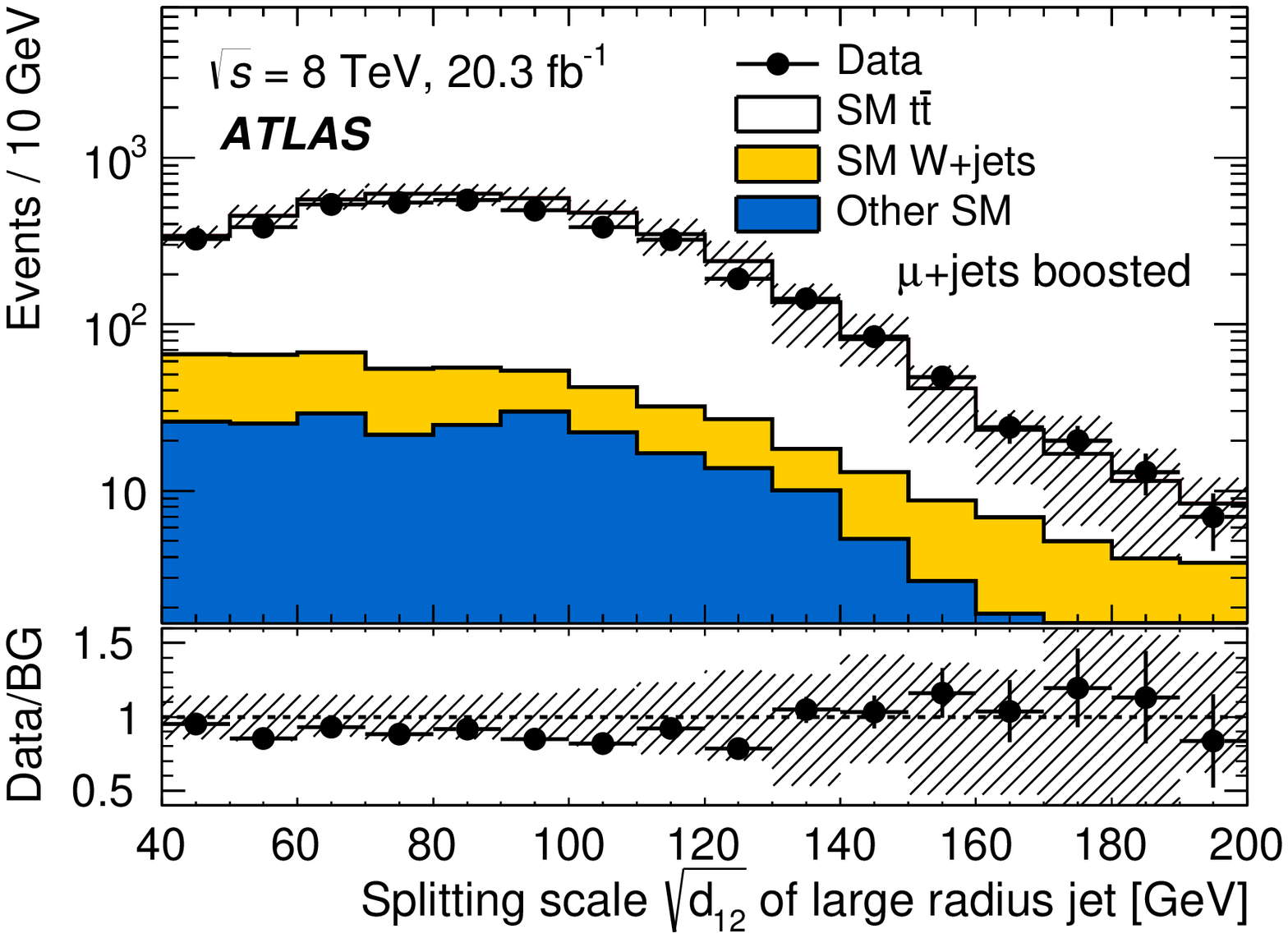}
}
\caption{
The transverse momentum, $p_{\mathrm{T}}$ , and first \kt\ splitting scale, \dsplit, of the large-radius jet after the boosted selection.  The SM background components are shown as stacked histograms. The shaded areas indicate the total systematic uncertainties. 
\label{fig:hadtopsplit_boosted} }
\end{center}
\end{figure}

Good agreement is observed between the data and the total expected background. 
Figure~\ref{fig:topmass_resolved} shows the reconstructed mass of the semileptonically and hadronically decaying top quark candidates, and the mass of the hadronically decaying $W$ boson candidate for the resolved selection. The equivalent distributions for the boosted selection are the mass of the semileptonically decaying top quark candidate and the mass of the large-radius jet; both are shown in figure~\ref{fig:topmass_boosted}.  Figure~\ref{fig:hadtopsplit_boosted} shows the distribution of the transverse momentum and first \kt{} splitting scale of the selected large-radius jets.

\subfigcapmargin = 0.5cm

\begin{figure}
\centering
\subfigure[Electron channel, resolved selection, both top candidates have $b$-tagged jets.]{
\includegraphics[width=0.43\textwidth]{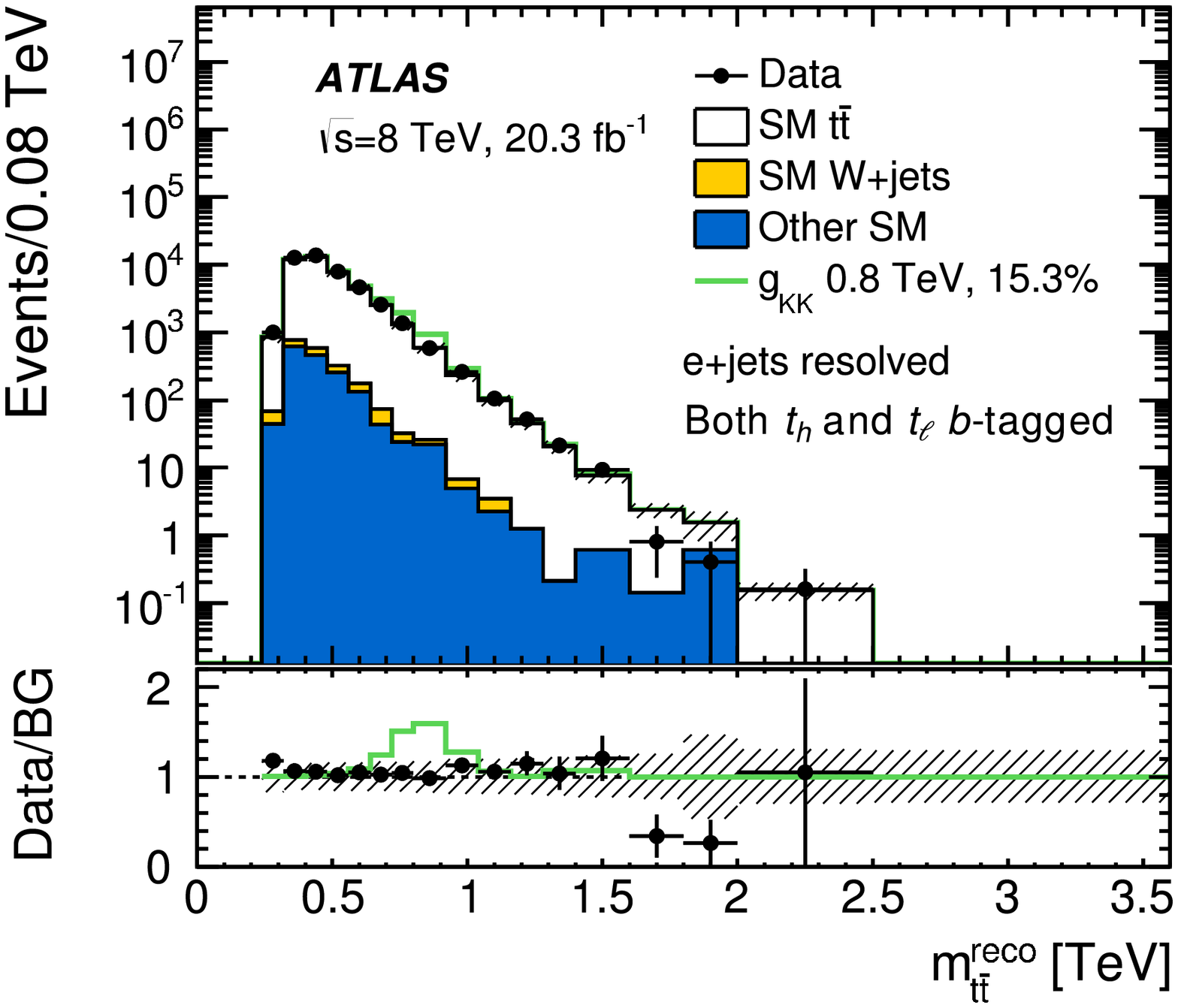}
}
\hspace{.5cm}
\subfigure[Muon channel, resolved selection, both top candidates have $b$-tagged jets.]{
\includegraphics[width=0.43\textwidth]{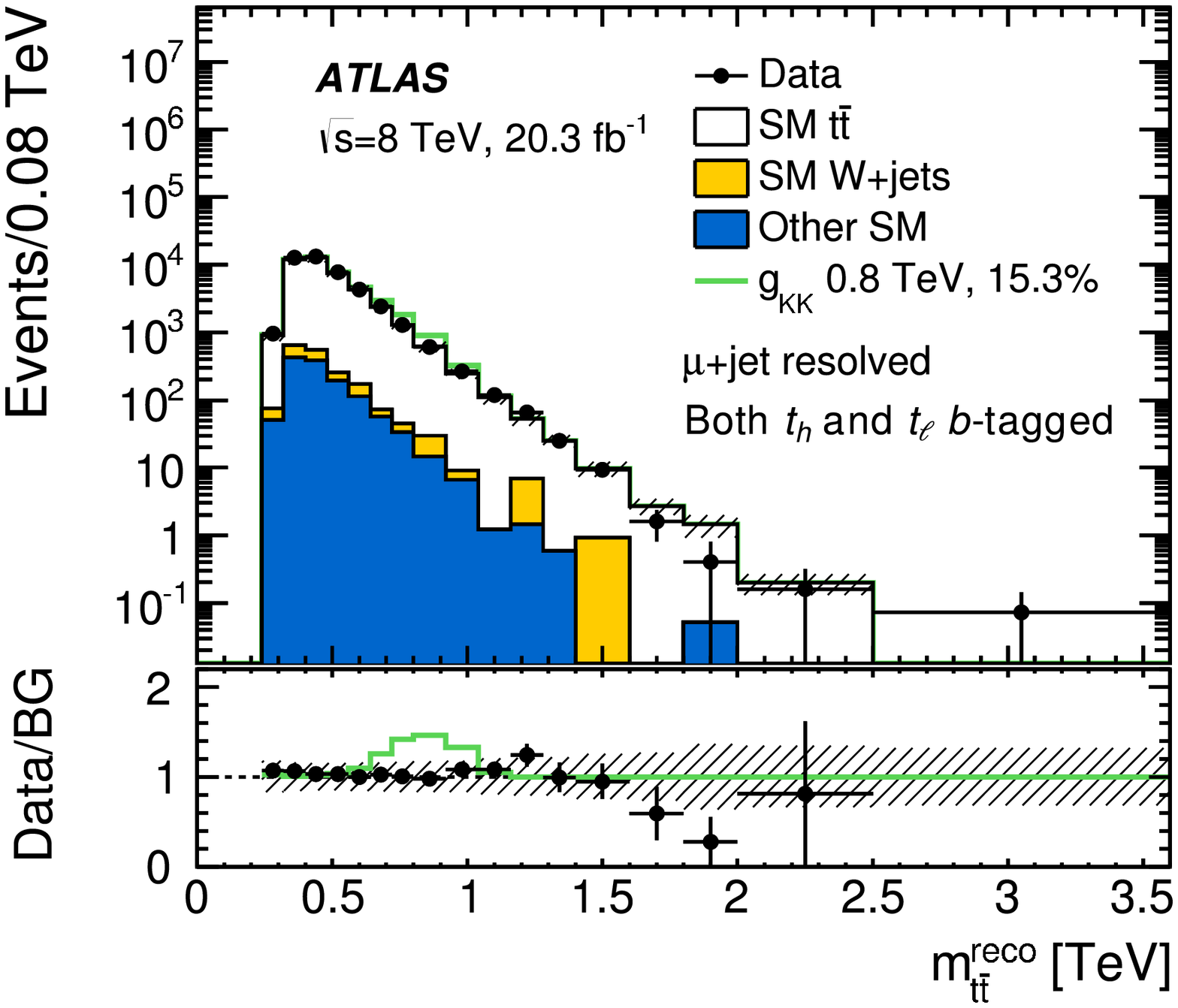}
}\\
\subfigure[Electron channel, resolved selection, only the hadronic top candidate has a $b$-tagged jet.]{
\includegraphics[width=0.43\textwidth]{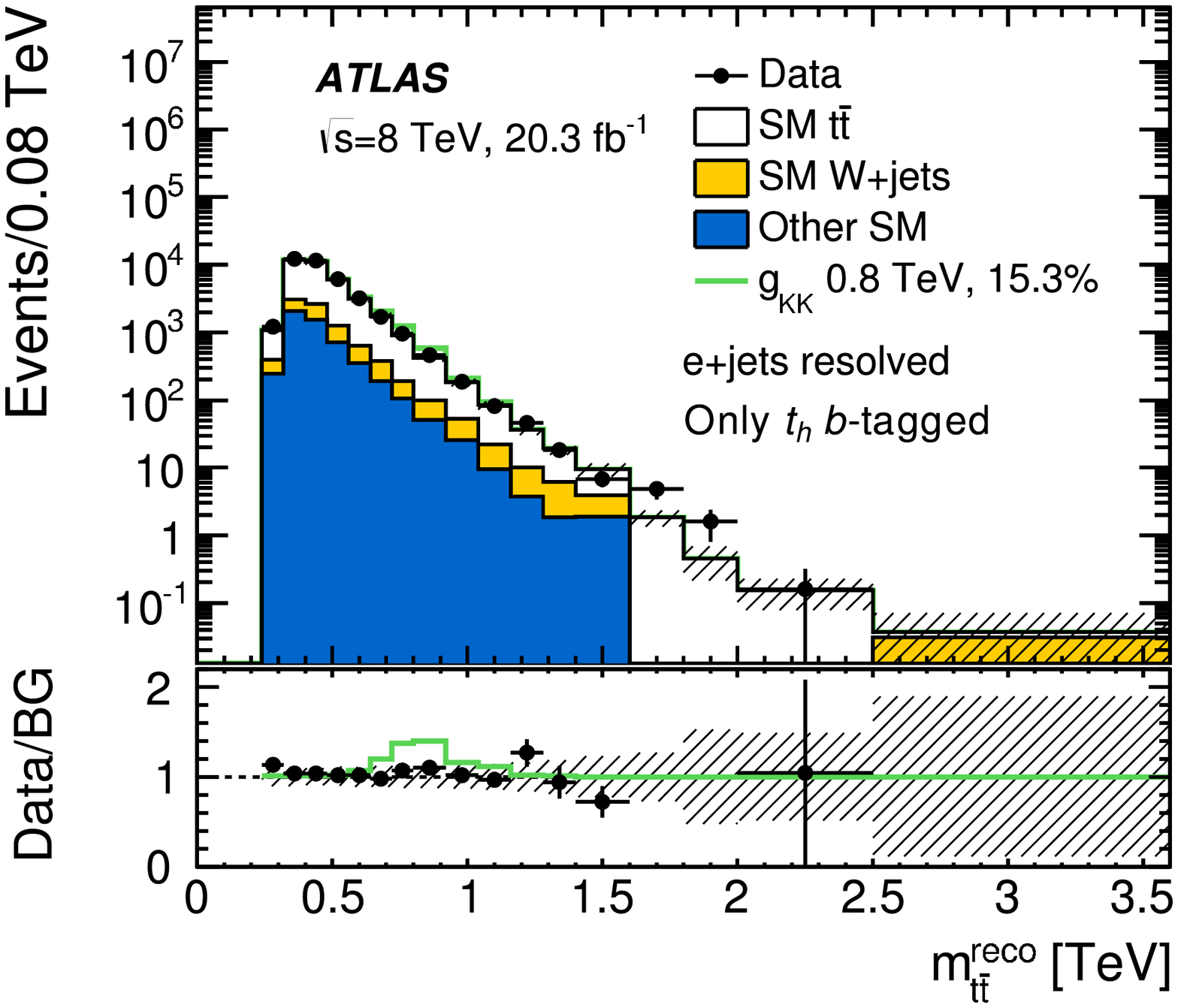}
}
\hspace{.5cm}
\subfigure[Muon channel, resolved selection, only the hadronic top candidate has a $b$-tagged jet.]{
\includegraphics[width=0.43\textwidth]{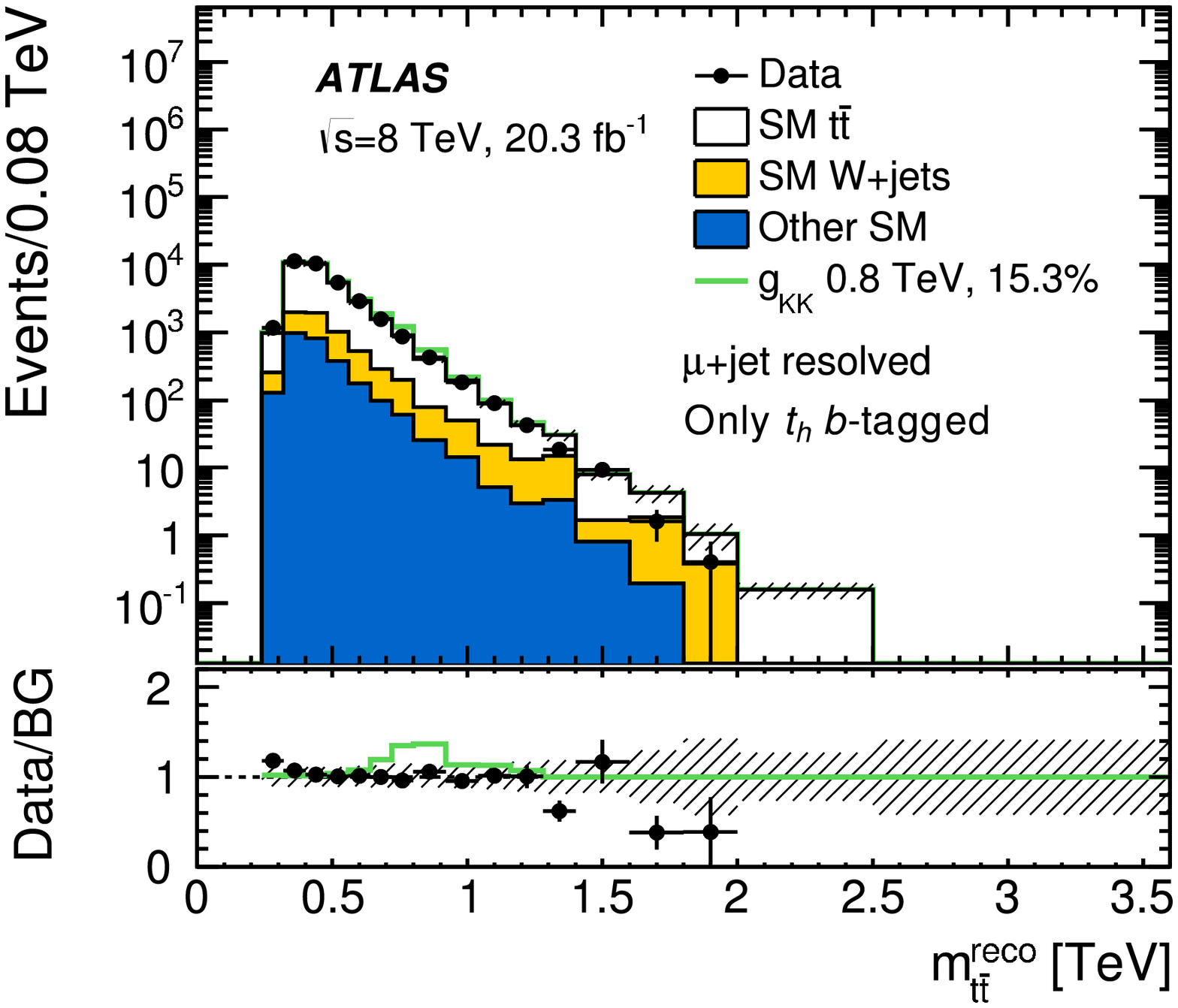}
}\\
\subfigure[Electron channel, resolved selection, only the semileptonic top candidate has a $b$-tagged jet.]{
\includegraphics[width=0.43\textwidth]{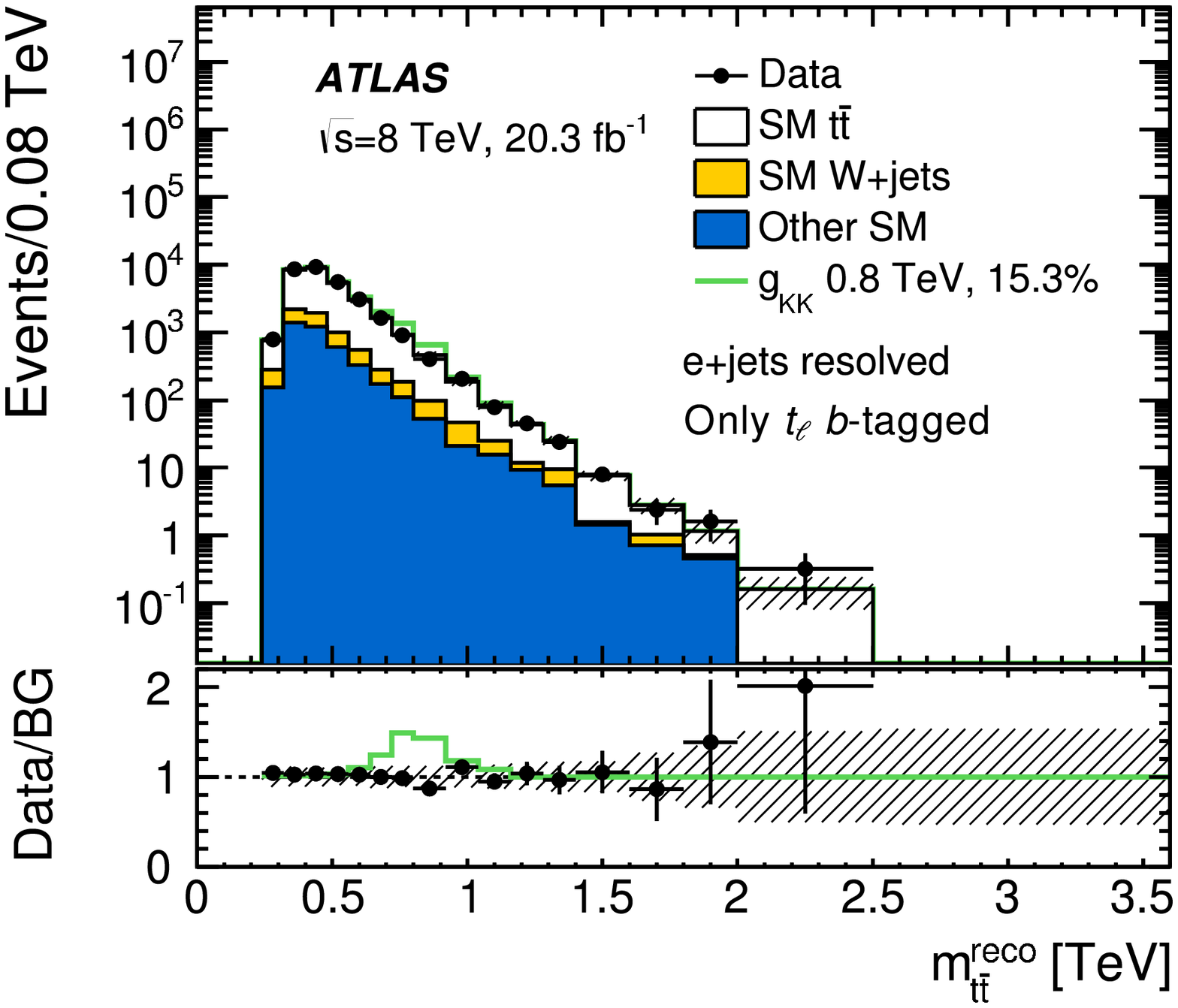}
}
\hspace{.5cm}
\subfigure[Muon channel, resolved selection, only the semileptonic top candidate has a $b$-tagged jet.]{
\includegraphics[width=0.43\textwidth]{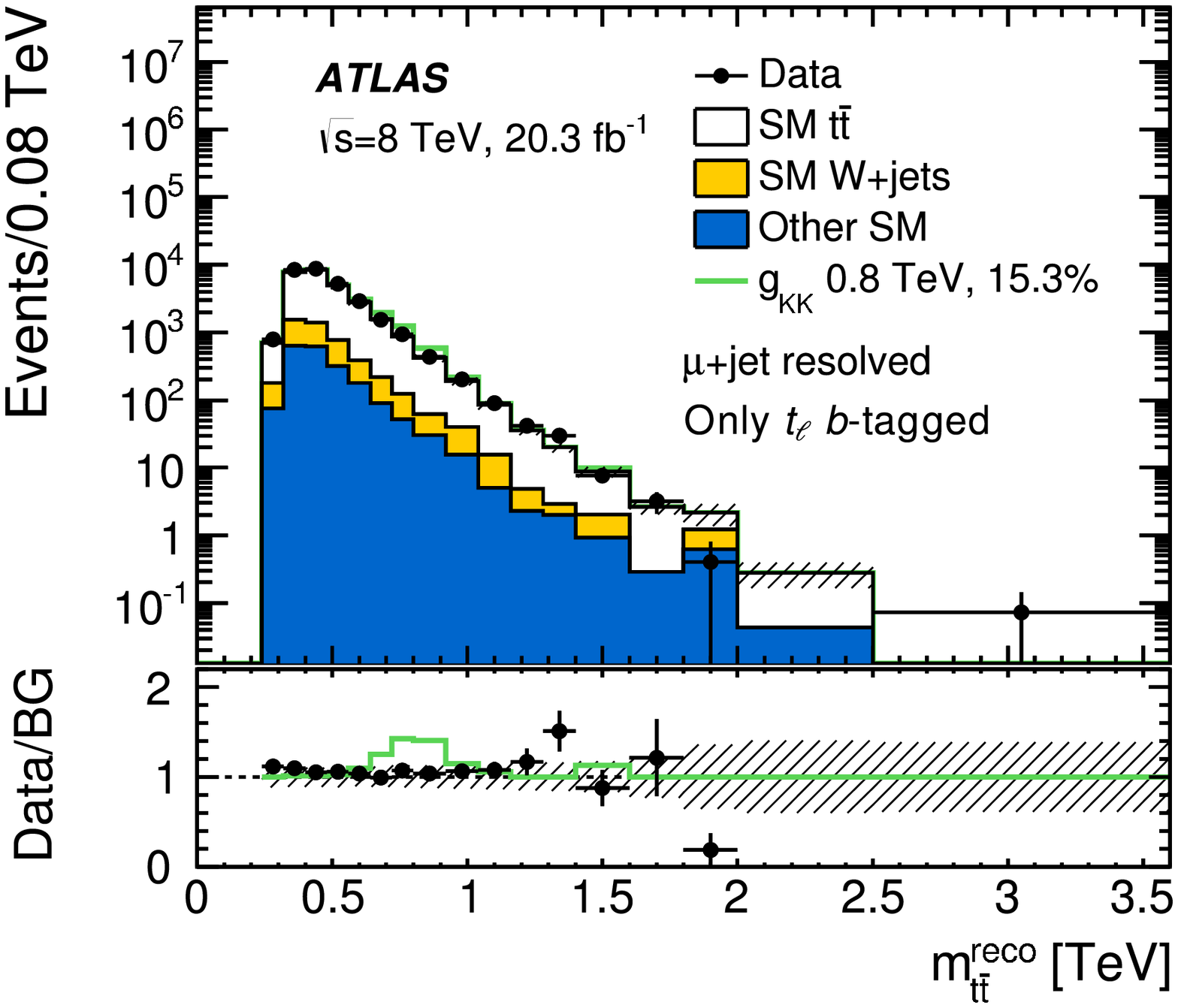}
}\\
\caption{The spectrum of the reconstructed ttbar 
invariant mass, \mttbarreco{}, for the different channels, before any nuisance parameter fit, after the resolved-topology selection. The SM background components are shown as stacked histograms. The shaded areas indicate the total systematic uncertainties. The green line shows the expected distribution for a hypothetical $g_{\mathrm{KK}}$ of mass 0.8 TeV, width 15.3\%.
\label{fig:mttlog_res}}
\end{figure}

\begin{figure}
\centering
\subfigure[Electron channel, boosted selection, both top candidates have $b$-tagged jets.]{
\includegraphics[width=0.43\textwidth]{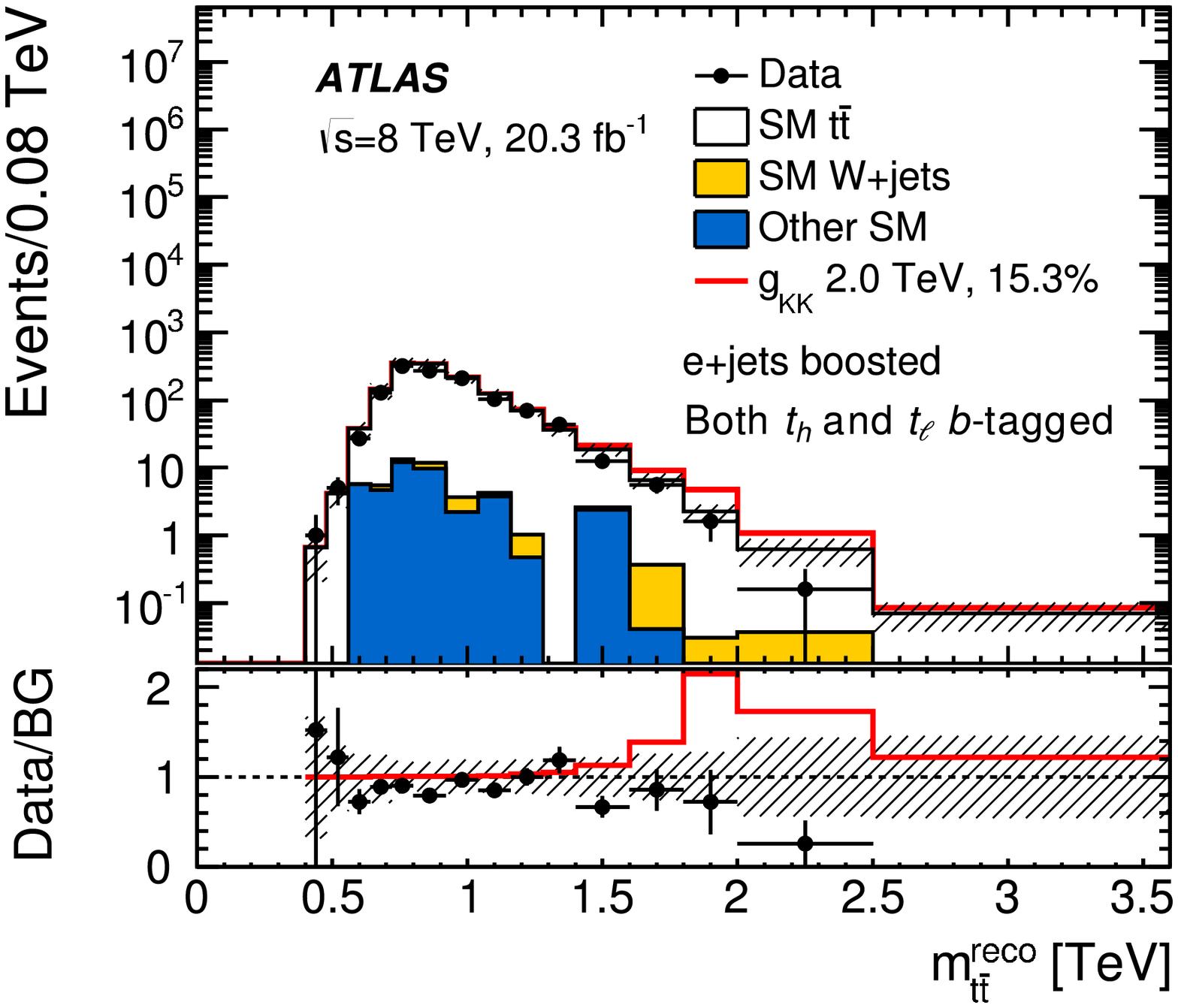}
}
\hspace{.5cm}
\subfigure[Muon channel, boosted selection, both top candidates have $b$-tagged jets.]{
\includegraphics[width=0.43\textwidth]{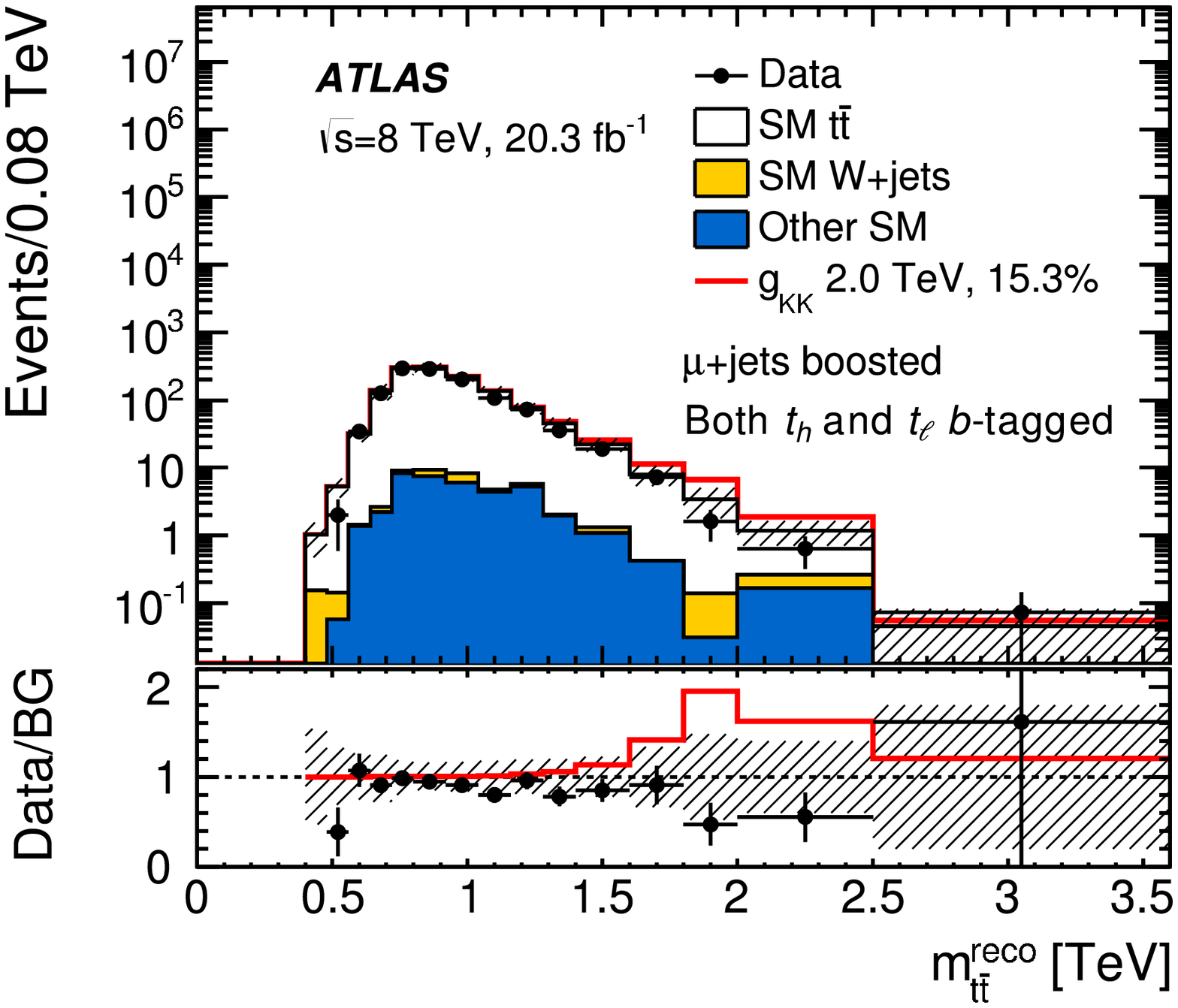}
}\\
\centering
\subfigure[Electron channel, boosted selection, only the hadronic top candidate has a $b$-tagged jet.]{
\includegraphics[width=0.43\textwidth]{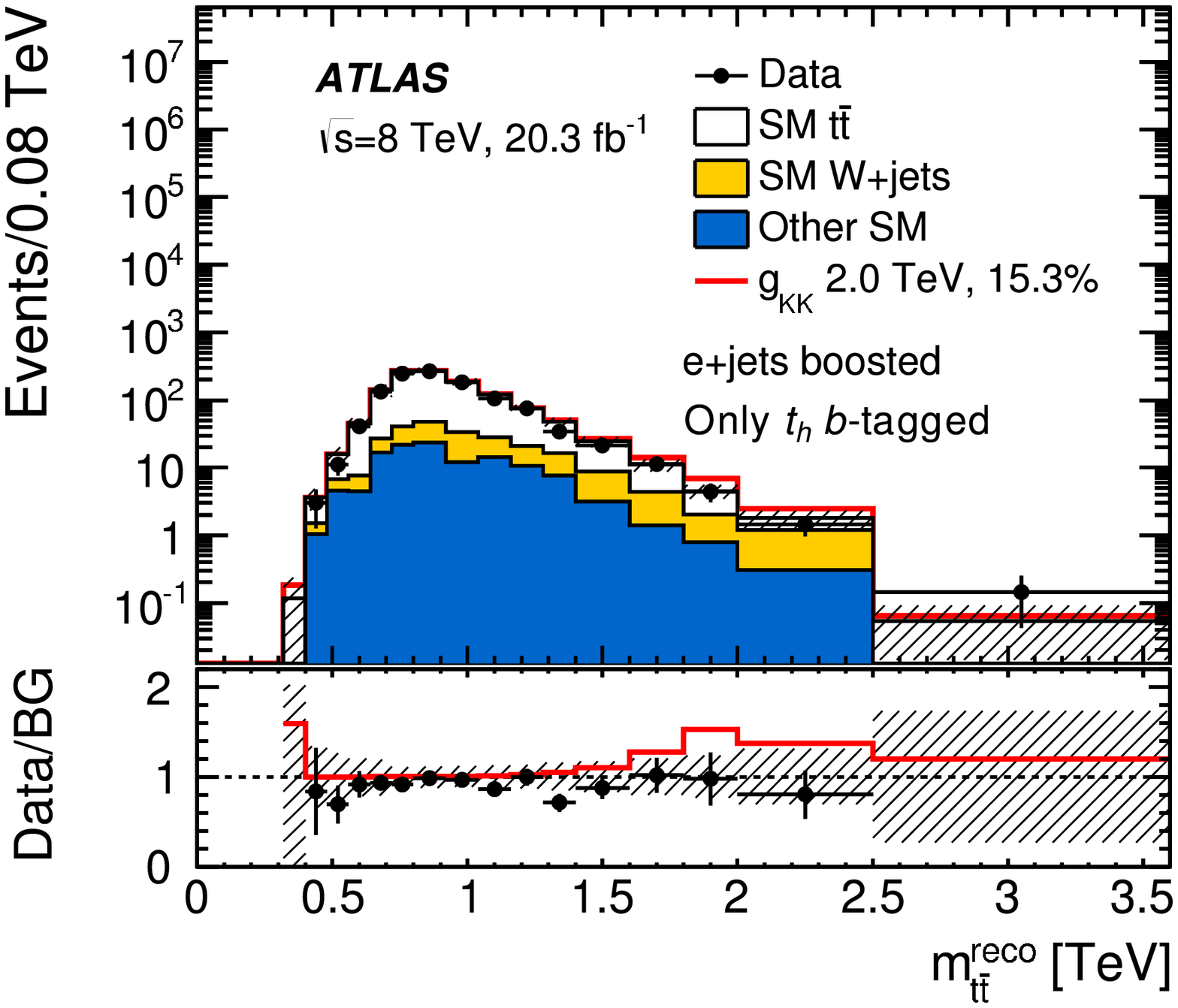}
}
\hspace{.5cm}
\subfigure[Muon channel, boosted selection, only the  hadronic top candidate has a $b$-tagged jet.]{
\includegraphics[width=0.43\textwidth]{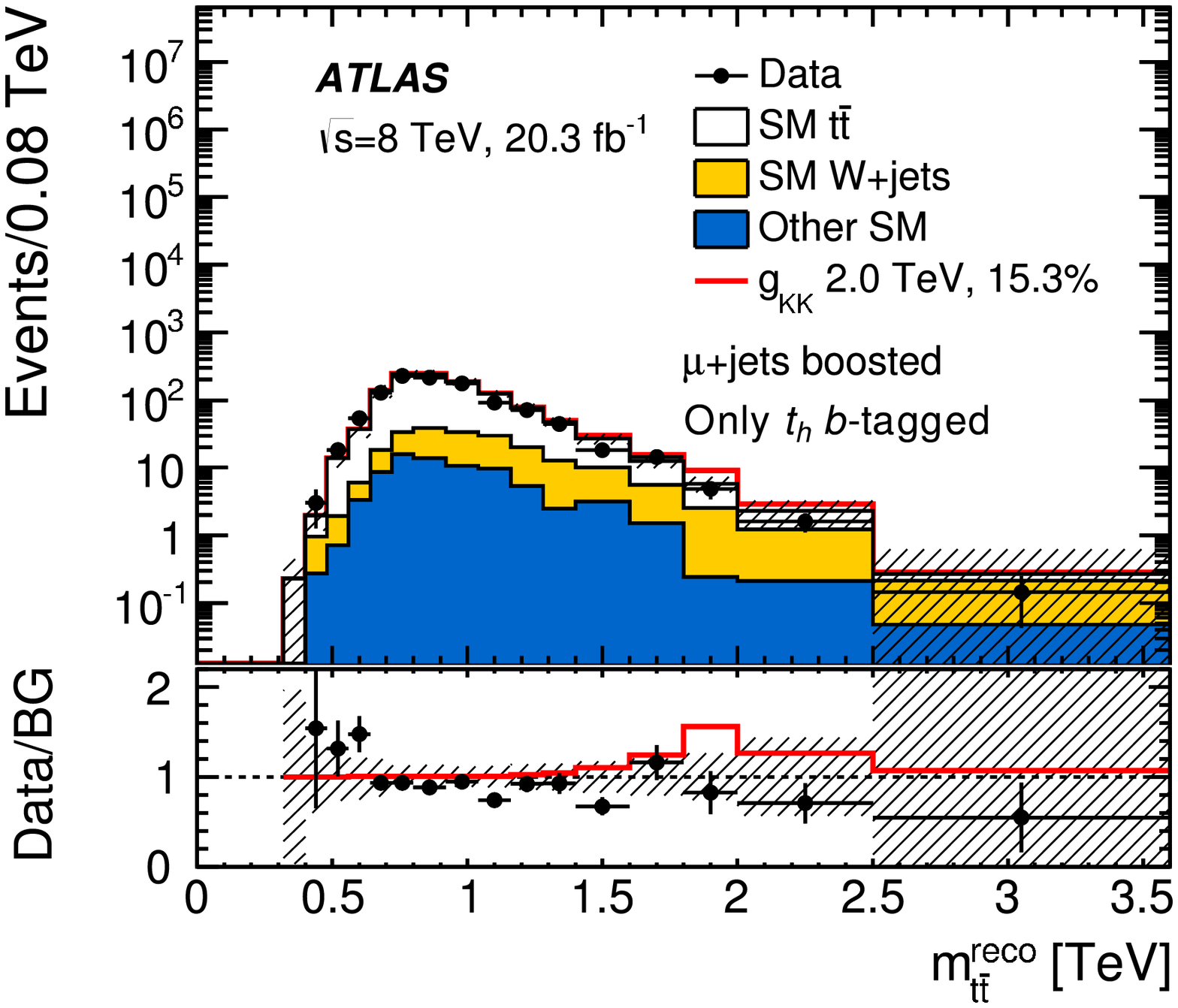}
}\\
\centering
\subfigure[Electron channel, boosted selection, only the semileptonic top candidate has a $b$-tagged jet.]{
\includegraphics[width=0.43\textwidth]{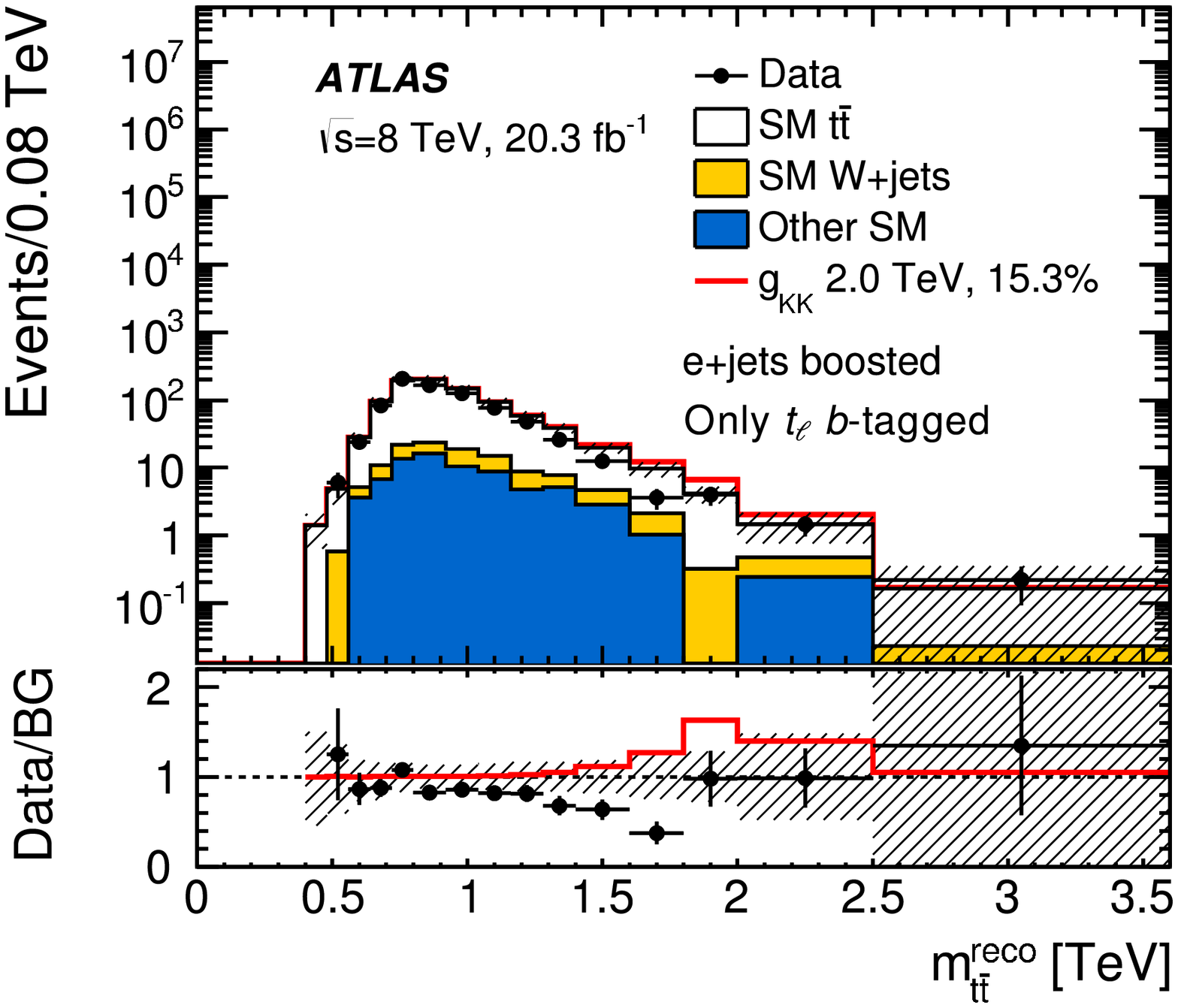}
}
\hspace{.5cm}
\subfigure[Muon channel, boosted selection, only the semileptonic top candidate has a $b$-tagged jet.]{
\includegraphics[width=0.43\textwidth]{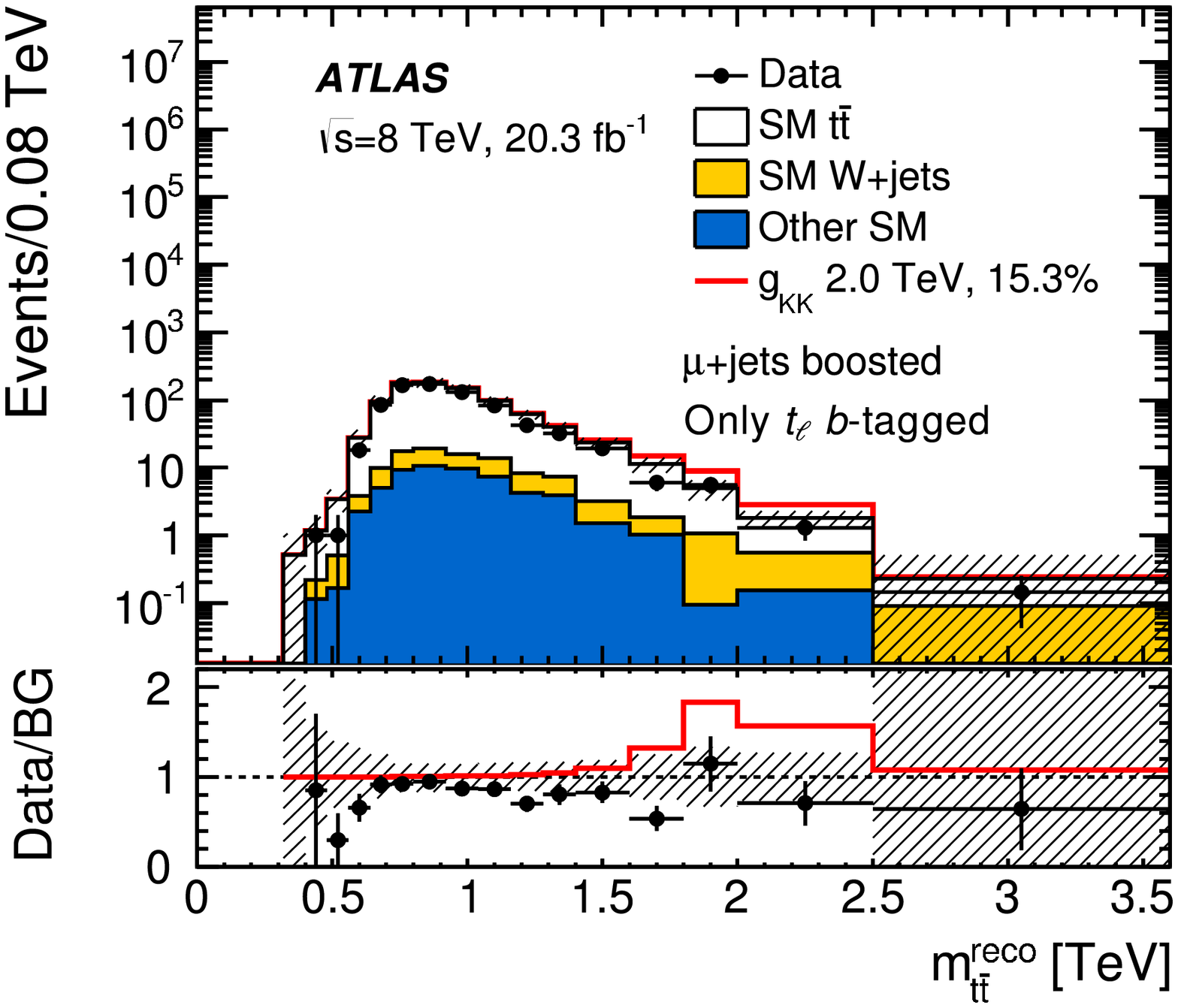}
}
\caption{The \mttbarreco{} spectrum for the different channels, before any nuisance parameter fit, after the boosted-topology selection. The SM background components are shown as stacked histograms. The shaded areas indicate the total systematic uncertainties. The red line shows the expected distribution for a hypothetical $g_{\mathrm{KK}}$ of mass 2.0 TeV, width 15.3\%.
\label{fig:mttlog_boo}}
\end{figure}

\begin{figure}
\centering
\subfigure[Boosted selections.]{
\includegraphics[width=0.45\textwidth]{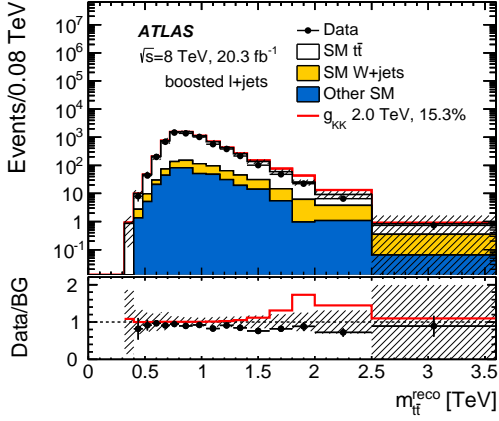}
}
\subfigure[Resolved selections.]{
\includegraphics[width=0.45\textwidth]{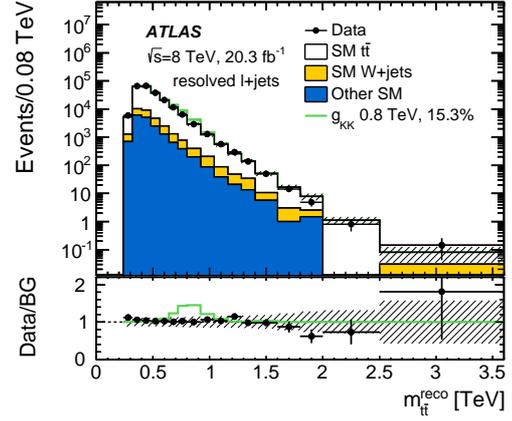}
}\\
\subfigure[All selections.]{
\includegraphics[width=0.45\textwidth]{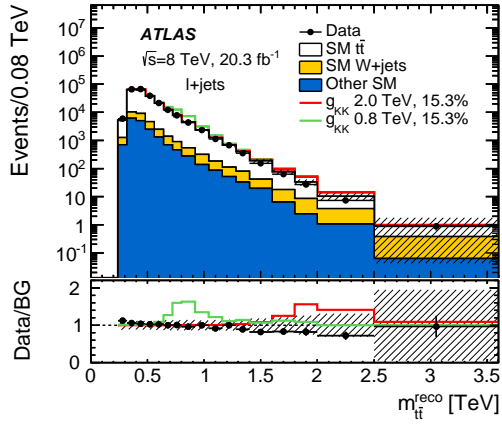}
}
\caption{The \mttbarreco{} distributions, before any nuisance parameter fit, summed over (a) all 6 boosted channels, (b) all 6 resolved channels, and (c) all 12 channels compared with data. The SM background components are shown as stacked histograms. The shaded areas indicate the total systematic uncertainties. The red (green) line shows the expected distribution for a hypothetical $g_{\mathrm{KK}}$ of mass 2.0 (0.8) TeV, width 15.3\%.
\label{fig:mtt_merged_prefit}}
\end{figure}

The \ttbar\ invariant mass spectra for the resolved and the boosted selections in the electron and muon channels, separated by $b$-tagging category, are shown in figures~\ref{fig:mttlog_res} and \ref{fig:mttlog_boo}. Figure~\ref{fig:mtt_merged_prefit} shows the \ttbar\ invariant mass spectrum for all channels added together. 
The data generally agree with the expected background with slight shape differences seen especially in the high-mass regions; these deviations are consistent with the uncertainties.

\subfigcapmargin = 0cm

\FloatBarrier
\section{Results}
\label{sec:results}

The final discriminating observables that are used to search for a 
massive resonance are the twelve \ttbar\ invariant mass spectra: three $b$-tag categories for two selections and two decay channels.
After the reconstruction of the \ttbar\ mass spectra, the data and expected background distributions are compared 
using {\sc BumpHunter}~\cite{Choudalakis:2011qn}, which is a hypothesis-testing tool that searches the data for local excesses or deficits compared to the expected background, taking the look-elsewhere effect~\cite{Lyons:1900zz} into account over the full mass spectrum. 
The search is performed on three combinations of the spectra: the six channels of the resolved selections, the six channels of the boosted selections and the twelve channels. 
The most significant deviation of the data from the expected background spectrum is required to appear at the same place in each of the channels of a combination.
After accounting for the systematic uncertainties, no significant deviation from the total expected background is found. 
Upper limits are set on the cross-section times branching ratio for each of the signal models using a profile likelihood-ratio test. 
The CL$_s$ prescription~\cite{Read:2002hq} is used to derive one-sided  95\% CL limits. 
The results are obtained using the {\sc{HistFitter}}~\cite{Baak:2014wma} framework with all spectra from the 12 channels,
excluding bins with few events with \mttbarreco\ below 400\,GeV in the boosted channels or above 2\,TeV in the resolved channels.  

\begin{figure} [htbp]
\centering
\subfigure[Boosted selections.]{
\includegraphics[width=0.45\textwidth]{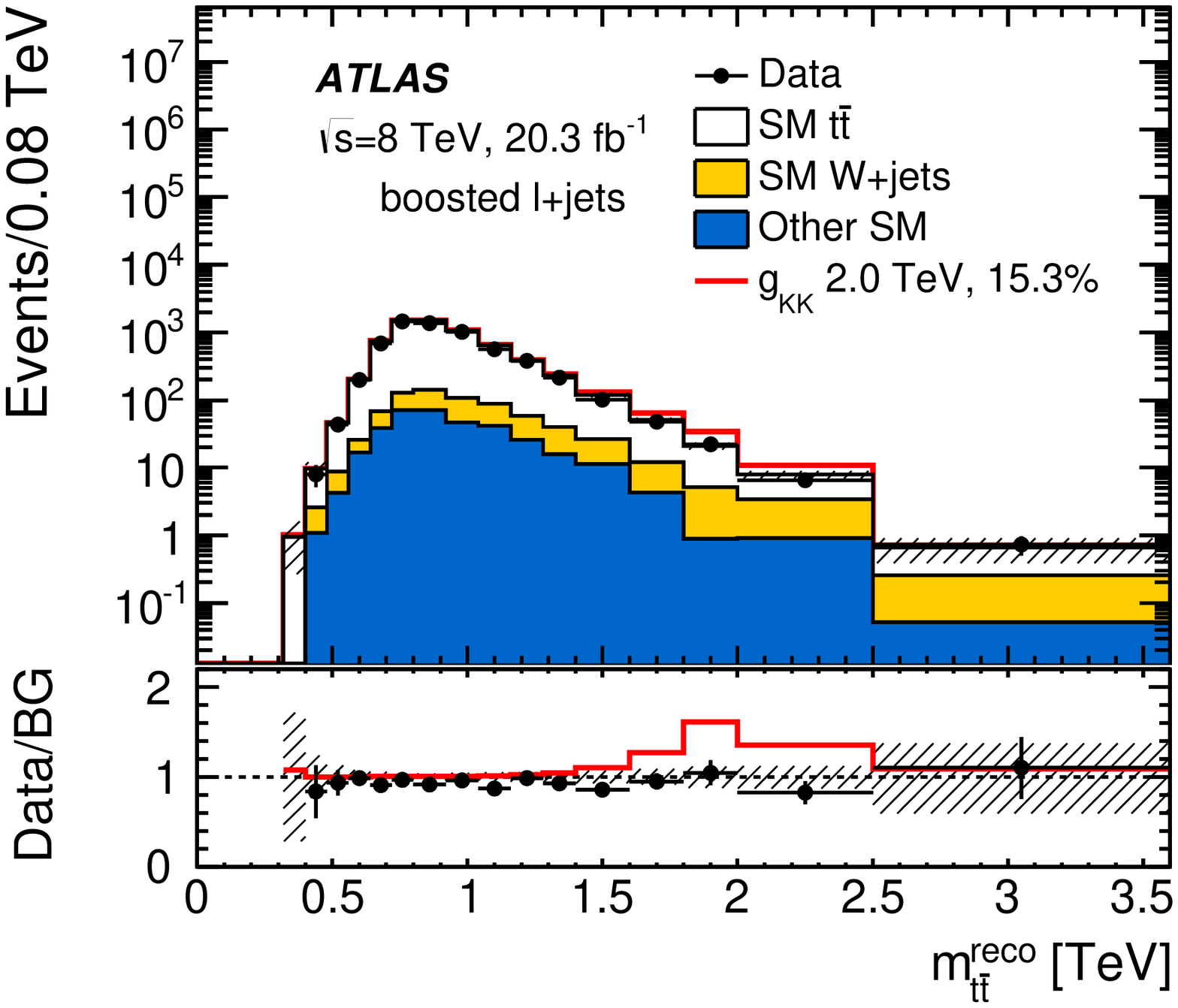}
}
\subfigure[Resolved selections.]{
\includegraphics[width=0.45\textwidth]{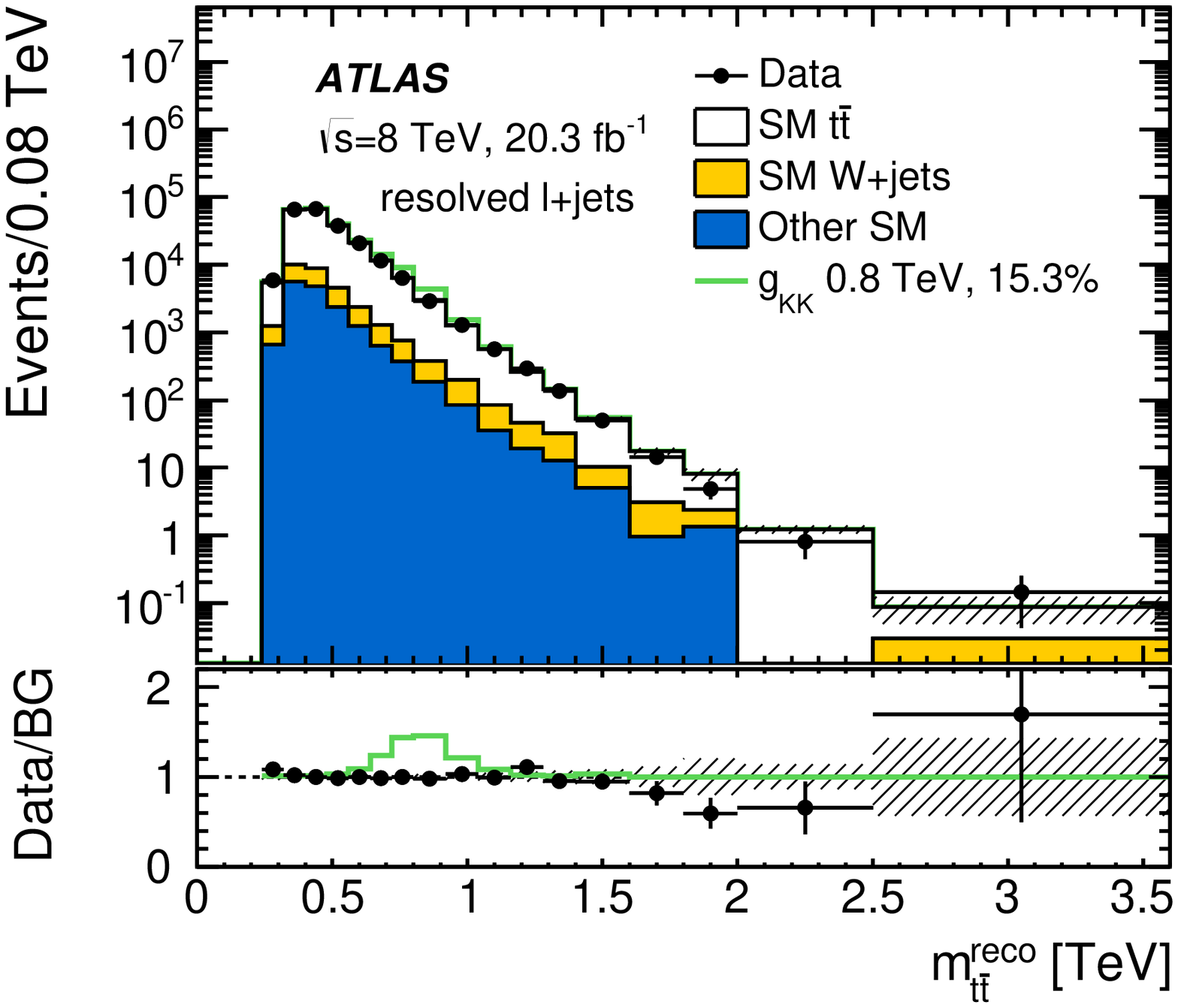}
}\\
\subfigure[All selections.]{
\includegraphics[width=0.45\textwidth]{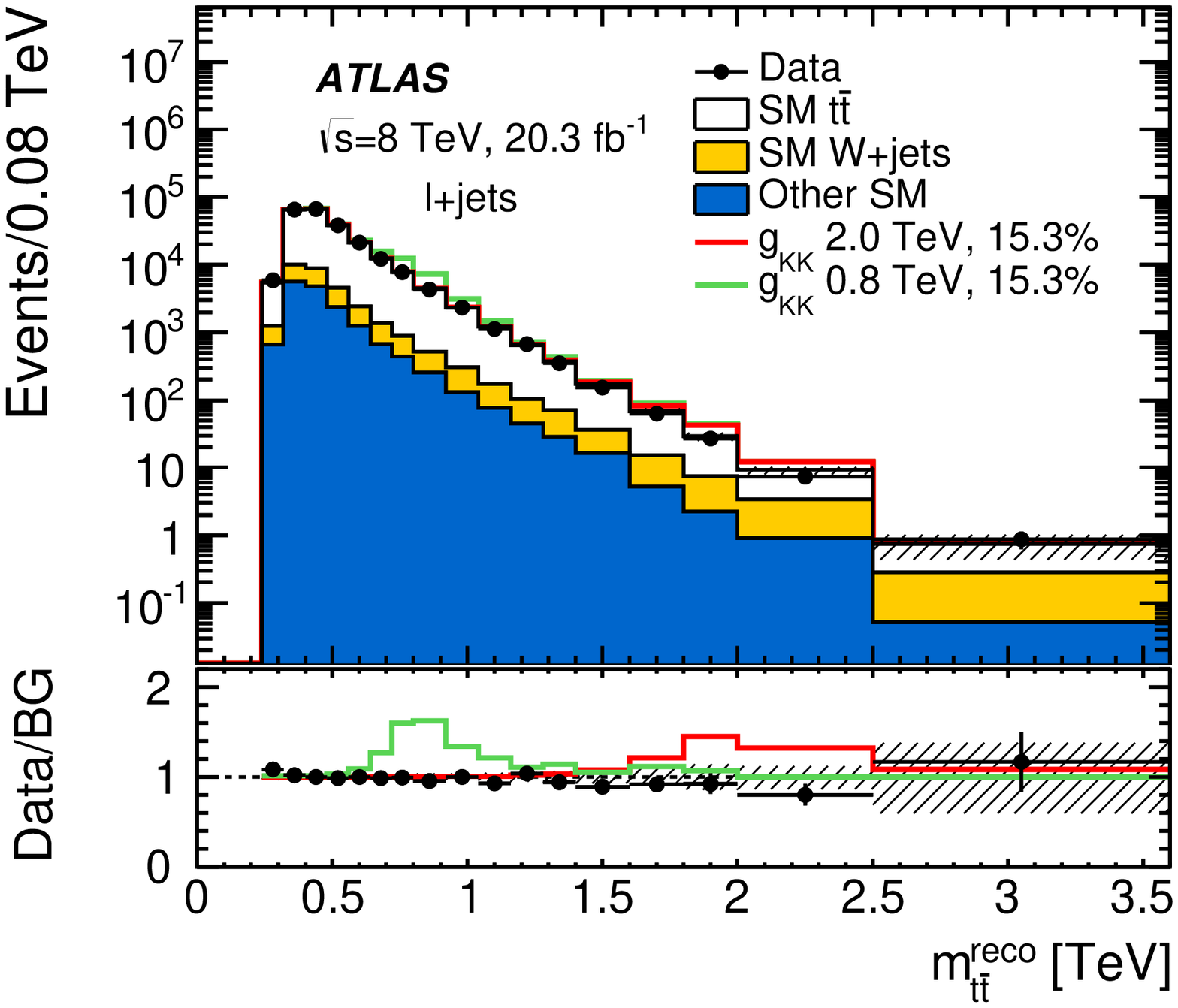}
}
\caption{The \mttbarreco{} distributions, after the nuisance-parameter fit under the background-only hypothesis, summed over (a) all 6 boosted channels, (b) all 6 resolved channels, and (c) all 12 channels compared with data. The SM background components are shown as stacked histograms. The shaded areas indicate the total systematic uncertainties.  The red (green) line shows the expected distribution for a hypothetical $g_{\mathrm{KK}}$ of mass 2.0 (0.8) TeV, width 15.3\%.
\label{fig:mtt_merged_postfit}}
\end{figure}

The statistical and systematic uncertainties on the expected distributions are included in this CL$_s$ procedure as nuisance parameters in the likelihood fits. 
The nuisance parameters for the systematic uncertainties are constrained by a Gaussian probability density function with a width corresponding to the size of the uncertainty considered. 
Correlations between different channels and bins are taken into account. 
The product of the various probability density functions forms the likelihood function that is 
maximised in the fit by adjusting the free parameter (the signal strength) and nuisance parameters. 
The expected \mttbarreco{} distributions are compared to data in figure~\ref{fig:mtt_merged_postfit} after a fit of nuisance parameters under the background-only hypothesis. 
It can be seen that the uncertainties are smaller than in figure~\ref{fig:mtt_merged_prefit} and that the procedure is able to produce a good-quality fit to the data.

\begin{figure}
\centering
\subfigure[\Zprime{}, resolved and boosted combination.]{
\label{F:limits_Zprime_chi2}
\includegraphics[width=0.45\textwidth]{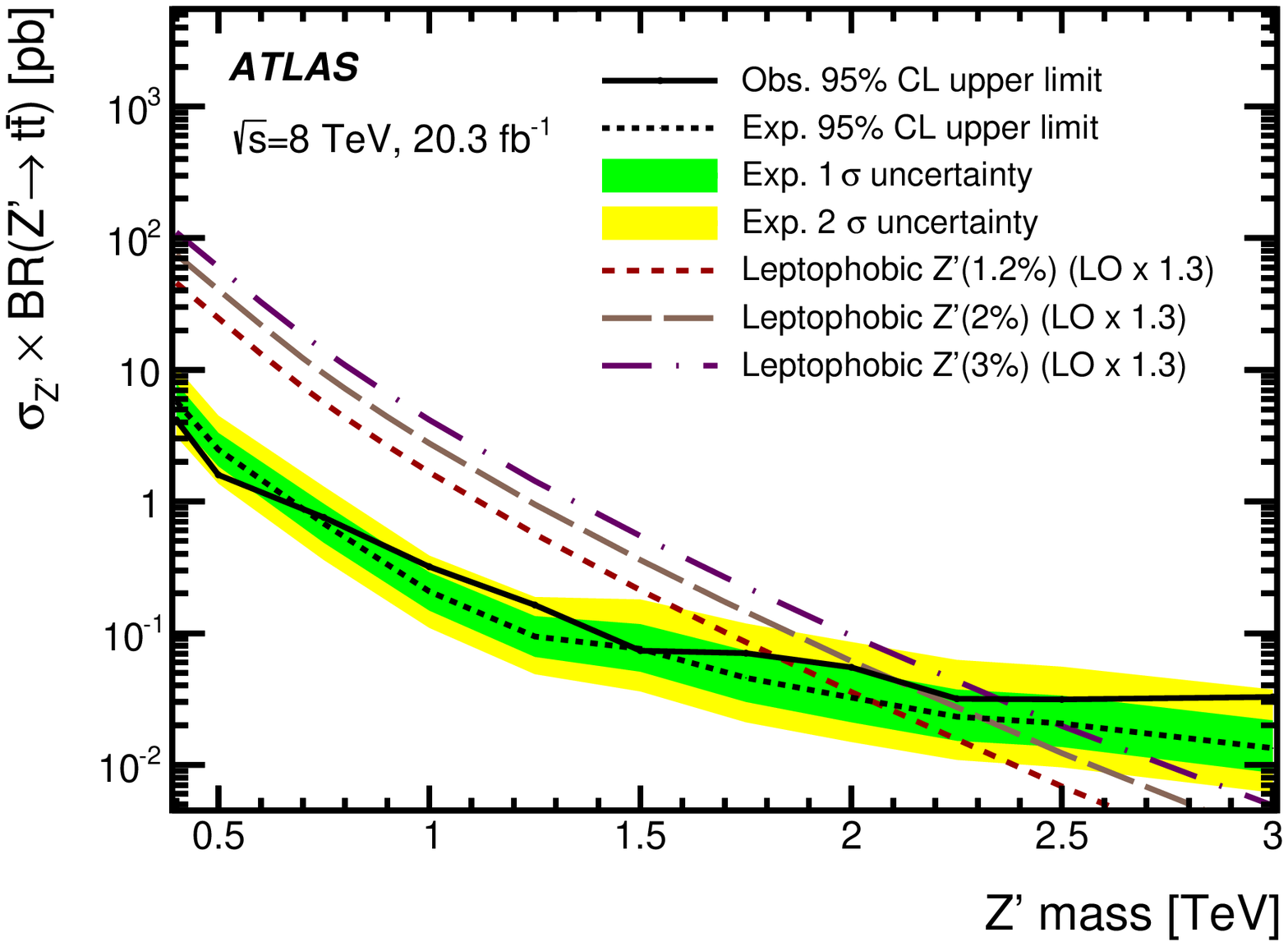}
}
\subfigure[$g_\mathrm{KK}$, resolved and boosted combination.]{
\label{F:limits_kkg_chi2}
\includegraphics[width=0.45\textwidth]{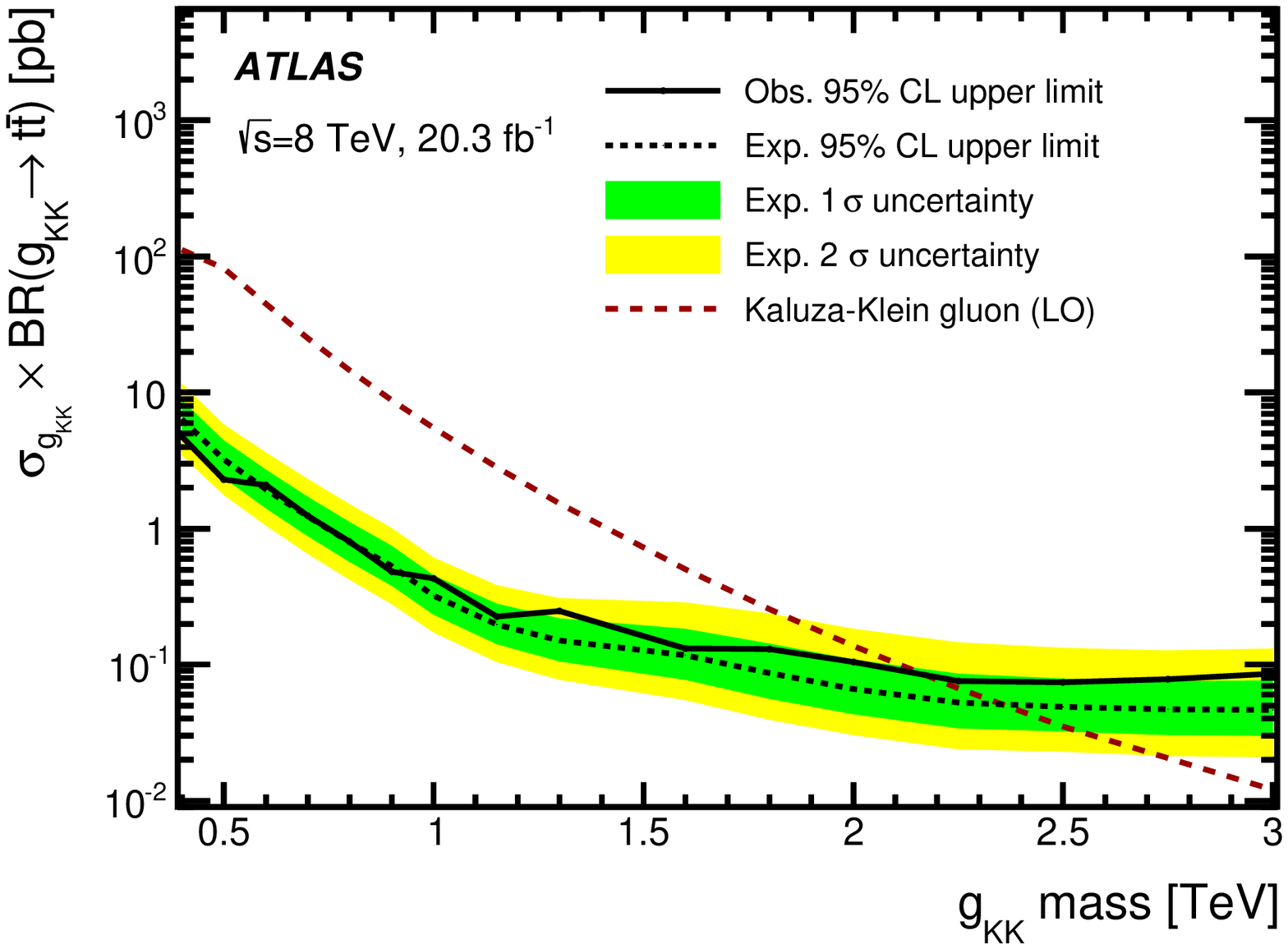}
}\\
\subfigure[$G_\mathrm{KK}$, resolved and boosted combination.]{
\label{F:limits_kkgrav_chi2}
\includegraphics[width=0.45\textwidth]{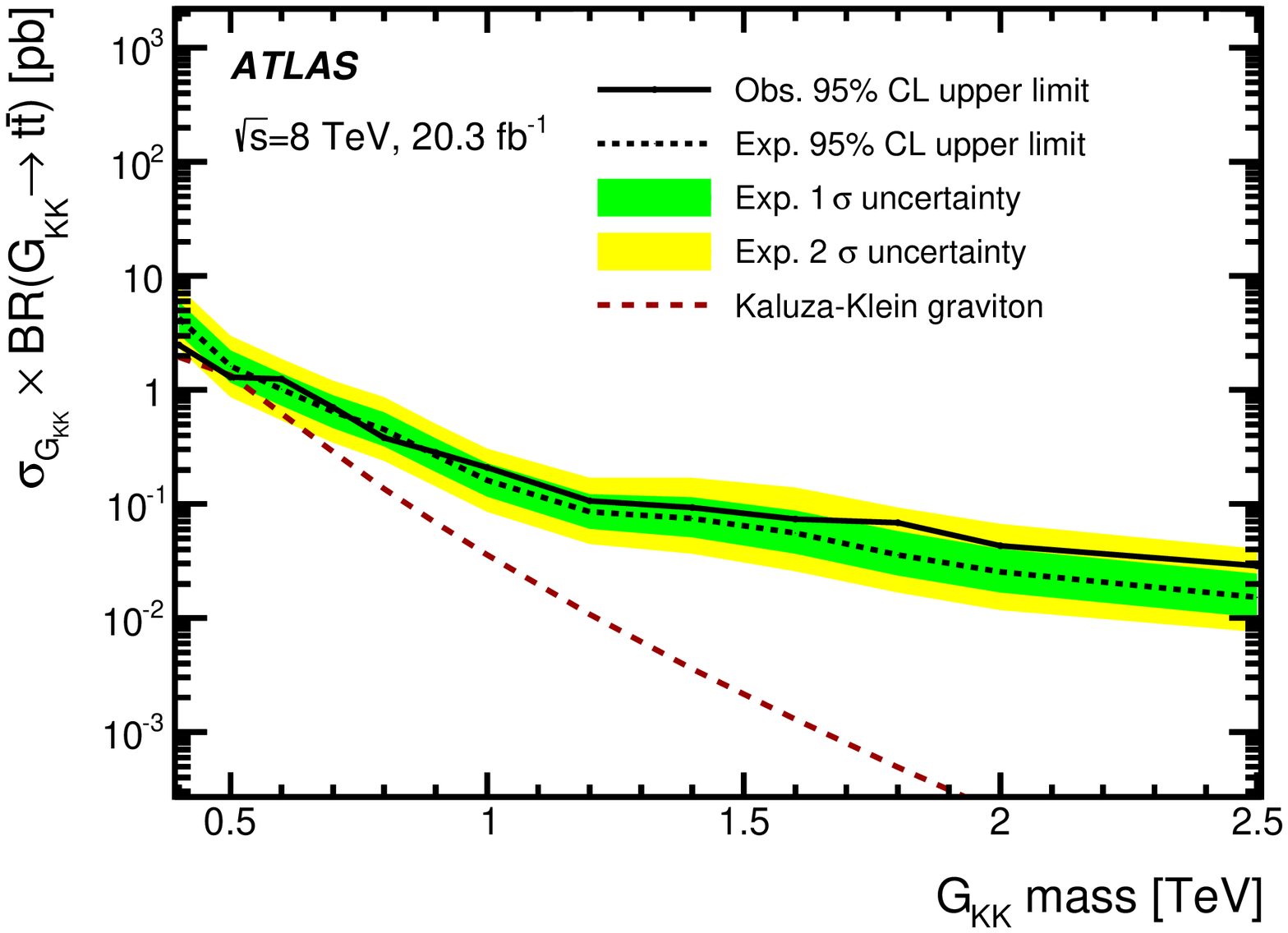} 
}
\subfigure[Scalar resonance, resolved and boosted combination.]{
\label{F:limits_hh_chi2}
\includegraphics[width=0.45\textwidth]{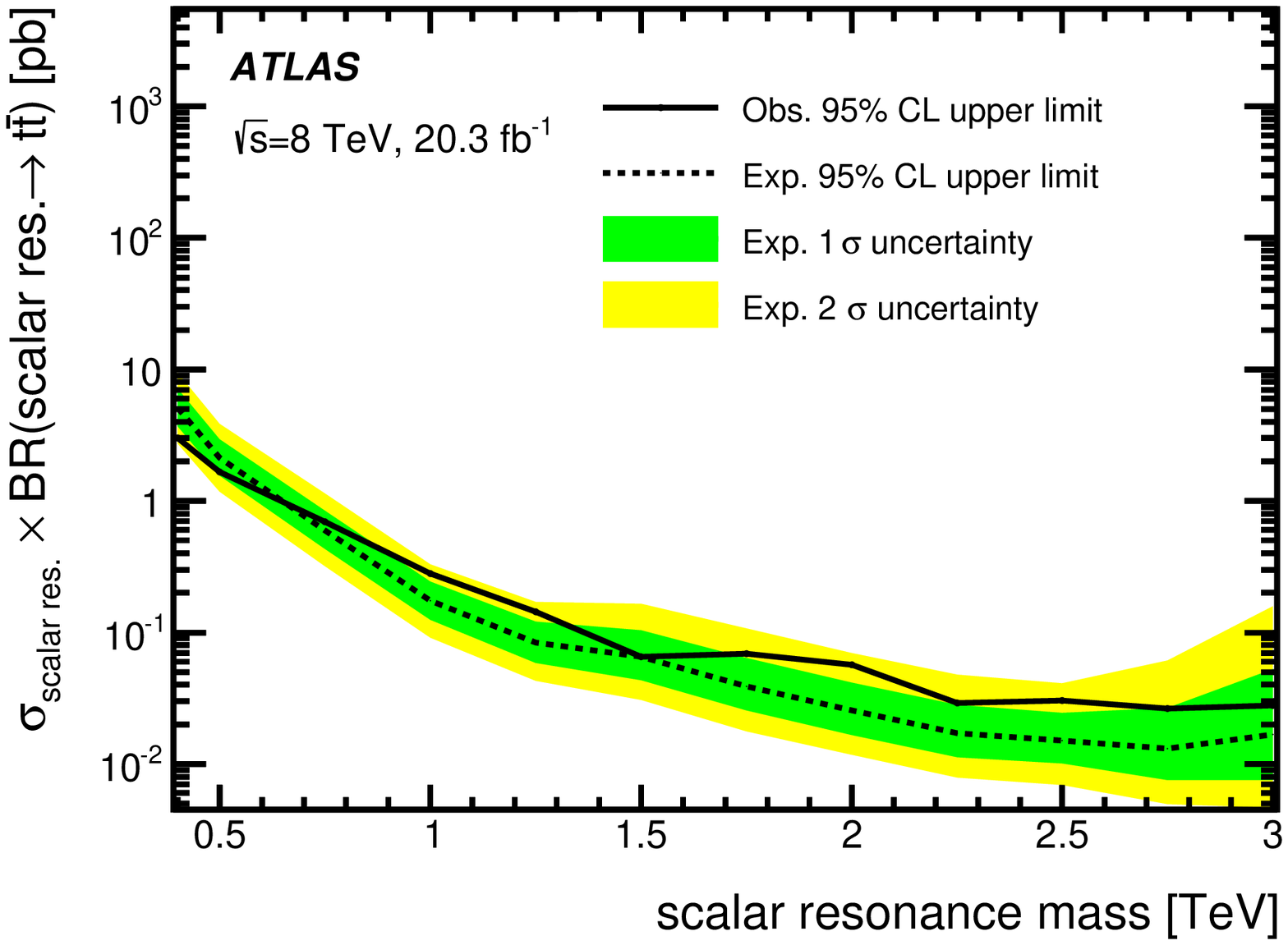} 
}
\caption{
Observed and expected upper limits on the production cross-section times branching ratio to \ttbar{} final states as a function of the mass of 
(a) Topcolour-assisted-technicolour $\Zprime_{\mathrm{TC2}}{}$, (b) Bulk RS Kaluza--Klein gluon, (c) Bulk RS Kaluza--Klein graviton, (d) scalar resonance. The expected limits are derived from nominal (pre-fit) background estimates. The theoretical predictions for the production cross-section times branching ratio at the corresponding masses are also shown.
\label{fig:limit_MassScan}}
\end{figure}

The expected and observed limits using different signal models are presented in figure~\ref{fig:limit_MassScan}. 
Here the expected limits are obtained by taking the nominal background estimates as the expected data.
For the $Z'_{\mathrm{TC2}}$ benchmark, 
limits on the production cross-sections vary from \Zpxseclow\  to \Zpxsechigh{} for masses from 0.4\,TeV to 3\,TeV. 
A $Z'_{\mathrm{TC2}}$ of width 1.2\%  is excluded for masses less than \ZpexcludedMass, while masses below \ZpexcludedMassExpected\ are expected to be excluded. 
The $Z'_{\mathrm{TC2}}$ mass limits are stronger for a width of 2\% (3\%), reaching 2.0\,TeV (2.3\,TeV). 
For the $g_\mathrm{KK}$ benchmark, limits on the production cross-sections vary from \kkgxseclow\ for a mass of 0.4\,TeV, 
to \kkgxsechigh\ for a mass of 3\,TeV. 
A $g_{\mathrm{KK}}$ of width 15.3\% is excluded for masses less than \kkgexcludedMass,  while masses below \kkgexcludedMassExpected\ are expected to be excluded. 
The cross-section limits in the $G_\mathrm{KK}$ model range from \kkGxseclow\ for a mass of 0.4\,TeV to \kkGxsechigh\ for 2.5\,TeV, with no mass range excluded in the benchmark scenario.
The cross-section limits on a narrow scalar resonance range from \scalarxseclow\ for a mass of 0.4\,TeV to \scalarxsechigh\ for 2.5\,TeV. 
The cross-section limits are generally stronger for the latter two benchmark models than those for the spin-1 resonances due to higher acceptance.

\begin{figure}
\centering
\includegraphics[width=0.7\textwidth]{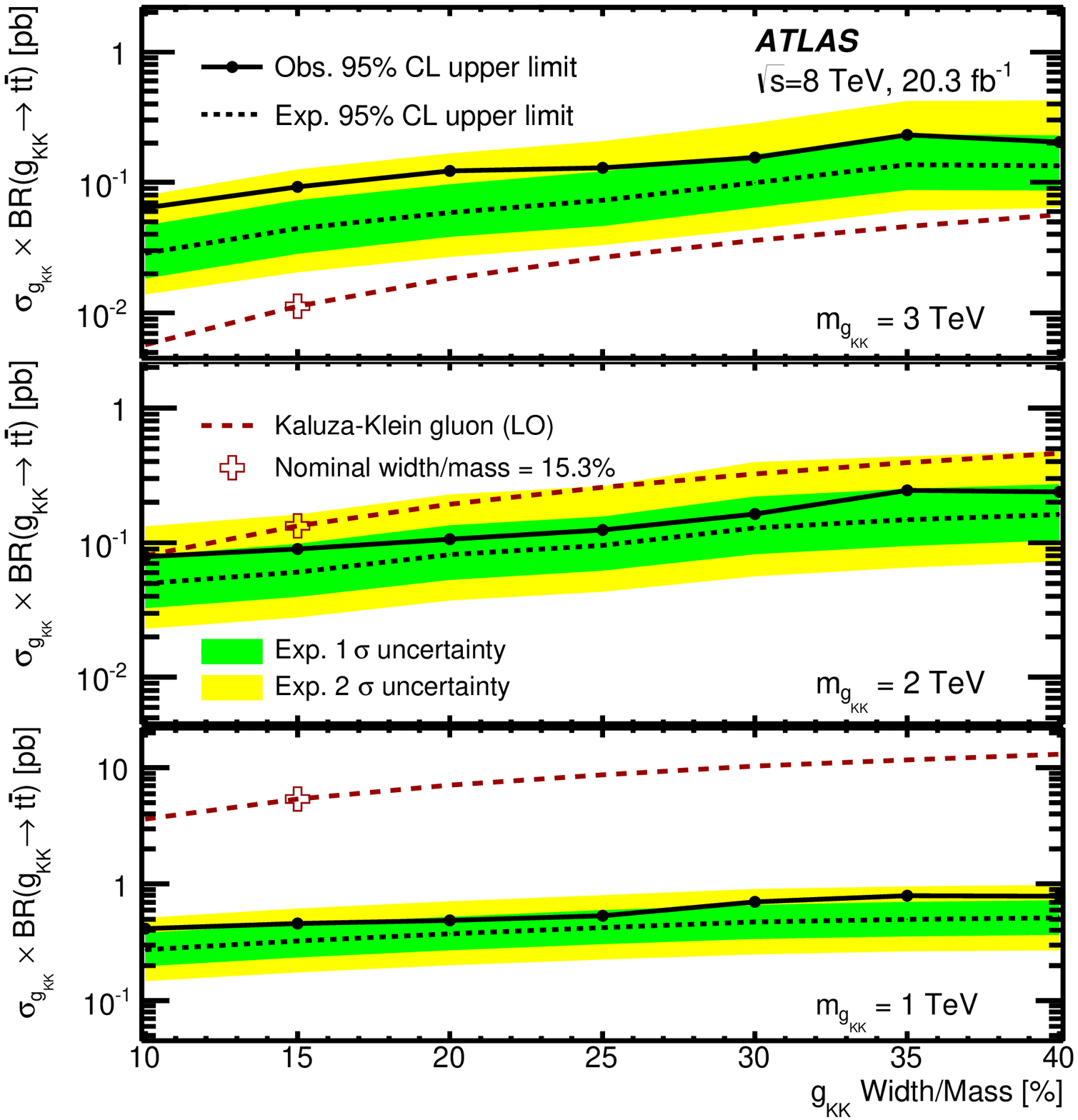}
\caption{
Observed and expected upper limits on the production cross-section for a Kaluza--Klein gluon times its branching ratio to $t\bar{t}$, as a function of its width, for three representative mass values. The expected limits are derived from nominal (pre-fit) background estimates. The theoretical predictions for the production cross-section times branching ratio at the corresponding widths are also shown.
\label{fig:limit_WidthScan}}
\end{figure}

The width dependence of the cross-section limits was also evaluated for the $g_\mathrm{KK}$ models. The results are presented in figure~\ref{fig:limit_WidthScan}. For a 1\,TeV resonance, the limits weaken by approximately a factor of two  as the width increases from 10\% to 40\%. The effect is stronger for 2\,TeV and 3\,TeV resonances,
 where the limits weaken by a factor of three over this width range.

The observed upper limits on the cross-section times \ttbar{} branching ratio are larger than the expected limits, 
especially for \ttbar{} resonance masses greater than 1.8\,TeV.  
This arises from the use of the profile likelihood method which allows the data to constrain the systematic uncertainties using the full \mttbarreco\ distribution, thanks to the abundant data. 
The maximisation of the likelihood can change the central values of the nuisance parameters and their associated uncertainties.
In the region of \mttbarreco\ above 1.5\,\TeV, the background prediction as seen in figure~\ref{fig:mtt_merged_prefit} is slightly higher than the data, 
which would lead to the anticipation that the observed limits should be slightly better than the expected ones.
However, the central values of some nuisance parameters are significantly shifted in the fit. 
In the high-\mttbarreco\  region one of the dominant uncertainties is 
the high-\pt{} jet $b$-tagging extrapolation uncertainty, as detailed in section~\ref{sec:systematics}.
This uncertainty is reduced to half of its original size by the fit, and the central value is also shifted downwards by approximately one pre-fit standard deviation.
As this uncertainty is associated with reconstructed physics objects, it has correlated effects on the predictions of BSM signal and SM background.
The 1$\sigma$ change in the central value of this nuisance parameter reduces the acceptance of high-mass signals in the boosted selection considerably: 
approximately 25\% for a \Zprime\ with a mass of 2\,TeV. 
As a consequence, the observed upper limits on the cross-section obtained from the fit to the data are larger than the expected ones fitted to the nominal background estimates. 
Constraints from the fit are also observed in the nuisance parameters associated with other major systematic uncertainties such as the PDF, 
the shape of $W$+jets background, the energy and mass scales for both the large-radius and small-radius jets, $b$-tagging efficiencies, and \ttbar{} MC modelling.
These constraints are understood through their impact on the fitted \mttbarreco\ distributions.
A few of the other nuisance parameters also have their central values changed slightly by the fit, 
but their impact on the signal acceptance and hence cross-section limits are much smaller than the one detailed above. 

Studies were done to understand the origin of the shifted nuisance parameter associated with the high-\pt{} jet $b$-tagging extrapolation uncertainty, 
in particular to investigate its potential origin as a mis-modelling of the \ttbar{} background. 
Indeed an equally good fit can be obtained by introducing an additional {\it ad hoc} 
uncertainty in the modelling of the SM \ttbar{} background with a similar dependency on \mttbarreco. 
Such a mis-modelling would not reduce the signal acceptance and results in more stringent limits on the cross-section times \ttbar{} branching ratio.
Ultimately, no method was found to unambiguously ascertain the origin of this mis-modelling.
The limits presented here are those with a 1$\sigma$ change in the fitted value of the nuisance parameter associated with the high-\pt{} jet $b$-tagging extrapolation uncertainty, 
which correspond to more conservative results. 
An upper cross-section limit was generated using expected data constructed from a background-only model built from the nuisance parameters fitted to real data. It excludes a $g_{\mathrm{KK}}$ of width 15.3\% for masses less than  2.1\,TeV, compared to 2.2\,TeV generated using nominal background estimates.

\FloatBarrier
\section{Summary}
\label{sec:summary}

A search for heavy particles decaying to $t\bar{t}$  in the lepton-plus-jets decay channel was carried out with the ATLAS experiment at the LHC. The search uses data corresponding to an integrated luminosity of \totlumi\ of proton--proton collisions at a centre-of-mass energy of 8 TeV. No excess of events beyond the Standard Model predictions is observed in the $t\bar{t}$ invariant mass spectra.
 Upper limits on the cross-section times branching ratio are set for four different signal models: a narrow ($ \le 3\%$ width) \Zprime\ boson, a broad (15.3\% width) Randall--Sundrum Kaluza--Klein gluon, a Bulk Randall--Sundrum Kaluza--Klein graviton, and a narrow scalar resonance. 
Based on these results, the existence of a narrow leptophobic topcolour \Zprime\ in the range $0.4$~\TeV\ $<$~$m_{Z'} <$~\ZpexcludedMass\ is excluded at 95\% CL.
A broad Kaluza--Klein gluon with mass between 0.4~\TeV\ and \kkgexcludedMass\ is also excluded at 95\% CL. These results probe new physics at higher mass than previous ATLAS searches for the same signature, and the results are applicable to a broader variety of heavy resonances.

\section*{Acknowledgements}
\label{sec:acknowledgements}

We thank CERN for the very successful operation of the LHC, as well as the
support staff from our institutions without whom ATLAS could not be
operated efficiently.

We acknowledge the support of ANPCyT, Argentina; YerPhI, Armenia; ARC,
Australia; BMWFW and FWF, Austria; ANAS, Azerbaijan; SSTC, Belarus; CNPq and FAPESP,
Brazil; NSERC, NRC and CFI, Canada; CERN; CONICYT, Chile; CAS, MOST and NSFC,
China; COLCIENCIAS, Colombia; MSMT CR, MPO CR and VSC CR, Czech Republic;
DNRF, DNSRC and Lundbeck Foundation, Denmark; EPLANET, ERC and NSRF, European Union;
IN2P3-CNRS, CEA-DSM/IRFU, France; GNSF, Georgia; BMBF, DFG, HGF, MPG and AvH
Foundation, Germany; GSRT and NSRF, Greece; RGC, Hong Kong SAR, China; ISF, MINERVA, GIF, I-CORE and Benoziyo Center, Israel; INFN, Italy; MEXT and JSPS, Japan; CNRST, Morocco; FOM and NWO, Netherlands; BRF and RCN, Norway; MNiSW and NCN, Poland; GRICES and FCT, Portugal; MNE/IFA, Romania; MES of Russia and NRC KI, Russian Federation; JINR; MSTD,
Serbia; MSSR, Slovakia; ARRS and MIZ\v{S}, Slovenia; DST/NRF, South Africa;
MINECO, Spain; SRC and Wallenberg Foundation, Sweden; SER, SNSF and Cantons of
Bern and Geneva, Switzerland; NSC, Taiwan; TAEK, Turkey; STFC, the Royal
Society and Leverhulme Trust, United Kingdom; DOE and NSF, United States of
America.

The crucial computing support from all WLCG partners is acknowledged
gratefully, in particular from CERN and the ATLAS Tier-1 facilities at
TRIUMF (Canada), NDGF (Denmark, Norway, Sweden), CC-IN2P3 (France),
KIT/GridKA (Germany), INFN-CNAF (Italy), NL-T1 (Netherlands), PIC (Spain),
ASGC (Taiwan), RAL (UK) and BNL (USA) and in the Tier-2 facilities
worldwide.

\clearpage

\bibliographystyle{bibtex/bst/atlasBibStyleWithTitle.bst}
\bibliography{bib/moriond.bib}

\clearpage

\newpage 
\begin{flushleft}
{\Large The ATLAS Collaboration}

\bigskip

G.~Aad$^{\rm 85}$,
B.~Abbott$^{\rm 113}$,
J.~Abdallah$^{\rm 151}$,
O.~Abdinov$^{\rm 11}$,
R.~Aben$^{\rm 107}$,
M.~Abolins$^{\rm 90}$,
O.S.~AbouZeid$^{\rm 158}$,
H.~Abramowicz$^{\rm 153}$,
H.~Abreu$^{\rm 152}$,
R.~Abreu$^{\rm 30}$
% ...
\bigskip
\\
% ...
$^{11}$ Institute of Physics, Azerbaijan Academy of Sciences, Baku, Azerbaijan\\
% ...
$^{30}$ CERN, Geneva, Switzerland\\
% ...
$^{85}$ CPPM, Aix-Marseille Universit{\'e} and CNRS/IN2P3, Marseille, France\\
% ...
$^{90}$ Department of Physics and Astronomy, Michigan State University, East Lansing MI, United States of America\\
% ...
$^{107}$ Nikhef National Institute for Subatomic Physics and University of Amsterdam, Amsterdam, Netherlands\\
% ...
$^{113}$ Homer L. Dodge Department of Physics and Astronomy, University of Oklahoma, Norman OK, United States of America\\
% ...
$^{151}$ Institute of Physics, Academia Sinica, Taipei, Taiwan\\
$^{152}$ Department of Physics, Technion: Israel Institute of Technology, Haifa, Israel\\
$^{153}$ Raymond and Beverly Sackler School of Physics and Astronomy, Tel Aviv University, Tel Aviv, Israel\\
% ...
$^{158}$ Department of Physics, University of Toronto, Toronto ON, Canada\\
% ...
\end{flushleft}

\end{document}